\documentclass[final,5p,times]{elsarticle}

\usepackage{newtxtext,newtxmath}

\usepackage[T1]{fontenc}
\usepackage{ae,aecompl}

\usepackage{totcount}

\usepackage{graphicx}	
\usepackage{amsmath}	
\usepackage{amssymb}	
\usepackage{verbatim}
\usepackage[colorlinks=true,linkcolor=black, citecolor=blue, urlcolor=blue]{hyperref}

\newtotcounter{citnum} 
\def\oldbibitem{} \let\oldbibitem=\bibitem
\def\bibitem{\stepcounter{citnum}\oldbibitem}

\journal{New Astronomy Reviews}

\begin{document}
\newcommand{\torus}{\textsc{torus}}
\newcommand\procspie{Proc.~SPIE}
\newcommand\mnras{MNRAS}  
\newcommand\aap{A\&A}                
\let\astap=\aap                          
\newcommand\aapr{A\&ARv}             
\newcommand\aaps{A\&AS}              
\newcommand\apj{ApJ}                 
\newcommand\apjl{ApJ}                
\newcommand\rmxaa{Rev. Mex. Astron. Astrofis.} 
\newcommand\nat{Nature}              
\newcommand\aj{AJ}                   
\newcommand\na{New~Astron.}          
\newcommand\nar{New~Astron.~Rev.}    
\newcommand\jqsrt{J.~Quant. Spectrosc. Radiative Transfer} 

\let\apjlett=\apjl   
\newcommand\pasa{Publ. Astron. Soc. Australia}  
\newcommand\pasp{PASP}               
\newcommand\pasj{PASJ}               
\newcommand\araa{ARA\&A}             
\newcommand{\angstrom}{\text{\normalfont\AA}}
\newcommand\apjs{ApJS}               
\let\apjsupp=\apjs                       
\newcommand\planss{Planet. Space~Sci.} 
\newcommand\hii{H\,\textsc{ii} }

\begin{frontmatter}

\title{Synthetic observations of star formation and the interstellar medium}

\author[label1]{Thomas J. Haworth}
\address[label1]{Astrophysics Group, Imperial College London, Blackett Laboratory, Prince Consort Road, London SW7 2AZ, UK}


\ead{t.haworth@imperial.ac.uk}

\author[label2]{Simon C. O. Glover}
\address[label2]{Universit\"at Heidelberg, Zentrum f\"ur Astronomie, Institut f\"ur Theoretische Astrophysik, \\ Albert-Ueberle-Strasse 2, 69120 Heidelberg, Germany}

\author[label3]{\\ and Christine M. Koepferl}
\address[label3]{University Observatory Munich, Scheinerstr. 1, D-81679 Munich, Germany}

\author[label4,label5]{Thomas G. Bisbas}
\address[label4]{Department of Astronomy, University of Florida, Gainesville, FL 32611, USA}
\address[label5]{Max-Planck-Institut f\"ur Extraterrestrische Physik, Giessenbachstrasse 1, D-85748 Garching, Germany}

\author[label6]{James E. Dale}
\address[label6]{Centre for Astrophysics Research, Science and Technology Research Institute, University of Hertfordshire, Hatfield, AL10 9AB, UK}

\begin{abstract}
Synthetic observations are playing an increasingly important role across astrophysics, both for interpreting real observations and also for making meaningful predictions from models. In this review, we provide an overview of methods and tools used for generating, manipulating and analysing synthetic observations and their application to problems involving star formation and the interstellar medium. We also discuss some possible directions for future research using synthetic observations.

\end{abstract}

\begin{keyword}
synthetic observations --- Interstellar medium (ISM) --- stars: formation --- astrochemistry --- radiative transfer
\end{keyword}

\end{frontmatter}


\tableofcontents

\section{\Large Part 1. Introduction and Scope}
  \subsection{What is a synthetic observation?}
  \label{sec:opening}

In this paper, we review the growing field of synthetic observations, with a particular focus on star formation and the interstellar medium (ISM). It therefore makes sense to begin by describing what we mean by a ``synthetic observation''. The majority of astronomy and astrophysics has been driven by the detection and manipulation of photons. That we can learn so much from light alone owes much to the fact that its nature is fundamentally determined by the conditions (temperature, density, velocity and composition) of the emitting source and its interaction with any intervening material. The photons we detect carry this information with them, and from them we can infer much about the source of the photons and the foreground material. However, doing so can be a serious challenge. 
    
There are means of comparing a theoretical model with observations that do not directly account for the details of photon emission from the system. For example, consider a numerical simulation of a molecular cloud collapsing to form stars. In this instance, one might compare the population of stars formed in the model with the observed initial mass function (IMF) or binary fraction \citep[e.g.][]{2012MNRAS.419.3115B}. Although this comparison relates a theoretical and an observed quantity, it would not be considered a synthetic observation. Comparisons of this kind are extremely useful, but do have limitations, particularly if one wants to learn about the properties of the gas and dust, rather than the properties of discrete objects such as stars or planets.

In practice, if we want to compare observations of the gas and the dust in a particular astrophysical system (e.g.\ a molecular cloud) with theoretical predictions for the behaviour of that system, we need to concern ourselves with the details of photon emission and absorption.\footnote{In principle, one could also produce synthetic observations of non-photon signals, such as gravitational waves or direct detection of cosmic rays, but this is outside of the scope of this review.} Because observations are generally limited in terms of resolution and sensitivity, and moreover give us information on projected quantities (column densities, line-of-sight velocities etc.), rather than the full three-dimensional distributions, deriving information on the underlying physical state of the system can be challenging and can produce ambiguous results. It is therefore often much better to compute the expected observational properties of the theoretical model in a way that can be compared as closely as possible with real observations. Therefore, \textbf{\textit {we define a synthetic observation to be a prediction, based on theoretical models, of the manner in which a particular astrophysical source will appear to an observer}}. Most commonly, we are interested in observing sources in emission, and the majority of our review deals with this case. However, in some circumstances (e.g.\ extinction mapping of molecular clouds), it is more interesting to observe the source in absorption, by looking at its effect on the light from a background object or collection of objects and so theoretical predictions of absorption maps should also be considered to be synthetic observations.

\subsection{Why synthetic observations?}
As we have already mentioned, there are quantities such as the stellar IMF that can be generated by theoretical models and compared with observational data to test the accuracy and predictive capability of the model, without us ever having to generate a synthetic observation. One might therefore ask whether synthetic observations add value beyond providing an image that looks similar to the observed data (particularly when using the same colour scheme). However, we argue here that there are many important reasons why one might want to generate a synthetic observation. These include:

\begin{enumerate}
    \item{\textbf{Observational limitations/complexity.} Since the optical depth of the ISM is highly frequency dependent and different lines probe different ranges of density and temperature, observations of only one or a few tracers do not provide us with all of the information available within a theoretical model. Furthermore, real observations are subject to other processes such as noise, resolution constraints and interferometric effects that may be significantly different to the limitations of a numerical model. The ISM is also geometrically complex and evolves on timescales beyond the human experience. Observers are therefore limited to a restricted view of a single snapshot in time. Comparing theoretical models directly with observations  without accounting for these effects may therefore be misleading \citep[indeed in practice even our example of measuring the IMF or binary fractions directly from a model might yield different results to the values an observer would infer when studying the same system, e.g.][]{2017ApJ...849....2K}. For example, the filamentary nature of much of the dense ISM has only been resolved in recent years, particularly with {\it Herschel} \citep[e.g.][]{2010A&A...518L.102A, 2010A&A...518L.100M, 2011A&A...529L...6A, 2013A&A...550A..38P}, despite being a feature of numerical models for some time \citep[e.g.][]{1994ApJ...420..692M, 2005MNRAS.361....2C, 2009MNRAS.397..232B}. Synthetic observations of the {\it Herschel}, pre-{\it Herschel} and perhaps also future instrumentation view of star-forming clouds may all yield distinct characteristics. }
    
    \item{\textbf{Observing mode/time estimates.} Synthetic observations are an incredibly useful tool for estimating the observational parameters (e.g.\ choice of mode, time) required to detect a given system. In particular, for highly oversubscribed facilities such as ALMA, where only around 30\,per cent of proposals were successful in cycle 4\footnote{https://almascience.eso.org/documents-and-tools/cycle4/c04-proposal-review-process}, a synthetic observation demonstrating that $t$\, minutes of observing time using antenna configuration $y$ really is essential to achieve the science goals adds valuable weight to a proposal.  }
 
     \item{\textbf{Test observational diagnostics.} Synthetic observations allow us to test diagnostics that are used by observers in controlled situations where the exact conditions (temperature, density, velocity, etc.) of the model are known. For example, running observational pipelines on synthetic data and checking the accuracy of the inferences. They can also be used to improve existing techniques and to develop new diagnostics.}
    
    \item{\textbf{Bespoke models for direct interpretation of real observations.} Synthetic observations can be used to interpret real observations of specific systems (so called backwards modelling). This use is particularly prevalent in the protoplanetary disc community, where the structure and kinematics are readily parameterised
    \citep[e.g.][]{2014ApJ...788...59W}, but is also used by the star formation/ISM community. For example, the Orion Bar PDR \citep{2009ApJ...693..285P,2017A&A...598A...2A}, Sgr B2 \citep{1990ApJ...356..195L,2016A&A...588A.143S} and Taurus Molecular Cloud 1 (TMC-1; see e.g.\ \citealt{2016ApJS..225...25G}) are just a few targets that have been the subject of bespoke modelling. A challenge with such modelling is that there can be degeneracies between the model, observations and reality. Large numbers of calculations may therefore be required.  }
    
    \item{\textbf{Additional predictive power.} Synthetic observations offer numerical modellers additional predictive power. For example kinematic signatures that might be detected in emission line profiles, or the spatial distribution of different emission sources. }

    \item{\textbf{Astrochemical probes.} Astrophysical systems act as laboratories for astrochemists to study the microphysics in extremes of density and temperature. Probing the composition of such a system requires computing the emission properties to compare with observations. }
    
    \item{\textbf{Designing new telescopes.} Synthetic observations can model and accurately predict the capabilities of new and forthcoming instruments in any wavelength regime. Such application is vital for guiding the development of (potentially very expensive) new equipment and also for developing and testing data processing pipelines. }
\end{enumerate}

Some members of the community refer to a theoretical model as having some ingredients, with synthetic observations acting as the ``taste test'' between theory and reality \citep{2011IAUS..270..511G}. 

Clearly then, synthetic observations have a broad range of uses. Given their ever increasing application across astrophysics this usefulness is being recognised.

\subsection{Types of synthetic observation}
Our definition of a synthetic observation in \ref{sec:opening} is somewhat generous in the sense that it permits simple measures of the emissivity to be considered a synthetic observation. In practice there are many layers of complexity that can be woven into a synthetic observation (which will be explored in more detail throughout this review). For now we define three basic classes of synthetic observation, although as we will see, the boundaries between these classes are not always clear:

\begin{enumerate}
    \item{\textbf{A simple emissivity measure.} The output of some microphysically simple model describing an astrophysical scenario (be it parametric or dynamical) is assumed to have the correct conditions (density, temperature) to inform an analytic computation of the emissivity. {An example of this type is the free-free radio continuum emissivity, which is a simple function of the temperature and electron density only and so can be estimated without any detailed radiative transfer}. This class requires some simple consideration of the microphysics but comes at almost no additional computational cost. }
 
    \item{\textbf{Detailed microphysics and radiative transfer.} This involves a more robust computation of the abundance and level population of the emitting species. Typically, this requires a solver for the chemical state of the gas/dust, and/or a treatment of radiation transfer to determine the level populations or the dust temperature. Radiative transfer is also used to produce the resulting synthetic observable (e.g.\ an image or spectrum). {An example would be explicitly computing the ionisation state of species in an H\,\textsc{ii} region using a photoionisation code, which coupled with the temperature and density can be used to compute, for example, forbidden line emission. } Often, the detailed microphysical state of the system is determined in a post-processing step applied to a dynamical simulation with a more approximate model for the chemistry and for the thermal behaviour. 
    In reality the dynamics, microphysics and radiative transfer (and magnetic fields) are all interlinked and modelling this inter-dependency is at the frontier of modern capabilities \citep{2016PASA...33...53H}. }
 
    \item{\textbf{Inclusion of instrumentational effects.} {This class extends either of the prior two to include observational/instrumentational effects} such as noise, beam size convolution and interferometric filtering (e.g.\ see \citealt{2017ApJ...849....3K}).}
\end{enumerate}

\begin{figure}
    \includegraphics[width=9cm]{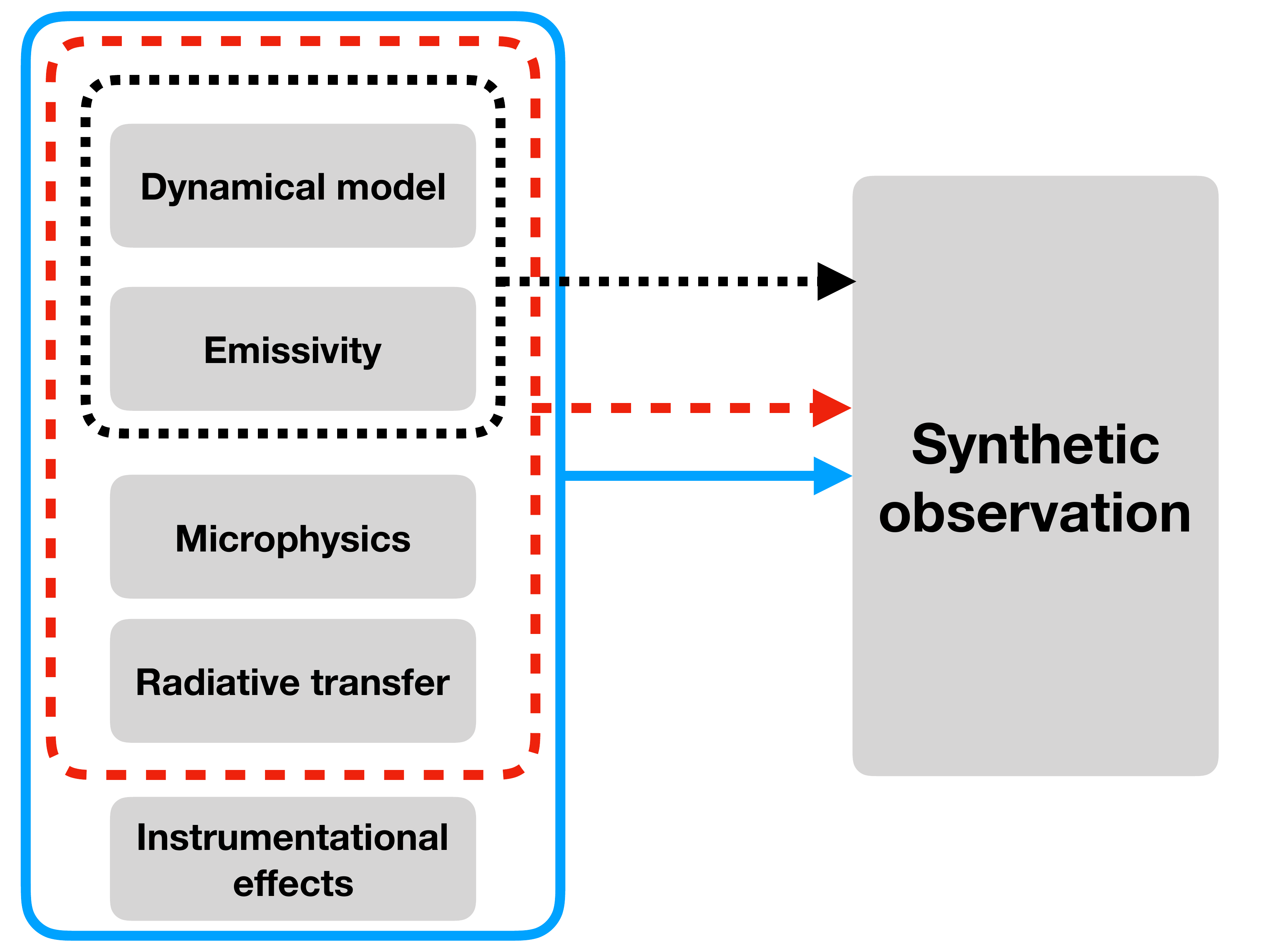}
    \caption{A schematic summary of the types of synthetic observations. At the most basic level, there is an estimate of the emissivity from some assumed (or computed) density, temperature and velocity structure. At the next level the composition and thermal structure are properly computed with radiative transfer and microphysics (e.g.\ a chemical network or ionisation structure solver) and the synthetic observation produced using radiative transfer. In the final level instrumentational (e.g.\ interferometric) effects are included. }
    \label{schematic}
\end{figure}

A schematic overview of these classes is given in Figure \ref{schematic}. Ultimately, although there are many layers of complexity behind modern synthetic observations, even the most sophisticated synthetic observations are still just theoretical representations of reality.

\subsection{Outline and scope of this review }
Given the almost universal value of synthetic observations in astrophysics, it is naturally a challenging topic to review. We therefore constrain our focus to problems in the context of resolved star formation and the ISM. We do not attempt to review the substantial body of work looking at star formation in an extragalactic context, nor do we propose to review stellar evolutionary models. We also do not consider supernovae, the circumstellar discs of material around young stellar objects or the outflows/jets from low mass stars. We leave assessment of other applications to future reviewers. With this in mind, the structure of the rest of the review is as follows. 

In section \ref{sec:nummeth} we briefly review the variety of numerical methods that can be used to produce synthetic observations in different observational and physical regimes. This includes computation of the physical conditions, radiative transfer and instrumentational processing. In section \ref{sec:apps} we review the application of synthetic observations to different topics related to star formation and the ISM in the Milky Way and its satellite galaxies, such as giant molecular clouds, turbulence, feedback, cores and filaments, and testing of observational diagnostics of the conditions. Finally in section \ref{sec:future} we discuss future prospects, including questions for synthetic observations to address, technical advances, increased dissemination of synthetic observations and the need for active communication in this interdisciplinary research field. We summarise a few key points in section \ref{sec:summary}.

\section{\Large Part 2: Numerical methods and resources }
\label{sec:nummeth}

To generate a synthetic observation, we need to know how the emitting or absorbing material is distributed in space. At a minimum, this means that we need a model of the density of the gas and its chemical composition, as well as the quantity and properties of the dust associated with it. In the case of emission, we also need to know the temperature of the emitting material. For gas-based tracers, this is generally the same as the kinetic temperature of the gas, while for dust-based tracers it is usually the dust grain temperature that is relevant. In addition, for observations of atomic or molecular lines, we also need a model for the velocity structure. With these {known over the entire domain}, it is possible to solve the {(non-local)} equation of radiative transfer to compute the radiation received by an observer at some external position. In this section of the review, we discuss methods by which the above are achieved. Note that we do not intend to describe techniques in detail. Rather we intend to provide an overview of modelling regimes and the tools available to study them.

\subsection{The basis for a synthetic observation: computing the density and velocity structure}
\label{sec:dynamics}
The density, velocity (and possibly the thermal) structure can be informed either by some form of parametric modelling (e.g. analytic or semi-analytic models) or by the output from full dynamical simulations. Parametric models have the great advantage that they are usually very quick to compute, permitting one to study large parameter spaces without a need for extensive computational resources. They are therefore often a good choice if one is trying to reproduce the properties of a particular astronomical object (so called ``backward modelling''). Their simplicity is also useful for helping to develop insight into the importance of different physical processes in a particular situation. Their main disadvantage is that they can often be {\em too} simple, particularly when it comes to the assumed geometry of the system and that they provide us with no information on the history of the system in question, unlike a dynamical model. Examples of useful parametric models include:
\begin{enumerate}
    \item{Bonnor-Ebert spheres (polytropic solutions to the Lame-Enden equation with $n=1$) that can be representative of cores \citep[e.g.][]{2011A&A...535A..49S, 2015A&A...582A..48S, 2017A&A...601A.113S} or globules \citep[e.g.][]{2005AJ....130.2166K,2015MNRAS.446.1098H}}
    \item{Parametric descriptions of filaments using, for example, a Plummer function \citep[e.g.][]{1911MNRAS..71..460P, 2008MNRAS.384..755N, 2016A&A...592A..90C}}
    \item{The expansion of photoionised H\,\textsc{ii} regions is well described in terms of the Str\"{o}mgren radius \citep{1939ApJ....89..526S}, the famous \textit{Spitzer} equation \citep{1978ppim.book.....S} and modern refinements of that description \citep{2006ApJ...646..240H,2012RMxAA..48..149R,  2015MNRAS.453.1324B,2018arXiv180509273W}}
    \item{Simple analytical models exist for energy-dominated and momentum-dominated stellar winds \citep{1977ApJ...218..377W,2013ApJ...765...43S}, while the more complicated details of the transition between phases can be treated semi-analytically \citep{2017MNRAS.470.4453R}}
    \item Although discs are beyond the scope of our review, there are simple and extremely useful parametric models of circumstellar matter \citep[e.g.][]{2014ApJ...788...59W}. 
\end{enumerate}

Dynamical models entail solving either the hydrodynamical or the magnetohydrodynamical fluid equations on some discretization of space, be it a fixed or moving grid, a set of Lagrangian particles or some other formulation. These models are often three-dimensional, although 1D and 2D models are also common in cases where the system of interest can be approximated as being either spherically or axially symmetric. A full discussion of the different methods that exist for solving the fluid equations and their various strengths and weaknesses is a review (or multiple reviews) in its own right \citep[e.g.][]{2010ARA&A..48..391S, 2015ARA&A..53..325T} and so in this review we will restrict ourselves to a few general comments.

One of the first decisions that must be made when constructing a dynamical model of a system is the amount of physics to include. In addition to pure hydrodynamics, many other physical processes can have important effects on the dynamical evolution of the system, including gravity, radiation (either directly, via radiation pressure, or indirectly, via radiative heating, photoionisation etc.), cosmic rays, chemistry, and magnetic fields, many of which are to some extent interlinked. Although good numerical treatments of all of these processes now exist, each has its own computational cost, which can often be large. Therefore, given a fixed amount of computing time, the more physical processes that are included, the smaller the number of resolution elements (e.g.\ grid cells, Lagrangian particles, etc.) that can be simulated. Consequently, there is always a trade-off between high numerical resolution (which minimizes the errors arising from the discretization of space) and high physics content. Synthetic observations can help us to manage this trade-off by showing us which physical processes are important for producing observations that look like the real ones, which are inessential.

A further complication when modelling the ISM is that dust and gas are not necessarily dynamically coupled. This effect has long been included in models of proto-planetary disks \citep[see e.g.][]{1996A&A...309..301S}. However, in models of larger-scale objects, such as cores, filaments or entire molecular clouds, it is typically not accounted for, although in recent years there has been some progress in this regard. For example, in H\,\textsc{ii} regions, \citet{2014A&A...563A..65O, 2014A&A...566A..75O} solved for steady state ``dust wave'' locations by balancing thermal pressure gradients and the radiation pressure force. \citet{2015MNRAS.449..440A, 2017MNRAS.469..630A} also studied H\,\textsc{ii} regions by solving the fully decoupled dust-gas dynamical equations in 1D. \cite{2011ApJ...734L..26V} have included decoupled dynamics for models of bow shocks about a fast-moving red supergiant.  Other studies include decoupled dust-gas dynamics in turbulent media \citep{2017MNRAS.471L..52T}, Bondi accretion \citep{2008ApJ...678.1099B} and collapse to the first hydrostatic core \citep{2017MNRAS.465.1089B}.

\subsection{Composition and thermal structure}
\label{sec:composition}
Given a means of computing the density and velocity structure of the system, the next consideration is how to model its thermal structure and chemical composition. This is a non-trivial problem, potentially requiring substantially different physics in different radiation and kinematic regimes. However, it is of central importance to producing synthetic observations, since the composition and thermal structure have a large influence on the emission properties of the system and potentially also its absorption properties.

As in the case of the densities and velocities, there are two main avenues open to us when determining the composition and temperature structure of the system: we can use a parametric model, or a more detailed simulation (which may or may not account for the dynamical evolution of the system). In this review, we will concentrate on the latter possibility, as even in cases where parametric models are used, the values adopted are often informed by the output of more detailed models (albeit ones with a reduced dimensionality). 

The next choice we face is whether to solve for the temperature structure, the chemical composition, or both together. The most prudent choice here is to solve for the temperature and composition simultaneously, since the two are often strongly coupled: chemical rate coefficients can vary exponentially with temperature and small changes in the chemical composition can have a large impact on the ability of the gas to cool. However, in some cases, where we know already that the coupling is weaker, it can be simpler to solve for the thermal structure and composition separately. In particular, if it is the temperature structure of the dust that we are interested in -- and not the gas -- then this is generally fairly insensitive to the chemical composition of the gas, and so dust-only calculations are reasonably common (see also Section~\ref{dustonly}).

Another important consideration is whether the radiation field plays an important or dominant role in determining the composition and temperature of the system. In scenarios where the radiation field is unimportant {(or at least of secondary importance, as in cold shielded zones)} the problem of determining the temperature and chemical composition {can be approximated as} a purely local one: the values at any given point in space do not directly depend on those at other points in space. This is a great advantage when trying to model such a system computationally, as it allows the problem to be parallelized extremely easily. On the other hand, in scenarios where the radiation field is important, the problem of determining the temperature and composition becomes inherently non-local, making it more difficult to handle numerically.

In the field of star formation and the ISM, we can identify five main compositional and thermal regimes, differing in the role played by radiation and hydrodynamical effects. 

In kinematically quiescent gas that is well-shielded from external sources of radiation, the temperature is low ($T \sim 10$~K) and the gas is typically highly molecular, with a chemistry that is dominated by ion-neutral and grain surface reactions. Dust in this regime is also cold and will often be coated with a thick layer of ice. 

If the gas is opaque to extreme ultraviolet (EUV) photons with energies $E > 13.6 \: {\rm eV}$, but is not well-shielded from lower energy EUV and far ultraviolet (FUV) photons, then we have what is known as a photodissociation region\footnote{or as a photon-dominated region} \citep[PDR][]{1985ApJ...291..722T, 1999RvMP...71..173H}. Gas in a PDR is warmer and less molecular than in the cold, dark regime, with the detailed behaviour controlled largely by the ratio of the UV photon flux to the gas density \citep[see e.g.][]{1999ApJ...527..795K,2014ApJ...790...10S}. Dust in a PDR is also warmer and the individual grains may have lost some or all of their ice coatings.

Alternatively, if the gas is well-shielded against EUV and FUV photons but is illuminated by a strong flux of X-rays, we have what is known as an X-ray dominated region (XDR). X-rays -- particularly hard X-rays -- can penetrate much larger column densities of gas and dust than UV photons, and so can provide heating and ionization deep within dense, high extinction molecular clouds. Gas temperatures in XDRs vary greatly, depending on the strength of the X-ray flux, but often occupy a similar range of values to the gas in PDRs \citep{1996ApJ...466..561M}. However, thanks to the ionization produced by the cosmic rays, there are clear chemical differences between the two regimes \citep[see e.g.][]{2005IAUS..235P..58M}. Regions illuminated by strong fluxes of cosmic rays (sometimes known as cosmic-ray dominated regions, or CRDRs) have also attracted attention in recent years \citep{2011MNRAS.414.1705P,2015ApJ...803...37B,2017ApJ...839...90B}, but behave in a similar fashion to XDRs.

If the gas is not shielded against photons with energies $E > 13.6 \: {\rm eV}$ and is also exposed to a large enough flux of these photons, then it will not only be photodissociated, it will also become photoionized. Typically, the mean free path of ionizing photons in the ISM is relatively small, so much photoionized gas is confined to regions close to bright UV sources (e.g.\ O stars), forming distinct \hii regions. However, some ionizing photons can propagate to large distances, particularly when travelling out of the Galactic plane, and so there is also a component of diffuse photoionized gas in the ISM \citep[see e.g.][]{2003ApJS..149..405H}. The chemical composition in photoionized regions is largely set by the balance between photoionization and recombination, and depends on the flux of ionizing photons, the density of the gas and the hardness of the ionizing spectrum. These factors also influence the gas temperature, although this is typically around $10^{4}$~K, to within a factor of two. Dust in photoionized gas is typically warmer than that in PDRs and grain temperatures can reach values as high as $\sim 100$~K \citep{2012ApJ...749L..21S}.

Finally, if some of the gas in the system is moving supersonically, shocks can be present. The impact that these have on the temperature structure and composition depend on the strength of the shocks and the fraction of the volume that is affected by them. In some cases (e.g.\ fast stellar winds, supernova remnants), shocks can heat the bulk of the gas to temperatures of $10^{6} \: {\rm K}$, far higher than can be reached by radiative heating alone. Even in cases where the shocks are less extreme (e.g.\ dissipation of turbulence in molecular clouds), they often result in excitation patterns that are clearly distinct from those in PDRs or \hii regions and they may also leave clear chemical imprints \citep[e.g.][]{2001A&A...370.1017V, 2004A&A...415.1021J, 2010A&A...518L.112C, 2012A&A...548A..77G, 2012ApJ...761...74A, 2013A&A...558L...2W}. Very strong shocks also have a profound effect on dust grains, reducing their mass by sputtering or even destroying them entirely \citep[see e.g.][]{1979ApJ...231..438D,1996ApJ...469..740J}.

In practice, there is not always a sharp boundary between these different regimes, so to some extent the distinction exists because in the past, numerical models were developed to study each of these regimes in isolation. As we will discuss below, in recent years new attempts at modelling all regimes simultaneously have been developed.

\begin{figure*}
    \centering
    \includegraphics[width=16cm]{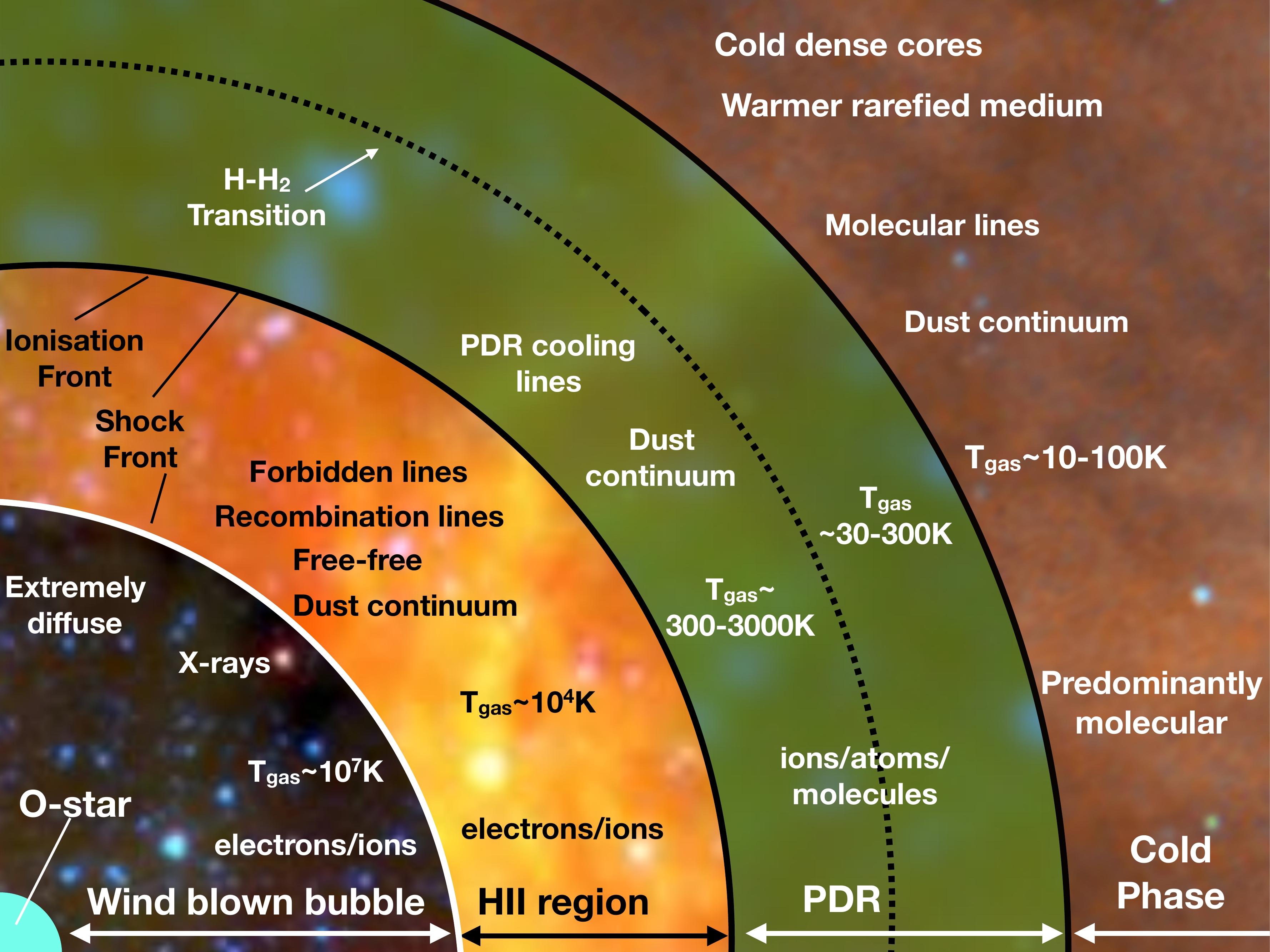}
    \caption{A schematic overview of the different microphysical regimes in the medium around an O star. Particles are driven away from the stellar atmosphere in a high velocity wind. This excavates a very hot, but also very diffuse bubble. Beyond this bubble, but still in close proximity to the star, there is strong EUV irradiation and the gas is predominantly photoionised. This higher pressure ionised region expands, resulting in a shock front that may affect the composition. FUV radiation can penetrate beyond the extent of the photoionised gas, creating a PDR. Where the medium is optically thick to the exciting O star radiation, we enter the cold phase regime where gas is either in cold dense clumps or the warmer more rarefied/turbulent medium. Note that the size of the different regions shown in the figure is not to scale {and that the dust temperature is generally decoupled from the gas temperature.}} 
    \label{fig:OstarRegimes}
\end{figure*}

A schematic of different microphysical regimes in the medium about an O star, including representative temperatures, composition and emission is given in Figure \ref{fig:OstarRegimes}.

In the rest of this section, we summarise the main components and methods for computing the composition and thermal structure of different astrophysical media, including the cold phase, PDRs, XDRs, photoionised gases and the dynamically-dominated regime. We begin with the question of where to obtain the necessary microphysical data. 

\subsubsection{Sources of microphysical data}
\label{sec:data}
Modelling of the species abundances, ionisation fractions, level populations, etc., of astrophysical gases relies on having the appropriate microphysical data available. For example, for molecular line emission, even if we know the molecular abundance, we still need data on the energies of the different rotational and vibrational levels, and the radiative and collisional rate coefficients linking different levels. Similarly, for atomic line emission, we need level energies and transition rate coefficients. To model the atomic and molecular abundances themselves, we need chemical reaction rates (both in the gas phase and on the surface of grains, the latter of which can be extremely complex). Finally, to model dust emission, we need data on the optical properties of different types of grains. To this end we are indebted to the efforts of the laboratory chemists who provide such data. The microphysical data procured through laboratory experiments and complex theoretical models  is now readily available online across a series of databases. Examples include:

\begin{itemize}

\item The Leiden Atomic and Molecular Database \citep[LAMDA\footnote{\url{http://home.strw.leidenuniv.nl/~moldata/}}, ][]{2005A&A...432..369S} includes  energy levels, statistical weights, Einstein A-coefficients and collisional rate coefficients for a series of 4 atomic and 36 molecular species. Linked to this database is the \textsc{radex} tool, which permits users to quickly carry out simple statistical equilibrium calculations \citep{2007A&A...468..627V}. 

\item The UMIST\footnote{\url{http://udfa.ajmarkwick.net/}} database for astrochemistry \citep{2000A&AS..146..157L, 2007A&A...466.1197W, 2013A&A...550A..36M} currently lists over 750 atomic and molecular species, between which it details many thousands of reactions and the corresponding rate coefficients. The online database also provides the time dependent evolution of species abundances for an example dark cloud model from \cite{2013A&A...550A..36M}. 

\item KIDA\footnote{\url{http://kida.obs.u-bordeaux1.fr/}} \citep{2012ApJS..199...21W} is another large database of atomic and molecular reactions (including gas-grain processes) which also includes a library of reaction networks that can be downloaded for use. These networks at the least provide an excellent starting point and can probably be used ``straight out of the box'' for many applications  (each network comes with a description of the environment in which it is applicable), though caution is required since a given network will not be universally applicable. KIDA also hosts the \textsc{nahoon} astrochemistry code \citep{2014ascl.soft09009W}.

\item{The NASA Ames PAH database\footnote{\url{http://www.astrochem.org/pahdb/}}} contains 700  computational and 75 experimental PAH infrared spectra. It also includes a number of online and offline tools  \citep{2010ApJS..189..341B, 2014ApJS..211....8B}

\item {The Basecol database\footnote{\url{http://basecol.obspm.fr/}}} contains a large number of ro-vibrational collisional (de-)excitation rate coefficients \citep{2013A&A...553A..50D}.

\item The Jena Database of Optical Constants for Cosmic Dust\footnote{\url{http://www.astro.uni-jena.de/Laboratory/OCDB/index.html}} (DOCCD) contains a comprehensive listing of optical constants for astrophysical grains. This includes silicates, oxides, sulfides and carbonaceous grains. 

\item{Heidelberg - Jena - St.Petersburg - Database of Optical Constants\footnote{\url{http://www2.mpia-hd.mpg.de/HJPDOC/}} (HJPDOC) also contains a large databse of optical constants for astrophysical dust grains \citep{1999A&AS..136..405H, 2003JQSRT..79..765J}}. 

\item The Cologne Database for Molecular Spectroscopy \citep[CDMS\footnote{\url{https://www.astro.uni-koeln.de/cdms}}][]{2001A&A...370L..49M, 2005JMoSt.742..215M,2016JMoSp.327...95E} contains spectroscopic data (transition frequencies, quantum numbers etc.) for many hundreds of molecules. 

\end{itemize}

Many of the above (and more) are tied into the ``Virtual Atomic and Molecular Data Centre'' (VAMDC\footnote{\url{http://www.vamdc.org/}}) consortium, which consolidates a broad range of atomic and molecular data resources.  

The sheer amount of microphysical data available, combined with the large number of groups working to provide such data, makes it challenging to give laboratory astrochemistry the credit it deserves. Without such data, compositional modelling and by extension synthetic observations would be impossible. The need to acknowledge this importance is, therefore, becoming more recognised. Recent efforts to automatically produce citation lists from reaction networks may help to improve the situation. For example, when running a code a bibliography file is produced with all of the appropriate papers that should be cited.

Good reviews of astrochemistry and the current availability of microphysical data can be found in e.g.\ \citet{2011IAUS..280..449V}, \citet{2012A&ARv..20...56C}, \citet{ObergReview}, \citet{Cuppen2017} and \cite{2017arXiv171005940V}.

\subsubsection{Cold, quiescent regime}
\label{cold-dark}
In kinematically quiescent environments, where there is little to no incident radiation, either because of an absence of sources or because the gas is completely shielded by dust absorption, some aspects of modelling the composition and temperature become simpler. In particular, in this regime, the chemical composition at any particular point in the gas depends only on the current conditions and previous history of the gas at that point and not on the current conditions elsewhere in the gas. 

Unfortunately, this simplicity is largely offset by the fact that the chemistry of these regions can be much more complicated than in irradiated regions. Despite the low temperatures, many gas-phase and grain surface reactions can proceed, and the absence of efficient destruction of molecules by photodissociation and/or collisional dissociation means that many different chemical species can reach abundances at which they are potentially observable. A further complication is that in these conditions, the chemical timescale for some species -- i.e.\ the characteristic timescale on which the abundance of that species reaches equilibrium -- can be long, so we cannot always assume that the chemistry has reached an equilibrium state. This is unfortunate, as when the chemistry is in equilibrium, the composition depends purely on the local conditions, but when it is out of equilibrium, it depends also on the previous history of the gas. Consequently, if we are dealing with equilibrium chemistry, we can determine the composition simply by considering the state of the gas at a single point in time, e.g.\ by post-processing a single snapshot from a hydrodynamical simulation. On the other hand, if we are dealing with non-equilibrium chemistry, we need some method for following the chemical evolution of the gas over time.

A further complication comes from the fact that in these conditions, grain surface processes can play an important role in determining the chemical composition. They influence the chemistry in two ways. First, they provide a means to form molecules than cannot be easily formed in the gas phase (e.g.\ H$_{2}$, methanol). Second, the formation of ice mantles on the grains alters the gas-phase chemical composition, as chemical species with high binding energies to the grain (e.g.\ H$_{2}$O, CO) freeze out much sooner than species with low binding energies (e.g.\ NH$_{3}$).

Several different approaches can be used to model the chemistry in this environment, depending on the level of accuracy required. The most straightforward approach is to assume chemical equilibrium. Given an appropriate chemical network, derived from e.g.\ the UMIST or KIDA databases or provided by some other package such as KROME \citep{2014MNRAS.439.2386G} or Grackle \citep{2017MNRAS.466.2217S}, the assumption of equilibrium reduces the problem of determining the chemical composition to one of solving a large set of linear equations. This can be time-consuming, but is well within the capabilities of modern computers, particularly if one takes advantage of the inherent sparsity of the system \citep[see e.g.][]{2013MNRAS.431.1659G}. The main disadvantage of this approach is that it will produce incorrect results if the chemistry is not actually in equilibrium.

In conditions where the gas is not in chemical equilibrium, it is necessary to find some way of following the evolution of the chemical composition over time. The most obvious approach is to solve the full set of evolution equations for the different chemical species as a function of time. Suppose we have some chemical species M with partial mass density $\rho_{\rm M}$. The evolution in time and space of $\rho_{\rm M}$ is described by an equation of the form:
\begin{equation}
\frac{\partial \rho_{\rm M}}{\partial t} + \nabla \cdot (\rho_{\rm M} \mathbf{v}) = C_{\rm M} - D_{\rm M},  \label{chem_evol}
\end{equation}
where $\mathbf{v}$ is the velocity field, and $C_{\rm M}$ and $D_{\rm M}$ are functions representing the formation and destruction of species $M$ by chemical reactions. In general, $C_{\rm M}$ and $D_{\rm M}$ will depend on the partial densities of some or all of the other chemical species present in the gas as well as the gas temperature, and potentially also the dust temperature and density. Since Equation~\ref{chem_evol} is essentially just a continuity equation with source and sink terms, when treating astrophysical flows numerically, it is common to split it into separate advection and reaction terms 
\begin{eqnarray}
\frac{\partial \rho_{\rm M}}{\partial t} + \nabla \cdot (\rho_{\rm M} \mathbf{v}) & = & 0, \\
\frac{{\rm d} \rho_{\rm M}}{{\rm d} t} & = & C_{\rm M} - D_{\rm M},
\end{eqnarray}
which are solved consecutively during each timestep. Solution of the advection term is straightforward, as it can be treated in the same way as the continuity equation for the total density (although care must be taken to ensure that the advection is carried out in a way which is consistent with the conservation laws for charge and the total abundances of the different chemical elements involved; see \citealt{1999A&A...342..179P} and \citealt{2010MNRAS.404....2G} for more details). 

In principle, the reaction equations are also simple to deal with: they form a set of simultaneous, first-order ordinary differential equations (ODEs) and many different algorithms exist for solving such sets of ODEs. In practice, however, the fact that the ODEs are typically stiff -- i.e.\ that they have a wide range of different chemical timescales associated with them -- presents a computational challenge. To preserve numerical stability, it is generally necessary to solve the stiff chemical rate equations using an implicit solver. Doing so, typically involves one or more matrix inversions; if we have $N$ equations, the total computational cost scales as ${\rm O}(N^{3})$. Since $N$ can easily be a few hundred or more even for a pure gas-phase chemical network \citep[see e.g.][]{2013A&A...550A..36M}, it means that the cost of solving the full set of equations is high. 

The high cost of solving the full chemistry network means that it is currently impractical to include such a network in a high-resolution 3D hydrodynamical model, as the required computational resources are simply too large. We are therefore faced with a choice. One option is to continue to use a full chemical network but to couple it to a 1D or 2D hydrodynamical model, in order to reduce the overall computational cost \citep[see e.g.][]{2008ApJ...674..984A}. This is a viable option when the symmetry of the problem allows it, but obviously is not useful when one is dealing with intrinsically three-dimensional flows, such as the turbulent flow of gas within a molecular cloud. 

The other option is to continue to use a 3D hydrodynamical model, but to simplify the treatment of the chemistry. The basic idea here is that we are often only interested in the abundances of one or a few chemical species, rather than the full set included in a modern astrochemical model. Therefore, if we can construct a reduced network capable of modelling the abundances of this small set of species accurately, we can simply include this reduced network in place of a full network. This reduced network can be constructed by including only a subset of the full range of possible chemical reactions and also by combining similar reactions and similar chemical species into artificial meta-reactions and meta-species (see e.g.\ the \citealt{1999ApJ...524..923N} reduced network, which uses the meta-species CH$_{x}$ to represent the hydrocarbons CH, CH$_{2}$, CH$_{3}$, CH$^{+}$ and CH$_{2}^{+}$). If the number of chemical species included in the reduced network is small, then this can result in large computational savings, thanks to the ${\rm O}(N^{3})$ cost of solving the chemical rate equations -- for example, if we can get away with including only 20\% of the species in the full network, the cost drops by a factor of over 100.

Examples of reduced networks designed for modelling molecular cloud formation can be found in \citet{1997ApJ...482..796N,1999ApJ...524..923N}, \citet{2010MNRAS.404....2G}, \citet{2010MNRAS.402.1625K}, \textbf{\cite{2015MNRAS.449.2643B}} and \citet{2017ApJ...843...38G}. Typically, the construction of these reduced networks is something of a dark art, but the general approach is to start with a large network and to remove components until the quantity of interest varies by more than a certain threshold (e.g.\ the resulting CO abundances becomes inaccurate at the 10\% level). Note, however, that the desired level of accuracy will generally only be achieved within the particular range of physical conditions considered when constructing the reduced network. Using a reduced network in physical conditions very different from those for which it was designed can lead to very large errors in the resulting abundances. Automated methods for deriving reduced networks from full networks exist and are widely used in some field, such as combustion chemistry \citep[see e.g.\ the review by][]{GM11}, but so far there have only been a few attempts to apply these methods to astrochemistry \citep{2002A&A...383..738R,2002A&A...381L..13R,2003A&A...399..197W,2004A&A...417...93S,2012MNRAS.425.1332G,2013MNRAS.431.1659G,2015MNRAS.451.2082G}.

The use of reduced chemical models in 3D hydrodynamical simulations is now becoming relatively widespread 
\citep[e.g.][]{2010MNRAS.404....2G,2012MNRAS.421..116G,2014MNRAS.439.2386G,2014MNRAS.441.1628S,2014MNRAS.444..919P,2015MNRAS.454..238W,2016MNRAS.459L..11S}. Most current treatments focus on following the subset of chemical species that are directly involved in determining the cooling rate of the gas in the ISM, namely H$^{+}$, H, H$_{2}$, C$^{+}$, C, O and CO (although some very simple treatments neglect C; see e.g.\ \citealt{1997ApJ...482..796N}), but models for studying other aspects of the gas chemistry are starting to become more common \citep[e.g.][]{2017arXiv170904013K}. 

The set of coolants followed in most of these models includes some important observational tracers, such as H, C$^{+}$ or CO. If one is interested in these particular tracers, the results of the simulations can be used directly to produce synthetic emission maps for them. However, if we are interested instead in tracers that are usually not included in this reduced set (e.g.\ N$_{2}$H$^{+}$, CS, H$_{3}^{+}$), then some form of post-processing of the simulation results is necessary. A promising approach here involves the use of Lagrangian tracer particles to follow changes in density and temperature over time. Information from these tracer particles can then be used to evaluate the chemical evolution of the fluid element represented by the tracer particle. This is particularly simple in the case of SPH simulations, as the SPH particles themselves can act as the required Lagrangian tracers, but in principle the same technique can also be used in grid codes. Recent examples of this approach can be found in \citet{2015EAS....75..391S} and \citet{2017MNRAS.472..189I}. One disadvantage of this method is that the sampling of regions of interest by the tracer particles can be poor. Douglas et al.~(in prep) have therefore developed a technique in which approximate paths for fluid elements can be computed from a series of data dumps (providing that they are closely spaced in time), without the need for tracer particles, mitigating this sampling issue.

Finally, as well as the chemical composition, it is also necessary to solve for the temperature of the gas. In the cold, quiescent regime, this is relatively simple. The main sources of heating are cosmic rays, adiabatic compression of the gas and in dense regions, collisional transfer of energy from dust grains to the gas (if $T_{\rm dust} > T_{\rm gas}$; otherwise, this process is responsible for cooling the gas). This is balanced by cooling due to atomic and molecular line emission \citep[e.g.][]{1993ApJ...418..263N,1995ApJS..100..132N}, which is often modelled using a large velocity gradient \citep[LVG;][]{1960mes..book.....S} approximation. Because the temperature is very low in this regime, the gas dynamics and chemical composition are often fairly insensitive to its exact value, meaning that even fairly crude approximations (e.g.\ the assumption of an isothermal equation of state) can often be adequate. Note, however, that this is not true if the observational tracer we are interested in is highly sensitive to temperature.

\subsubsection{PDRs and XDRs}

\begin{figure}
    \includegraphics[width=0.95\linewidth]{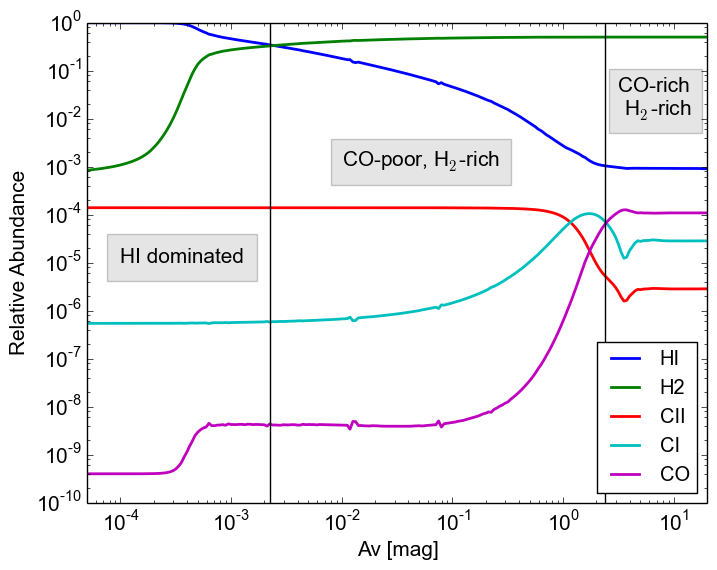}
    \caption{ Milky-Way typical relative abundances of species for a one-dimensional uniform-density distribution with total H-nucleus number density of $n=10^3\,{\rm cm}^{-3}$ irradiated by $\chi/\chi_0=1$ \citep[normalized according to][]{1978ApJS...36..595D}. The cosmic-ray ionization rate is taken to be $10^{-17}\,{\rm s}^{-1}$ and the metallicity is solar. The H{\sc i}-to-H$_2$ transition (blue and green lines on top) and the carbon phases (red, cyan, magenta) are shown. Note that there is a range of visual extinction in which the cloud is H$_2$ rich but CO-poor. Changing any of the assumed quantities would shift these transitions accordingly and would extend or shrink the CO-poor region (see \S\ref{ssec:traceh2} for techniques measuring the molecular gas).}
 \label{fig:pdr}
\end{figure}

If the gas is well-shielded from ionizing photons, but not from far-ultraviolet (FUV, 912$ <\lambda < $2400\AA) radiation, then we are in the PDR regime. Gas in this regime is primarily atomic in regions where photodissociation dominates over other chemical processes, or largely molecular, in regions where the radiation field is less important but not negligible. Consequently, a key thermal and chemical transition in the PDR regime is the point at which the gas moves from being predominantly atomic to predominantly molecular hydrogen (the H-H$_2$ transition).  Figure~\ref{fig:pdr} shows a typical PDR regime representing conditions commonly found in Milky Way.
 
PDRs are important both because of the rich array of emission lines that they produce and because the temperatures within them can be dynamically significant (ranging from tens to a few thousand Kelvin). Since far-ultraviolet (FUV) photons have a lower optical depth than extreme ultraviolet (EUV) photons, PDRs are often found at the boundary of H\,\textsc{ii} regions. Often these are associated with massive O and B type stars. However, an FUV source need not always be a star. FUV photons (and also X-rays) are also produced in extreme environments, such as the accretion disks around supermassive black holes in active galactic nuclei (AGN), or during a supernova explosion, creating extended PDRs. Studying the emission from PDRs is therefore crucial for revealing the physical and chemical structure of the dense ISM and for determining the thermal balance in local objects and distant galaxies. Indeed, most of the mass in the ISM of our Galaxy can be considered as being in a PDR \citep{1999RvMP...71..173H}.

Many of the considerations of the previous section also apply to modelling the chemistry of PDRs, although the fact that chemical timescales are often short in highly irradiated regions means that an equilibrium treatment is frequently adequate. However, when modelling a PDR, we face two additional complications that we do not face in the cold, quiescent regime. First and foremost, we need to model the attenuation of the UV field through the PDR and the fact that this depends on the chemical composition of the gas (due to e.g.\ H$_{2}$ self-shielding or shielding due to C photoionization) means that the problem is no longer purely local: the chemical composition at one point in the PDR is coupled to the composition at many other points in the PDR. Second, because the gas temperature is higher, its value has more impact on the dynamical and chemical behaviour of the gas and the strength of the emitted radiation. However, the temperature also depends strongly on the chemical composition of the gas, meaning that it is often necessarily to solve iteratively for the local UV field strength, temperature and composition until one can arrive at a self-consistent solution.

These considerations mean that solving for the 3D structure of a PDR can be  computationally expensive. For this reason, most numerical codes used to model PDRs are confined to a 1D geometry. Most of these models consider a slab of material that is irradiated from one direction by an FUV field. The slab is often considered to be semi-infinite, so that photons can only escape in a direction opposite to that of the incoming radiation. Alternatively, some models \citep[e.g.][]{2006A&A...451..917R} consider a spherical geometry and an isotropic FUV field. In either case, these models have the advantage that a range of different UV fields and density structures can be modelled relatively quickly.  

{Most 1D codes also assume that the medium is smooth or uniform, though the clumpy, fractal, nature of the ISM can be approximated by considering ensembles of 1D calculations. For example, \cite{2008A&A...488..623C} approximated the clumpy nature of the global Milky Way to compare with COBE FIR data, using a collection of 1D spherical models that sample from a physically motivated clump mass--radius relationship.   }

A few codes are also available that adopt a two-dimensional geometry, which is particularly useful for modelling protoplanetary disks \citep{2009A&A...501..383W, 2010A&A...510A..18K, 2015ApJ...808...46M}.

Nevertheless, in recent years there have been a number of developments in the three-dimensional modelling of PDRs. \citet{2007ApJS..169..239G,2007ApJ...659.1317G} and \citet{2010MNRAS.404....2G} carried out hydrodynamical simulations of turbulent gas in the PDR regime, using a six-ray approximation to model the attenuation of the external radiation field \citep[see also][]{1997ApJ...482..796N}. In this approximation, the absorption of UV radiation by H$_{2}$ and CO molecules and by dust is followed along a set of rays chosen to be parallel to the Cartesian axes of a cubical simulation volume. The hydrodynamical simulations were carried out using a uniform Cartesian grid and the rays were chosen so that each grid cell was sampled by three rays (one per direction). Since the attenuation was modelled in both directions along a ray, the result is that each grid cell sees six different values for the local strength of the radiation field, each one corresponding to $2\pi / 3$ steradians on the sky. This approach has the benefits of speed and simplicity, particularly in the context of a uniform grid calculation, but the low angular resolution with which the radiation field is modelled renders the method prone to shadowing artifacts. In addition, it is difficult to generalize the method to work with adaptive mesh refinement (AMR) codes.

An improved method for accounting for UV attenuation in hydrodynamical simulations is the \textsc{TreeCol} algorithm developed by \citet{2012MNRAS.420..745C}. This algorithm constructs for each computational element (e.g.\ a grid cell or SPH particle) a $4\pi$ steradian map of the column density distribution seen by that element. This map is computed using information on the spatial distribution of the gas stored in an oct-tree structure and is mapped onto the sphere using the \textsc{HEALPIX} pixelation scheme \citep{2005ApJ...622..759G}. Given a map of the column densities, a corresponding map of the UV radiation field strength can then be calculated and since \textsc{HEALPIX} uses equal-area pixels, it is then straightforward to calculate local photochemical rates such as the H$_{2}$ photodissociation rate. One big advantage of this approach is that the required oct-tree structure is already available in any code that uses an oct-tree algorithm for computing gravitational forces (e.g.\ {\sc gadget, arepo, flash}). In such a code, the column densities can be calculated at the same time that the tree is being walked to compute the gravitational accelerations, avoiding the need for a separate communication step, hence potentially offering a large time saving. In addition, the angular resolution of the attenuated UV field is typically significantly larger than with the six-ray approximation; for example, the simulations by \citet{2012MNRAS.426..377G} and \citet{2015MNRAS.448.1607G} use 48 pixels for their \textsc{TreeCol} maps. Disadvantages of this method are that it can be highly memory-intensive and that it is not simple to extend it to handle embedded sources of radiation. \textsc{TreeCol} was originally developed for use with the {\sc gadget} SPH code, but has subsequently been ported to other codes, such as {\sc arepo} \citep{2014MNRAS.441.1628S} and {\sc flash} \citep{2015MNRAS.454..238W,2017arXiv170806142W}. In addition, \citet{2014A&A...571A..46V} have developed a tree-based method for use with the {\sc ramses} AMR code that is not a direct port of \textsc{TreeCol} but that is very similar in spirit.

One general drawback of these hydrodynamical models of three-dimensional PDRs is that they are typically forced to adopt highly simplified chemical models, for reasons of computational efficiency (see Section~\ref{cold-dark}). Therefore, efforts have also been made to develop static three-dimensional PDR models that sacrifice the ability to follow the hydrodynamical evolution of the gas for the ability to follow its chemistry and excitation state in much greater detail.   
Three-dimensional models can be built up by considering a series of one-dimensional calculations \citep[see e.g.\ the models of][which consist of a series of spatially distributed 1D spherical clumps]{2017A&A...598A...2A}, but it is unclear how well these models capture the interaction between different substructures. A more flexible approach was presented by \citet{2012MNRAS.427.2100B} in the \textsc{3d-pdr} code, who use a \textsc{healpix}-based scheme to model the attenuation of UV radiation coming into the PDR and also the escape of far-infrared and sub-millimetre line emission out of the PDR. 
\textsc{3d-pdr} performs a ray-tracing calculation along a set of rays centred on and orthogonal to the \textsc{healpix} pixels. One limitation of the original version of \textsc{3d-pdr} was the need to specify in advance the form of the incident UV field (e.g.\ planar, isotropic, or point-source). However, more recently, \textsc{3d-pdr} has been coupled with the \textsc{torus} Monte Carlo radiative transfer code \citep{2015MNRAS.454.2828B}, allowing it to model arbitrary distributions of sources and to account for the effects of photoionization as well as photodissociation (see also Section~\ref{photoion} below).

Finally, a few PDR codes are also capable of modelling the impact of high energy X-rays on the chemical and thermal structure of the gas, allowing them to be used to study XDRs \citep[e.g.][]{2005A&A...436..397M,2011A&A...526A.163A,2017arXiv170510877F}. In addition to the physics included in a standard PDR code, an XDR code must also be able to model the attenuation of the X-ray flux due to absorption by H, He, metals and dust; the direct photoionization of the chemical constituents of the gas by X-rays and their secondary ionization by energetic photoelectrons; secondary photoionizations and photodissociations caused by UV photons produced by electronic excitation of H, He and H$_{2}$ by these same photoelectrons; and the presence of much higher than usual abundance of vibrationally excited H$_{2}$, which can significantly impact the chemical evolution of the gas. Most codes capable of modelling XDRs are 1D, but there are starting to be attempts to implement the same physics in 3D models (e.g.\ \citealt{2013ApJ...771...50A}; {\citealt{2018arXiv180310367M}}).

\subsubsection{Photoionised gas}
\label{photoion}
Photons of sufficiently high energy can liberate electrons from atoms, thereby photoionising them. A gas irradiated by a large enough flux of such photons will become predominantly photoionised (usually referred to as an H\,\textsc{ii} region -- the term used to describe ionised hydrogen). If the energy of the photon is higher than the ionization potential of the atom in question, the excess energy is carried away as kinetic energy by the electron and the ion, with most of it quickly being converted into heat. Therefore, photoionized gases tend to be hot, with typical equilibrium temperatures in the range $8\times10^{3} - 10^4$\,K for solar metallicity gas and somewhat higher for lower metallicity gas. Photoionised gasses are hence typically of high enough temperature and pressure to be dynamically important, for example drastically altering the medium in the vicinity of massive stars \citep[][]{Dale:Review:2015}. Excellent general resources on photoionised gasses include \cite{2006agna.book.....O} and \cite{1978ppim.book.....S}.

Photons with energy $\geq13.6\,$eV (912\AA, the Lyman limit) are capable of ionising atomic hydrogen -- ionised hydrogen is the dominant constituent of photoionised gas in the universe. Working out the ionisation state of a gas composed of pure atomic hydrogen is relatively straightforward -- in equilibrium it is set by the balance of the photoionisation rate and recombination rate of protons and electrons. Therefore, if one knows the local photoionising flux, the ionisation fraction is easily found. Furthermore, when dealing with stellar sources of photoionising radiation, it is often a good approximation to consider only a single energy bin for the ionising photons (at the Lyman limit for a hydrogen only gas). A further widely used simplification is the ``on-the-spot'' approximation, i.e.\ the assumption that the ionising photons created by recombination directly into the ground state of the hydrogen atom are absorbed locally and hence can be accounted for simply by modifying the recombination rate coefficient. Finally, when including ionising radiation in a hydrodynamical calculation, another common approximation is the adoption of a single representative temperature for fully ionized gas, $T_{\rm ion}$, often taken to be $10^{4}$\,K. This allows one to avoid having to solve for the actual temperature at each point in the H\,\textsc{ii} region, giving a significant speed-up, while introducing a typical error of only a few tens of percent. When using this approximation, it is also common to set the temperature of partially ionized cells based on their ionisation fraction $\eta$ (e.g.\ $T(\eta) = T_{\rm neut} + \eta\left[T_{\rm ion}-T_{\rm neut}\right]$, where $T_{\rm neut}$ is the temperature of the neutral atomic gas). {Although this is not as accurate as a full photoionisation calculation, and does not capture, for instance, the rise in temperature near the H\,\textsc{ii} region boundary, it does still allow the dynamical evolution of the H\,\textsc{ii} region to be modelled accurately \citep{2015MNRAS.453.2277H}}.  Radiation hydrodynamical models that use this approach include \citet{2010ApJ...723..971G}, \citet{2009A&A...497..649B}, \citet{2012MNRAS.427..625W} and \citet{2013MNRAS.430..234D}. 

For examining the dynamical impact of photoionisation, these simplifications are generally adequate, but for almost all (synthetic) observational applications they are insufficient. For example, key observational tracers towards H\,\textsc{ii} regions are forbidden lines from species such as carbon, oxygen, neon and sulphur. In order to correctly compute the strength of these lines, it is necessary to solve for the ionisation state of each of these elements, rather than simply for the ionisation state of hydrogen. However, calculating the ionisation state of multiple species is computationally expensive and so is not ideal for dynamical calculations. Furthermore, for systems where the X-ray flux is important (e.g.\ for discs around young low mass stars, supernova remnants and active galactic nuclei) temperatures can be much higher than for gases heated only by EUV photons. In this case, higher ionisation states of heavier elements (e.g. Fe, Al, Mg) become important. In such a regime, a single high energy photon can also liberate many electrons, making solution of the ionisation state more difficult to compute as the number of liberated electrons has to be treated probabilistically using Auger yields \citep[e.g.][]{1998PASP..110..761F, 2008ApJS..175..534E}. In view of the complex microphysics involved, the treatment of H\,\textsc{ii} regions at this level of detail generally involves the post-processing of a radiation hydrodynamic simulation or the use of an analytic description of the density structure and photon sources in the system. 

Situations where the dynamical behaviour is highly sensitive to accurately modelling the temperature, or where the ionisation state of the gas is far from equilibrium are particularly difficult to model because of the strong coupling between the dynamics and the composition. In addition to the methods discussed earlier for modelling complex chemical networks in 3D hydrodynamical simulations, another approach that has proved successful is the use of a parameterised cooling curve that accounts for the effects of cooling from the photoionised metals in a simplified way \citep[e.g.][]{2010MNRAS.403..714M, 2012A&A...538A..31T}.

There are many codes capable of photoionisation modelling. These range from hydrodynamical codes that account for photoionisation in a highly approximation fashion (e.g.\ \textsc{seren}; \citealt{2009A&A...497..649B, 2011A&A...529A..27H}), or with a greater degree of microphysical accuracy (e.g.\ \textsc{Enzo} or \textsc{flash}; see \citealt{2011MNRAS.414.3458W} or \citealt{2015MNRAS.454..380B}, respectively), to codes designed to compute the photoionisation structure in full microphysical detail, for comparison with real observations. The latter type of code includes \textsc{cloudy} \cite{2013RMxAA..49..137F, 2017arXiv170510877F}, \textsc{mocassin} \citep{2003MNRAS.340.1136E, 2005MNRAS.362.1038E, 2008ApJS..175..534E}, \textsc{torus} \citep{Harries:2000,2012MNRAS.420..562H,2015MNRAS.448.3156H} and the code of \cite{2004MNRAS.348.1337W}.

\subsubsection{Shocks}
Shocks dissipate kinetic energy as heat in a relatively small volume of gas. They can therefore produce local gas temperatures that are much higher than those typically found in the cold ISM. This can have a profound impact on the chemical composition of the gas and also on the emission that it produces. 

As the interstellar medium is magnetised and hence has three different characteristic velocities (the Alfv\'en velocity and the fast and slow magnetosonic velocities), a wide variety of different types of shock are possible \citep[see e.g.][for a useful review]{1993ARA&A..31..373D}. From an observational point of view, however, one of the main distinctions is that between a single fluid and a multi-fluid shock. When the coupling between ions and neutrals in the ISM is strong, both components move with the same velocity and the fluid behaves as if the neutral component were directly affected by the magnetic field\footnote{In reality, of course, the neutral atoms and molecules are not directly affected by the Lorentz force and the effect of the magnetic field on the neutrals is indirect, mediated by collisions between ions and neutrals.}. In this single fluid regime, a shock corresponds to a discontinuity in the velocities and densities of both ions and neutrals. On the other hand, if the coupling between ions and neutrals is weak, as can occur when the fractional ionisation is low, then the ion velocity and the neutral velocity can become significantly different. In this multi-fluid regime, the Alfv\'en velocity of the ions is generally very large and the velocity and density of the ionised component will therefore often remain continuous through the shock. The behaviour of the neutral component then depends on the velocity of the shock and the ability of the neutrals to cool. If the shock is weak and/or cooling is efficient, then the neutrals may remain supersonic throughout the shock. In this case, their velocity, temperature and density all remain continuous through the shock. A shock of this type is termed a ``C-type'' shock. On the other hand, if the shock velocity is high, or cooling is inefficient, then the neutral fluid will undergo a discontinuous change in its velocity, density etc.\ as in a single fluid shock; such shocks are termed ``J-type'' shocks. 

Far upstream and far downstream, single fluid and multi-fluid shocks obey the same jump conditions (provided they have similar post-shock temperatures), so for many dynamical applications, it doesn't matter whether we have a single fluid shock, a J-type shock or a C-type shock. However, these different types of shock have different temperature and density profiles within the shock transition region itself and hence have different consequences for the emission produced within this region and for the chemical evolution of the gas. For example, J-type shocks are typically associated with much higher temperatures than C-type shocks. Hence, the former can excite atomic and molecular transitions that the latter cannot. 

Since the ISM is supersonically turbulent, a given region will often contain many different shocks with varying strengths and geometries. Ideally, therefore, one would like to use a 3D MHD approach to model the chemical effects of these shocks and to make predictions for the associated emission. Unfortunately, this is currently impractical if we are interested in simulating a representative volume of the ISM, owing to the short length scales associated with shocks. For example, at a number density of $n = 100 \: {\rm cm^{-3}}$, which is at the low end of the range of values typically found in molecular clouds, the thermal relaxation region behind a $15 \, {\rm km \, s^{-1}}$ J-type shock -- i.e.\ the region where the temperature remains out of equilibrium owing to the energy dissipated by the shock -- has a thickness of only $\sim 30$~AU \citep{2013A&A...550A.106L}. Resolving this region with e.g.\ 10 grid cells while at the same time simulating a region with a size comparable to that of a small molecular cloud, say 5~pc, therefore requires an effective resolution of $400000^{3}$ grid cells. Such high effective resolutions have been achieved in AMR simulations of gravitational collapse \citep[see e.g.][]{2017ApJ...846....7K}, where high resolution is needed in only a small fraction of the computational volume, but are impractical when modelling turbulence, which fills a much larger fraction of the volume \citep{2006ApJ...638L..25K}. Therefore, most current attempts to model shock chemistry and shock emission have focused on the construction of detailed 1D shock models.

When constructing a 1D shock model, a common approach is to assume that the shock is steady, i.e.\ that the time-dependent terms in the fluid equations can be set to zero. This allows one to solve in a straightforward fashion for the density, temperature and velocity structure of the shock, with the main remaining difficulty being the modelling of the close coupling between the chemistry and the cooling of the gas. Steady shock models of this type are relatively quick to compute and hence can be used to explore a broad parameter space. Examples of 1D steady-state shock codes include the Kaufman and Neufeld shock code \citep{1996ApJ...456..611K}, the Paris-Durham shock code\footnote{\url{https://ism.obspm.fr/?page_id=151}} \citep{2003MNRAS.343..390F} and the UCLCHEM code\footnote{\url{https://uclchem.github.io/}} \citep{2017AJ....154...38H}. However, steady shock models are not always appropriate -- for example, if the time taken for the gas to flow through the shock and the thermal relaxation region is comparable to the overall evolutionary timescale of the system -- and in this case, a fully dynamical approach is necessary.

A further complication that is also sometimes encountered is the fact that in some cases, it is necessary to account for the effects of radiation as well as the dynamical heating produced by the shock. Very fast shocks generate this radiation internally: the radiation emitted by the hot, post-shock gas includes photons capable of ionising atomic hydrogen and dissociating H$_{2}$ and other molecules in the pre-shock gas \citep{1979ApJS...41..555H}. Alternatively, the radiation can be from an external source, such as the interstellar radiation field \citep{2013A&A...550A.106L} or a nearby star \citep{2017A&A...602A...8G}.

Models of astrophysical shocks in the CNM and in dense molecular clouds have identified several clear chemical signatures of the presence of shocks. In the CNM, shock heating of the gas enables the endothermic reactions
\begin{equation}
{\rm C^{+} + H_{2}} \rightarrow {\rm CH^{+} + H},    
\end{equation}
and
\begin{equation}
{\rm S^{+} + H_{2}} \rightarrow {\rm SH^{+} + H},    
\end{equation}
to proceed rapidly, enabling the gas to form much larger abundances of CH$^{+}$ and SH$^{+}$ that is possible purely with cold ion-neutral chemistry \citep{1978ApJ...222L.141E, 1986MNRAS.221..673M}. In denser gas, a more important effect is the fact that shocks can liberate molecules from grain surfaces, greatly enhancing the abundances of species that would otherwise rapidly freeze-out. Sputtering of grains in shocks also recycles material from the grain substrate to the gas phase. In particular, if silicate grains are present, this process can greatly enhance the gas-phase silicon abundance, leading to the formation of large quantities of SiO. Consequently, the presence of SiO emission is often considered to be diagnostic of the presence of shocks \citep[see e.g.][]{2008A&A...482..809G,2009ApJ...695..149J}.
Key instances of shock chemistry in the context of this review are hot cores, which will be discussed further in section \ref{sec:cores}.

\subsubsection{Dust properties for radiative equilibrium and continuum synthetic observations}
\label{dustonly}
{Dust plays an important role in setting the chemical and thermal structure in many astrophysical environments. In this review we do not touch upon the details of dust relating to grain surface reactions, which are themselves extremely complicated and the subject of ongoing research \citep[e.g.][]{2017SSRv..212....1C}. Nor do we explore in detail polycyclic aromatic hydrocarbons, which are particularly important for heating in many regimes and require a substantial review in themselves, as given by \cite{doi:10.1146/annurev.astro.46.060407.145211}. Rather we focus on the properties of dust relevant to radiative transfer modelling for dust radiative equilibrium computation and synthetic observations in dust continuum emission. With this in mind, what is required is just the grain abundance (canonically the dust--to--gas mass ratio is $\delta=10^{-2}$ in the ISM), size distribution and composition.} 




The most commonly used grains size distribution has a size distribution 
\begin{equation}
    \frac{dn(a)}{da}\propto a^{-q}
\end{equation}
between some maximum and minimum grain size, and typically $q\approx 3.5$ \citep{1977ApJ...217..425M}.

Calculations of this type are frequently carried out using Monte Carlo radiation transport \citep[e.g.][]{Lucy:1999, Harries:2000, 2006A&A...459..797P, 2009A&A...497..155M, 2011A&A...536A..79R, 2012ascl.soft02015D, 2016A&A...593A..87R}. 

There is a large range of possible dust compositions which also affect the opacity and hence the emission properties \citep[see e.g.][]{2003ARA&A..41..241D,Draine:2007}. Optical constants for a large number of grain types are available from, for example, the Jena dust database (see section \ref{sec:data}). Other than the composition, the main complication that might arise in dust-only calculations comes from the spatial distribution of dust. For example the dust-to-gas mass ratio and size distribution may vary. The grain distribution may therefore either need to be solved for dynamically (see section \ref{sec:dynamics}) or a spatially varying grain distribution {(minimum/maximum grain size, and power law of the distribution)} imposed.



\subsection{Radiative transfer}
As discussed above, in many environments radiative transfer is of central importance for determining the chemical composition of the gas and the temperature of both gas and dust. Furthermore, the production of a synthetic observation itself requires computation of the emission of photons and their propagation through matter. Radiative transfer is hence of sufficient importance that methods for computing it warrant a brief discussion in this review. {For further information we direct the reader to the textbooks of \cite{1960ratr.book.....C}, \cite{1979rpa..book.....R} and \cite{2001irt..book.....P} }

At first glance the radiative transfer equation appears simple:
\begin{equation}
    \frac{dI_{\nu}}{d\tau_{\nu}} = S_{\nu} - I_{\nu}
    \label{eq:radeq}
\end{equation}
where $I_{\nu}$ is the specific intensity at frequency $\nu$, $\tau_{\nu}$ is the optical depth, and $S_{\nu}$ is the source function. However this is a nonlinear and multi-dimensional problem. The radiation field is computed over three spatial dimensions, with direction, frequency, three polarisation intensities and possibly time dependence. Furthermore, the source function, which is the ratio of the emission and absorption coefficients, is a function of the matter properties (density, temperature, composition) which themselves are sensitive to the radiation field strength. {In addition, if the dynamical state, composition or thermal structure of the system change substantially on a timescale that is comparable to or shorter than the light travel time across the system, then the influence of the finite speed of light cannot be neglected and one must solve the time-dependent radiative transfer equation. This is usually of importance in radiation hydrodynamics applications, with the finite speed of light being particularly important in supernova \citep[e.g.][]{2012MNRAS.424..252H} and cosmological modelling \citep[e.g.][]{2011MNRAS.411.1678C}. }

Radiative transfer is hence a formidable coupled problem that frequently has to be solved iteratively. An excellent review of radiative transfer methods in the context of dust is given by \cite{2013ARA&A..51...63S}; many of the methods discussed therein are also relevant for line transfer. Also see \cite{2009A&A...498..967P} for benchmarking of continuum radiative transfer codes. 


{A wide variety of different approaches to solving (or approximating) the radiative transfer equation exist, including ray tracing, Monte Carlo radiative transfer, flux limited diffusion \citep{1981ApJ...248..321L}, escape probability estimators such as  the large velocity gradient (LVG) approximation \citep[which the famous {RADEX} code is based on, ][]{1960mes..book.....S, 2007A&A...468..627V} and refinements such as the Coupled Escape Probability method \citep[CEP,][]{2006MNRAS.365..779E}.  The primary methods employed when dealing with synthetic observations in modern astrophysics are ray tracing and Monte Carlo techniques. We therefore examine both of these in more detail below.}

\subsubsection{Ray tracing}
Ray tracing involves directly solving the radiative transfer equation along a given direction. The ``long characteristics'' approach involves doing this along the same trajectory over the entire computational domain. For example, a ray originating from some point in a cell will have an initial intensity set by the local emissivity. As we move along the ray, this will be attenuated by absorption or increased by further emission.  If the emissivity and opacity through a medium are known, the intensity towards an observer is easily calculated with such an approach, since rays need only to be traced in a single direction. Consequently, long characteristic methods are often used to produce synthetic observations based on known distributions of density, temperature, composition and excitation (which may also have been computed with the help of a ray-tracing scheme, or using some other approach). Examples of radiative transfer codes that adopt this approach include {\sc radmc-3d} \citep{2012ascl.soft02015D} and {\sc hyperion} \citep{2011A&A...536A..79R}.

On the other hand, if this technique is used to e.g.\ compute the cooling rate of the gas, or the ability of UV radiation to propagate through it, then a much larger number of rays is necessary, in order to sample the full 4\,$\pi$ steradians of solid angle. This can be time-consuming, particularly when one is dealing with multiple sources. In particular, if every computational element is potentially a source of radiation, then the cost of solving the radiative transfer equation using a long characteristics approach scales as O$(N^{5/3})$, where $N$ is the number of computational elements; for comparison, solving the equations of hydrodynamics has a cost that scales only as O$(N)$. In addition, long characteristics methods can be difficult to parallelize efficiently, owing to the inherently sequential nature of the way in which the calculation proceeds along a ray. Fortunately, some of these efficiency concerns can be addressed using techniques such as adaptive ray splitting or ray merging \citep{2002MNRAS.330L..53A}, or hybrid characteristics (\citealt{2006A&A...452..907R};\citeauthor{2010ApJ...711.1017P}~2010a),
High efficiency can also be obtained by using hybrid methods for problems involving a single strong point source plus an extended source with a much lower radiation temperature (e.g.\ an H\textsc{ii} region ionized by a central O star). In a hybrid method, the radiation from the central point source is followed using ray-tracing, while the cooler extended emission is modelled using a cheaper but more diffusive technique, such as flux-limited diffusion \citep[see e.g.][]{2010A&A...511A..81K,2017JCoPh.330..924R}.

An alternative ray-tracing approach makes use of so-called ``short characteristics''. In this method, the specific intensity and source function are discretized on a set of rays at each cell centre in the simulation volume. The radiative transfer equation is then solved along each short ray, using initial conditions interpolated from the rays in neighbouring grid cells. This approach is computationally cheaper than the long characteristics approach, since it avoids the need to repeatedly trace different rays through the same region and it is also easier to parallelize. The main disadvantage is that it is inherently more diffusive than a long characteristics calculation, owing to the repeated use of interpolation to determine the initial intensity in each grid cell. This makes it a sub-optimal method for situations dominated by a single bright point source. On the other hand, it can be a very powerful method for situations dominated by diffuse emission, or when acting as part of a hybrid solver. An example of the use of this method in astrophysics can be found in \citet{2012ApJS..199....9D}. The SimpleX algorithm (\citealt{2010A&A...515A..79P,2010A&A...515A..78K,2017arXiv171102542J}), which models radiative transfer by transporting photons from cell to cell on an  unstructured Delaunay grid, can also be considered to be a form of short characteristics method.

\subsubsection{Monte Carlo radiative transfer}
In recent years one of the most popular approaches to solving the radiative transfer problem is the Monte Carlo approach \citep{Lucy:1999}. There are variations in the implementation, but broadly the approach is as follows. 
The energy output from photon sources is discretised into a series of packets. For example a star of luminosity $L$ emits $N$ photon packets each with energy $\epsilon=L\Delta t/N$ for an experiment of duration $\Delta t$. Each photon packet has a frequency associated with it and since each packet has the same energy, photon packets with different frequencies have different numbers of photons associated with them. The frequency of each packet is generally randomly sampled from the stellar spectrum, and the packet is emitted with a direction {set by sampling from a probability distribution (e.g.\ at the most basic level in the case of a point source this is a random direction with uniform probability)}. Each photon packet then undergoes a random walk of absorption and scattering through the computational domain, much like photons in reality.

What is the value of this random walk? Whenever a photon packet traverses some discretisation of space (e.g.\ a grid cell), it contributes to an estimate of the mean intensity (or equivalently the energy density) there. That is, in a cell of volume $V$ the mean intensity is just a sum over path lengths $\ell$ through that cell
\begin{equation}
    \frac{4\pi J_{\nu}}{c} d\nu = \frac{\epsilon}{c\Delta t}\frac{1}{V}\sum \ell. 
    \label{equn:MCgetJ}
\end{equation}
This quantity is required for computing dust radiative equilibrium temperatures, photoionisation balance, UV fields for photodissociation calculations and so on. Often, the value of the mean intensity will affect the composition and thermal properties of the system, which in turn will affect the opacities and/or emissivities. In this case, the random walk and consequent update to the state of the gas and dust need to be computed iteratively until the system converges.

With the composition and thermal structure known, Monte Carlo radiation transport (MCRT) can also be used to compute synthetic images or spectra in a very similar way. The absorption and emission properties of the star(s) and matter are known, so the emission and propagation of photon packets can be computed and the number that reach the observer can be recorded. An array of bins (pixels) collecting photon packets can be used to build up a synthetic image. Of course, doing this with completely random photon trajectories would be very computationally inefficient as only a small fraction would make it to the observer. This problem can be avoided by using the peel-off technique \citep{1984ApJ...278..186Y}. {This artificially directs photon packets towards the observer after each interaction, and weights the packet so that the total energy reaching the observer remains the same as it would be if the photon trajectories were completely random.} This technique significantly improves the signal-to-noise of MCRT that can be achieved for a given computational cost. 

MCRT has many strengths. It is relatively straightforward to implement (although parallelizing it efficiently over distributed domains remains a hard problem). It is flexible, handling polychromatic radiation and the diffuse radiation field with ease. It also naturally handles multiple stellar sources and complex distributions of matter. Moreover, since MCRT is basically sampling a series of independent random events, it also scales incredibly well up to large numbers of cores, given a suitable domain decomposition. 

One problem with MCRT is that in extremely optically thick regions, photon packets can get trapped in an extremely long random walk, where they propagate only tiny distances before being absorbed and re-emitted. This problem is easily alleviated by identifying such regions and applying a flux limited diffusion approach there in a hybrid scheme. However, it is a sympton of the true weakness of MCRT, which is that large numbers of photon packets are required to ensure that the entire computational domain is well sampled and hence not subject to noise-induced artifacts. The signal-to-noise scales with the square root of the number of photon packets, meaning that the number of packets required can be huge, particularly for multi-frequency problems.
Nevertheless, for many applications, the efficient scaling of MCRT alleviates this problem. It should also be noted that when producing synthetic observations, often only a moderate level of accuracy is required. For example, a 10\% error in the dust luminosity -- on the high side for an MCRT calculation -- corresponds to an error of only a few percent in the dust temperature, a level of accuracy which is almost never reached with a real observation.

In general, therefore, the strengths of MCRT often outweigh its weaknesses, making it one of the most popular methods of radiative transfer at present. Implementations exist for regular Cartesian grids \citep{Lucy:1999, 2003MNRAS.340.1136E, 2004MNRAS.348.1337W, 2011MNRAS.416.1500H,2012ascl.soft02015D, 2011A&A...536A..79R, 2015A&C.....9...20C}, SPH distributions \citep{2016MNRAS.461.3542L} and Voronoi meshes (\citealt{2016MNRAS.456..756H,2016A&A...593A..87R} and the extension of \textsc{Hyperion} originally published in \citealt{2011A&A...536A..79R}).

\subsection{Backgrounds and foregrounds}
\indent Part of the purpose of synthetic observations is to encourage the community of Computational Astrophysics to think not just about what their simulations \textit{do}, but about how they would \textit{appear} to an observer if they were real objects at some chosen distance. Simulations often consider isolated objects, such as individual molecular clouds. In reality of course, observations of many classes of object are hampered by the presence of foreground and background emission which may be difficult or impossible to disentangle from that coming from the target {and may itself be (marginally) optically thick}. When making a synthetic observation, careful thought should be given to whether such contamination is an issue in observing the real counterparts of the systems being modelled. If it is, some attempt to model the unwanted emission should be made.\\
\indent Diffuse contamination could be modelled simply by thinking of it in terms of noise and adding a uniform background emissivity at a suitable level to every image pixel. A more realistic approach would be allow the contamination to fluctuate statistically across the image. However, care must be taken here because any attempt to recover structural information from the resulting image (e.g.\ column density PDFs, correlation functions) will be affected by the size scales of the fluctuations introduced and choosing these arbitrarily may not be appropriate.\\
\indent An alternative, utilised by \citet{2017ApJS..233....1K} with the help of the \textsc{FluxCompensator} tool (\citealt{2017ApJ...849....3K}; see also Section~\ref{sec:singledish}), is to add the synthetic emission to a suitably chosen observed background field (i.e.\ one which is itself not too richly structured). In this case, the exercise immediately produced the sobering result that the chosen simulated cloud, when observed in all five \textit{Herschel} bands, was too faint to be extracted from the background in the two or three bands at the longest wavelengths. Compare the rows in right hand column of Figure \ref{fig:instrument_effects}, where we display the background effects for the five \textit{Herschel} bands. This is an excellent example of the kind of sanity checks that can only be done by accounting for observational contaminants.\\

\subsection{Instrumentation}
\label{sec:instrumentation}
The radiation field produced by a simulation, and any non-observational imposed contamination (extinction, distance, redshift, etc.), is in itself of interest, since it represents what an ideal observatory could observe. However, in many cases, the final data product can be significantly modified by the instrument used to collect the radiation, due to e.g.\ the smoothing caused by a finite beam size, or the filtering of large-scale structures inherent to interferometric observations. Computing the simulated radiation field in a realistic way (i.e.\ accounting for these and other instrumentational effects) is key if the application of the synthetic observation is to interpret some real observation and we want to maximise the consistency between real and simulated data. It can also be important if we want to understand the limitations of observational diagnostics with a given instrument. 

Given the above, the final step in making synthetic observations as realistic as possible (assuming that the prior steps are performed optimally) is therefore to choose (if this was not already implicit in the wavebands modelled) a telescope/ instrument combination and incorporate its inherent response. Theoretically, this is perhaps the most straightforward step in the process, as these observational effects are well understood physically. In practice, most of the difficulty in this stage of handling synthetic data arises because detailed knowledge of the observational framework may be required. For example processing pipelines (to clean, convolve, calibrate data etc.). This is hence often a facet of synthetic observations that requires an interdisciplinary approach between the theoretical and the observational communities (typically on an instrument-by-instrument basis).

\begin{figure*}
    \centering
    \vspace{-1.1cm}
    \includegraphics[width=0.88\textwidth]{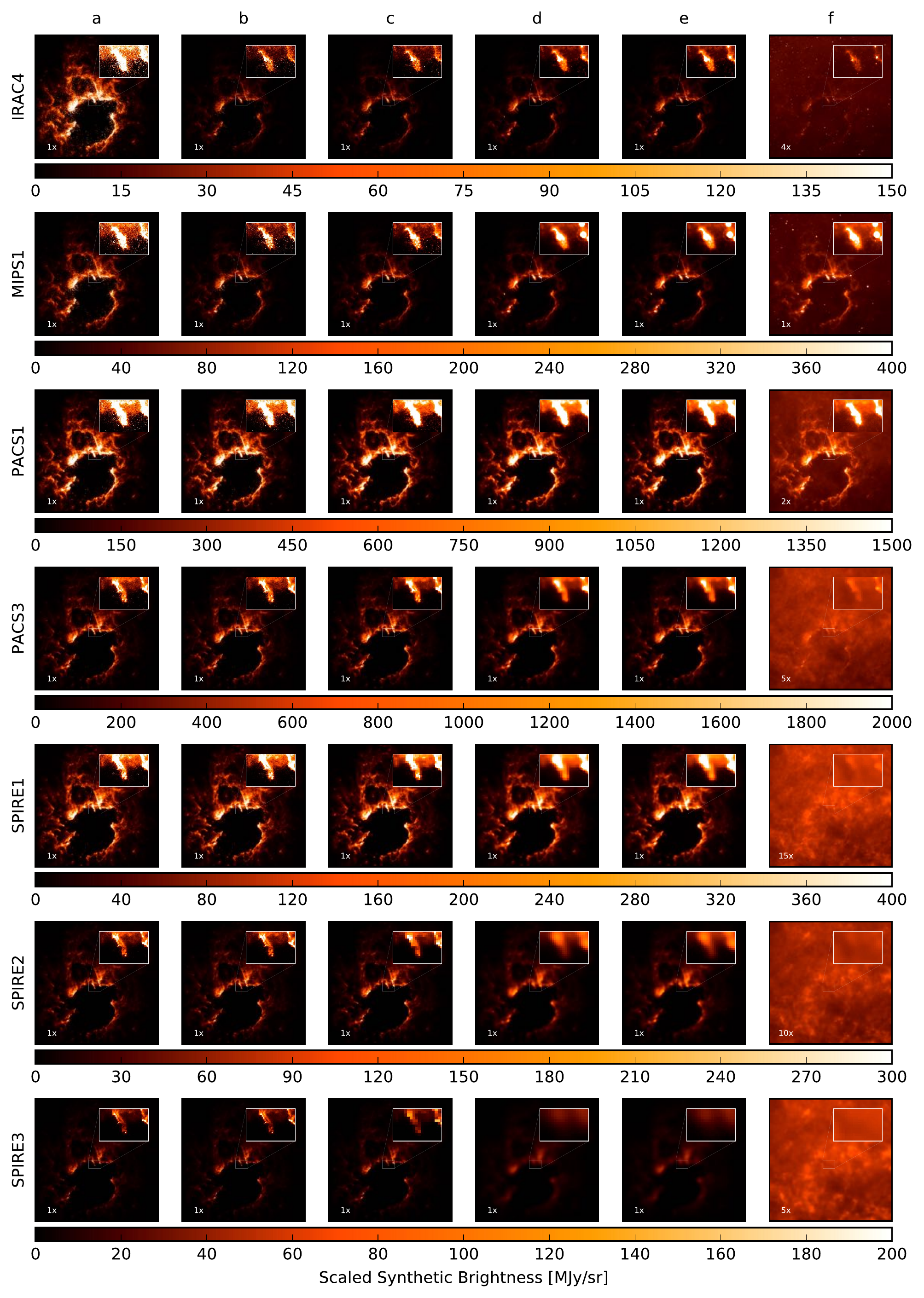}
    \caption{The effects of interstellar reddening and observational limitations on the ``observed" synthetic brightness modelled with the \textsc{FluxCompensator} tool \citep{2017ApJ...849....3K}. Numbers in the panels indicate the additional stretch with the colorbar. Column (a) shows the ideal radiative transfer output after filtering with the spectral transmission. The remaining columns show the impact of successively including additional effects: (b) interstellar reddening; (c) regridding of the image to the pixel resolution of the corresponding detector; (d) convolution with the PSF; (e) the addition of thermal noise; and finally (f) combination with real background observations.}
    \label{fig:instrument_effects}
\end{figure*}

It is important to bear in mind that there are also non-observational\footnote{These are the effects that distance, extinction and redshift have on the radiation field, which can all be important when modelling the radiation field one would expect to detect from a real system.} effects which can alter the images significantly. Apart from these non-observational effects, some of the most important observational effects that can arise are
\begin{itemize}
    \item \textit{Spectral transmission:} Real telescopes and instruments are sensitive only to a limited range of wavelengths and are not uniformly sensitive within this range. For ground-based instruments the atmospheric conditions and hence transmission can also be time varying (e.g.\ the precipitable water vapour, ``pwv'', level can have a significant impact and can now be accounted for in synthetic observations using software like \textsc{casa}, which we discuss in section \ref{sec:multidish}).
    \item \textit{Convolution:} The diffraction at the telescope opening for single dish telescopes is commonly described mathematically by a convolution with the point-spread function (PSF) of the observational device. Note that interferometers have no PSFs as they have no single telescope opening, but an array. Their distortion of the ``virtual" opening of the array can be approximated by a Fourier transform in the $uv$ plane.
    \item \textit{Finite pixel resolution}: The real observable pixel size is usually much larger than the output pixel size of the ideal radiative transfer image.
    \item \textit{Noise:} The thermenolgy includes both, the Gaussian random errors introduced by e.g.\ thermal noise, and also various kinds of non-Gaussian noise, such as speckles in coronographic imaging or side-lobe noise in radio observations. 
    \item \textit{Interferometric effects:} More extended configurations of interferometric telescopes (such as ALMA, the SMA and the EVLA) have higher resolution at the expense of larger scale sensitivity. Accounting for the beam size is there non-trivial, in the sense that simple convolution with a Gaussian neglects the possible decrease in sensitivity on larger spatial scales. However, software such as \textsc{casa} (discussed below, section \ref{sec:multidish}) has excellent resources for modelling interferometric response.
\end{itemize}
The effects that dominate single dish continuum observations are summarised in \cite{2017ApJ...849....3K}. 


In Figure~\ref{fig:instrument_effects} we show an example of how the spectral transmission, the PSF convolution and the treatment of noise can alter the appearance of a synthetic observation extracted from a radiative transfer calculation and virtually observed with parameters appropriate to several different infrared telescopes. We used the ideal synthetic observation created by \cite{2017ApJS..233....1K} with the radiative transfer code \textsc{Hyperion} \citep{2011A&A...536A..79R} from the hydrodynamical simulation of \cite{DaleBoth:2014}. In column a) the filtered image with the respective spectral transmission at 3 kpc distance is shown. Column b) shows the effect of accounting for interstellar reddening ($A_V=50$~mag\footnote{For illustration due to color stretch}). The impact of regridding to the pixel size appropriate for each detector (column c), convolving with the telescope PSF (column d) and adding thermal noise at a level that is 20\% of the  median intensity in each image (column e) are displayed next. Finally, in the right column we show the effect of combining with the image with background emission based on a real observation from the GLIMPSE \citep{2003PASP..115..953B}, MIPSGAL \citep{2009PASP..121...76C} or HIGAL \citep{2010PASP..122..314M} surveys, as appropriate. The modeling of the telescope limitations and the extinction was performed with the \textsc{FluxCompensator} \citep{2017ApJ...849....3K} software (see Section~\ref{sec:singledish}).

Of course, a first order approximation can be provided for a single dish telescope by approximating the PSF with a 2D Gaussian, the spectral transmission by a box filter and the noise in the camera by Gaussian statistical noise. A useful tool for such tasks is the open-source Astropy software \citep{2013A&A...558A..33A}. In particular, the \verb|astropy.convolution| module can help with the convolution of common kernels for a first order estimate. However, when doing delicate analysis of the synthetic observations it is often feasible to use actual PSFs and optical information provided by the observatories that the synthetic observations should mimic. To this end, there is a large amount of ``virtual observatory software" currently available (much of it open source) which can simulate the effects introduced by the observational process. In the following, we give an overview of what is currently available, but we note that as the field is rapidly growing, this  list may not be complete.

\subsubsection{Mimicking Single Dish Observatories}
\label{sec:singledish}
Here we discuss some software that is useful for accounting for observational effects from single dish instruments (though some of the functionality of these tools is also useful for interferometers too). 

The Chandra Interactive Analysis of Observations (\textsc{ciao}) software\footnote{\url{http://cxc.harvard.edu/ciao/}} \citep{2006SPIE.6270E..1VF}, although constructed with a focus on the Chandra X-ray space telescope, has many useful tools that are more generally applicable, such as a function for convolving 2D/3D images or datacubes to account for a circular beam (\verb|aconvolve|). 

{The \textsc{Tiny Tim} software computes the PSF for Hubble Space Telescope observations \citep{TinyTim}}.

\textsc{starlink}\footnote{\url{http://starlink.eao.hawaii.edu/starlink}} is a long running astronomical data processing package \citep[e.g.][]{1982QJRAS..23..485D, 2013ASPC..475..247B,  2014ASPC..485..391C}, which has useful resources such as functions to add noise. It also includes the graphical astronomy and image analysis tool \textsc{gaia}. 

The image reduction and analysis facility \textsc{iraf}\footnote{\url{http://iraf.noao.edu/}} also hosts a range of useful tools for simulating instrument response and manipulating synthetic images in the manner of real data. These include filtering, convolution and the addition of noise. 

The open-source \textsc{FluxCompensator}\footnote{\url{https://github.com/koepferl/FluxCompensator}} software package \citep{2017ApJ...849....3K}  has been especially designed to apply simulated observational effects to the ideal synthetic observations that are output by radiative transfer codes, allowing one to more easily produce realistic synthetic observations. It was originally designed as an extension the 3D continuum Monte Carlo radiative transfer code \textsc{Hyperion} \citep{2011A&A...536A..79R}, but it is written in a generic way and so the outputs from other radiative transfer codes can also be used. Arbitrary PSFs, spectral transmission curves, finite pixel resolution, noise, and and also non-observational effects such as reddening can be modelled with the \textsc{FluxCompensator}. It is Python based and has a built-in database which provides the relevant information for many common continuum single-dish observatories, such as 2MASS, WISE, IRAS, \textit{Spitzer}, or \textit{Herschel} (see e.g.\ Figure~\ref{fig:instrument_effects}, which has been produced with the \textsc{FluxCompensator} code). It is possible to combine the resulting realistic synthetic observation with the real observation from the FITS file for optimal comparison (see e.g.\ Figure~3 of \citealt{2017ApJ...849....3K} for an example or the right column of  Figure~\ref{fig:instrument_effects}). The \textsc{FluxCompensator} has currently pre-computed interfaces to the all-sky surveys from 2MASS and WISE \citep{2006AJ....131.1163S,2010AJ....140.1868W} and the Galactic surveys GLIMPSE \citep{2003PASP..115..953B}, MIPSGAL \citep{2009PASP..121...76C} and HIGAL \citep{2010PASP..122..314M} and can be extended further by the user. 



\subsubsection{Mimicking Interferometers}
\label{sec:multidish}
To mimic the limitations of telescope arrays, we need to account for not only the properties of the individual telescopes but also the configurations taken by the whole array. In particular, if all we have is interferometric data, without the corresponding ``zero spacing'' measurements, then structure in the observational data on scales larger than some maximum angular scale that depends on the array configuration will be filtered out by the interferometer.
In addition, data reduction techniques for dealing with interferometric data (e.g.\ cleaning, which requires a Fourier transform in the uv plane) tend to be more complex than those for dealing with single-dish data, and yet their complexities must be accounted for if we want to produce realistic synthetic observations.

Thankfully for interferometers such as ALMA, the SMA, the EVLA etc., these observational effects can be accounted for using specialist software. The common astronomy software applications package \textsc{casa}\footnote{\url{https://casa.nrao.edu/}} provides an array of tools for simulating instrument response, particularly for interferometric observations \citep[e.g.][]{2012ASPC..461..849P}. For example \verb|simobserve| permits one to simulate observations made with these arrays down to the level of detail of the exact observing time (i.e.\ the date, observing mode and time on source) and conditions (i.e.\ atmospheric precipitable water vapour). A major strength of this software is that it is commonly used to process real observed data too. 
Exactly the same data reduction techniques can therefore be applied to both real and synthetic data. \textsc{casa} also has tools to produce outputs such as moment maps and position-velocity diagrams. Although \textsc{casa} is predominantly used for interferometric applications it can also be applied to single dish instruments. A very useful feature when comparing synthetic observations with real observations from ALMA is that the configuration file for modelling the effect of observational limitations on the synthetic observations can be extracted from the real observations with CASA.

Overall, there are ample well documented resources for including instrumental limitations and processing in ``realistic" synthetic observations, for both single dish as well as interferometric observations. We encourage scientists producing synthetic observations to strongly consider making use of these tools.

\subsection{Methods and resources for comparing real and synthetic observations}

There are a small number of ``standard objects'' which continue to be used as laboratories for comparing models and reality (e.g.\ the Orion Bar PDR). However, most simulations are aimed at modelling generic objects or systems having representative values of important parameters (e.g.\ mass, extent). This is largely because most astrophysical systems (particularly in the geometrically complex and often non-equilibrium ISM) are sufficiently complex both in terms of structure and time evolution that we can never hope to fully reproduce them.

Even for objects that we can observe in great detail, there are limits on how much of their behaviour we can reasonably expect to reproduce with simulations. For example, without precise knowledge of the initial distribution and densities and turbulent velocities within a molecular cloud, we cannot predict the exact state of the cloud at later times. Instead, the best that we can do is to make predictions about statistical properties of the cloud that are not highly sensitive to the exact initial conditions.

It follows that the majority of comparisons between simulated and observed entities must be statistical in nature. Unfortunately, there are a bewildering variety of tools available for comparing multidimensional datasets with each other, particularly when one is dealing with three-dimensional data, and it is often unclear what we can learn from a particular statistic, or which method is the ``best'' one to use. In this section of our review, we briefly summarize some of the most widely used statistical techniques for characterising and comparing synthetic and real observations, and discuss what we have learnt about when we should apply them.

\subsubsection{Probability density functions }
\label{sec:pdfs}
One of the simplest ways to characterise an observation (real or synthetic) is by computing the empirical probability density function (PDF) for some aspect of the observed emission or absorption. In principle, we can construct PDFs for a wide variety of different observed quantities; for example, \citet{2013ApJ...777..173B} consider PDFs of peak brightness temperature and line width when comparing synthetic $^{12}$CO and $^{13}$CO emission maps with observations of the Perseus molecular cloud. However, most commonly the quantity of interest is something closely related to the column density, such as the dust extinction or the integrated intensity of a presumed optically thin molecular tracer such as $^{13}$CO or C$^{18}$O.

The reason for the particular focus on tracers of the column density is the fact that numerical simulations of interstellar turbulence have long demonstrated that gas dominated by supersonic turbulence should have a log-normal volume density PDF \citep{1997ApJ...474..730P,2000ApJ...535..869K,2001ApJ...546..980O}. Since the column density is simply a line-of-sight integral of the volume density, this in turn implies that the column density PDF should also have a log-normal form, albeit with a narrower dispersion owing to the averaging caused by the integration along the line-of-sight \citep[see e.g.][]{2001ApJ...546..980O,2010MNRAS.405L..56B,2012ApJ...755L..19B}. On the other hand, in regions where runaway gravitational collapse dominates over turbulence, one expects a power-law PDF instead of a log-normal, while in regions where turbulence dominates on large scales and gravity dominates on small scales (as in e.g.\ many models of molecular clouds), we should find a log-normal PDF with a power-law tail at high densities \citep{2000ApJ...535..869K}. It is clearly important to test these predictions with observations of real clouds. Moreover, simulations have also shown that there should be a clear relationship between several properties of the turbulent flow (the Mach number, the magnetic field strength, and the relative importance of compressive versus solenoidal modes) and the width of the log-normal distribution \citep[see e.g.][]{2008ApJ...688L..79F,2012MNRAS.423.2680M}. Again, testing these prescriptions is clearly important.

From an observational point of view, the main problem we face when trying to test these predictions is that we do not observe the column density of the gas directly. Instead, we are limited to observational tracers that we hope are related in a straightforward fashion to the column density. By creating synthetic observations, we can test how well these tracers perform and understand any biases that they may introduce. For example, since we expect that on scales larger than a protoplanetary disk, dust and gas should generally be well coupled and the dust-to-gas ratio should not vary much within gas of the same metallicity, it is natural to use observations of the dust to constrain the gas column density. Dust extinction measurements provide the cleanest measure of the column density \citep{2001A&A...377.1023L,2013A&A...549A..53K,2017ApJ...834...91A} but require one to have deep near IR and/or mid IR maps of the background stellar population in order to detect the extinction. If these are unavailable, the technique cannot be used. Dust emission measurements have therefore attracted interest as an alternative tracer of the column density \citep[see e.g.][]{2010A&A...518L..98P,2013ApJ...766L..17S,2015A&A...578A..29S}. However, the thermal emission that we measure for the dust along a particular line of sight depends not only on the total column density but also on the temperature structure of the dust, since the dust emissivity varies strongly with the dust temperature. Therefore, in order to recover the column density from measurements of the dust emission, we need to also determine the dust temperature. This is typically done by SED fitting of dust emission measurements at different wavelengths. However, studies where this same technique is applied to synthetic dust emission maps show that it does not give an accurate measurement of column density when there is significant variation in the dust temperature along the line of sight \citep{2009ApJ...696.2234S,2011A&A...530A.101M,2017ApJ...849....1K,2015ApJ...809...17W}. 
Instead, in starless clouds it tends to systematically over-estimate the dust temperature and hence under-estimate the column density, while in star-forming clouds, a wider range of behaviour is possible. Synthetic observations have also been used to develop and benchmark improved methods for simultaneously constraining dust temperatures and column densities, such as the \textsc{ppmap} method of \citet{2015MNRAS.454.4282M,2016MNRAS.461L..16M,2017MNRAS.471.2730M}.

Synthetic observations have also been used to study other issues that may confound measurements of the column density PDF using dust, such as the impact of foreground or background contamination \citep{2015MNRAS.449.4465B,2015A&A...575A..79S,2017ApJ...849....1K,2016A&A...590A.104O} and edge effects arising from the finite size of the observational maps used for these studies \citep{2017A&A...606L...2A}. { In addition, they were used by \citet{2016A&A...590A.104O} to explore the impact of noise on the inferred PDF. Noise is found to have a large impact on the low-density tail of the PDF \citep[c.f.][]{2015A&A...576L...1L}, but does not significantly affect the peak or the high-density portion of the PDF unless the noise level is high ($\sigma_{\rm N} \sim 0.5 N_{\rm peak}$ or larger, where $N_{\rm peak}$ is the peak column density in the PDF).}

Some effort has also been devoted to studying the potential of molecular line tracers for this kind of study, although in practice, the limited dynamical range of common column density tracers such as $^{13}$CO or C$^{18}$O renders these methods of dubious utility \citep{2016A&A...587A..74S}.

In addition to measurements of the column density PDF, a few efforts have also been made to determine the volume density PDF \citep{2013ApJ...779...50G,2014Sci...344..183K} using observational data, with synthetic observations used to help calibrate the methods. However, these methods generally require one to supplement the observational data with additional assumptions about the structure of the gas, and as yet their usage is not widespread.

\subsubsection{Structure functions}
Structure functions provide a means of determining how strongly a field is correlated with itself on a particular scale or ``lag'' $l$. The structure function of order $p$ for a field $x(\mathbf{r})$ and a lag $l$ is given by
\begin{equation}
S_{p}(l) = \left \langle \left[x(\mathbf{r+l}) - x(\mathbf{r}) \right]^{p} \right \rangle, 
\end{equation}
where $l \equiv |\mathbf{l}|$ and the average is taken over all possible positions $\mathbf{r}$ and all orientations of $\mathbf{l}$ with respect to $\mathbf{r}$. Structure functions can be computed for any desired observed or derived field (integrated intensity, column density, velocity dispersion, etc.), but they are most commonly applied to the velocity field, as measured by e.g.\ the centroid velocity. The reason for this is that for a turbulent velocity field, we expect the velocity structure functions to be power-law functions of the lag, $S_{p}(l) \propto l^{\zeta(p)}$. For Kolmogorov turbulence without intermittency, we expect that $\zeta(p) = p/3$; on the other hand, statistical models of turbulence that account for intermittency \citep[e.g.][]{1994PhRvL..72..336S,2002ApJ...569..841B} predict different scalings.

\subsubsection{Delta variance}
\label{sec:deltavariance}
The $\Delta$-variance statistic was developed by \citet{1998A&A...336..697S} and \citet{2001A&A...366..636B} as a means of analysing the structure as a function of scale within an observational dataset. Suppose we have a 2D map\footnote{In principle, we can apply the $\Delta$-variance method to data with an artbitrary number of dimensions, but in practice it is most often applied to 2D datasets.} of some quantity $f(\mathbf{r})$. The $\Delta$-variance of this map on a scale $l$ is then given by 
\begin{equation}
\sigma_{\Delta}^{2}(l) = \left \langle \left(f({\mathbf{r}}) * \odot_{l}({\mathbf{r}}) \right)^{2} \right \rangle,
\end{equation}
where the average is taken over the whole of the map, the symbol $*$ indicates a convolution, and $\odot_{l}$ is a filter function known as the French-hat filter. This is defined by
\begin{equation}
\odot_{l}({\mathbf{r}}) = \frac{4}{\pi l^2} \left \{ \begin{array}{lr}
1 & \hspace{50pt} |{\bf r}| \leq \mbox{ }l/2 \\
-1/8 & l/2 < |{\bf r}| \leq 3l/2 \\
0 & |{\bf r}| > 3l/2 
\end{array} \right.
\end{equation}
and consists of a positive core surrounded by a negative annulus. The convolution of the French-hat filter with the function $f(\mathbf{r})$ produces a convolved map that only retains structure on scales close to $l$, and so the $\Delta$-variance is a measure of the structure in the map on that scale. By varying $l$ and examining how $\sigma_{\Delta}^{2}(l)$ varies, we can explore how the amount of structure in the image varies as a function of scale. 

The main advantage of the $\Delta$-variance relative to other ways of computing the structure as a function of scale, such as the use of structure functions, is that it is highly robust to the artifacts introduced by e.g.\ gridding of the observational data or the finite size of the map. Moreover, the method was extended by \citet{2008A&A...485..917O} to improve the treatment of edge effects and to allow it to more robustly deal with data with a varying noise level. 

Finally, note that while the method is often applied to maps of integrated intensity (see Section~\ref{sec:turb} for some examples), it can also be applied to maps of other quantities, such as the centroid velocity (see e.g.\ \citeauthor{2015MNRAS.451..196B}~2015b).

\subsubsection{Principal component analysis}

Principal component analysis (PCA) is a statistical technique designed to take a set of possibly correlated variables (e.g.\ the brightness temperatures in a position-position-velocity [PPV] cube of spectral line data) and to derive from them a set of orthogonal linear combinations of the variables, the so-called principal components. These combinations are constructed so that the first principal component has the largest possible amount of variance (i.e.\ it accounts for as much of the variation in the data as possible), the second principal component has the largest possible variance consistent with it remaining orthogonal to the first principal component, etc. This requirement ensures that for most datasets, most of the variation in the data is accounted for by the first few principal components. Therefore, if only the first few principal components are considered, PCA offers a means of significantly reducing the dimensionality of a dataset without introducing a large error; in other words, we can think of it as a type of data compression.

Mathematically, the application of the PCA method to a PPV datacube is straightforward. We write the datacube as $T(x_{i}, y_{i}, v_{j}) = T(r_{i}, v_j) = T_{ij}$, where $r_{i} = (x_{i}, y_{i})$ is the position on the sky and $v_{j}$ is the velocity channel. We then construct the covariance matrix $\mathbf{S}$, whose components are defined by
\begin{equation}
S_{jk} = \frac{1}{N} \sum_{i=1}^{N} T_{ij} T_{ik},    
\end{equation}
where $N$ is the total number of spatial positions. Using $\mathbf{S}$, we then solve the eigenvalue equation
\begin{equation}
\mathbf{S} \mathbf{u} = \lambda \mathbf{u}.
\end{equation}
Each eigenvalue $\lambda_{l}$ represents the amount of variance projected onto the corresponding eigenvector $\mathbf{u}_{l}$, where $l$ labels the eigenvalues in order of decreasing variance. Finally, we can compute the corresponding eigenimage for each eigenvalue by projecting the original datacube onto the appropriate eigenvector:
\begin{equation}
I^{l}(r_{i}) = \sum_{j=1}^{M} T_{ij} u_{lj},
\end{equation}
where $M$ is the number of velocity channels. These eigenimages are the principal components of the original datacube.

PCA was invented over a century ago by \citet{PCA1}, but was first applied to astronomical spectral line data by \citet{1997ApJ...475..173H}. They showed that the method could be used to study the scale dependence of velocity fluctuations in the spectral data. They did this by computing the autocorrelation function (ACF) for each eigenimage and eigenvector. The characteristic width of the ACF for the eigenimage yields a length scale, $\delta l$, while the characteristic width of the ACF for the eigenvector yields a velocity scale, $\delta v$. Therefore, we recover a pair of values $\delta v$ and $\delta l$ for each principal component. \citet{1997ApJ...475..173H} subjected observations of several different molecular clouds to PCA and showed that the recovered values of $\delta v$ and $\delta l$ obeyed a power-law relationship, $\delta v \propto \delta l^{\alpha}$. \citet{2002ApJ...566..276B} improved on the \citet{1997ApJ...475..173H} approach by accounting for the effects of noise and finite resolution. They also investigated the relationship between $\alpha$ and the properties of the three-dimensional velocity field. They created synthetic emission maps using a fractal Brownian motion technique and used these to demonstrate that the $\alpha$ index recovered by PCA is related to the slope of the energy spectrum of the turbulence, $\beta$, by $\alpha \simeq 0.33 \beta$. The link between $\alpha$ and $\beta$ was also investigated by \citet{2011ApJ...740..120R}, who used synthetic emission maps based on high-resolution simulations of isothermal turbulence to derive a very similar relationship.  \citet{2013MNRAS.433..117B} subsequently derived a slightly steeper relationship, $\alpha = 0.36 \beta$, using an analytical approach, although this is consistent with the earlier work once one accounts for the scatter in the empirical relationship and the approximations made in the analytical derivation.

\subsubsection{Dendrograms}
\label{sec:dendrograms}

A dendrogram is a method of graphically representing the hierarchical structure in a 2D image or 3D datacube (e.g.\ a position-position-velocity cube of molecular line emission). They were first introduced into astronomy by \citet{1992ApJ...393..172H} under the name of {\it structure trees}, but remained largely overlooked as a method for characterising astronomical images prior to the work of \citet{2008ApJ...679.1338R}, who were the first to refer to them as dendrograms. 

The basic idea underlying this technique is straightforward. Suppose we have a 3D datacube that consists of a set of intensity values at different positions and in different velocity channels. We can use these intensity values to construct a set of isosurfaces in the cube, i.e.\ surfaces on which the value of the intensity is constant.\footnote{Isosurfaces are the 3D analogue of the more familar 2D contours.} The shape of these isosurfaces depends on the distribution of intensities, but typically isosurfaces associated with very high values of the intensity tend to be broken up into multiple distinct closed surfaces surrounding high intensity regions in the datacube, while isosurfaces associated with low intensities are more likely to form a single connected surface. \citet{2008ApJ...679.1338R} give the analogy of a submerged mountain chain: when the water is high, only a few distinct peaks are visible, but when the water is low, the peaks merge together into a larger object (the mountain chain). 

Suppose we now ``threshold'' this dataset at some fixed value of the intensity (i.e.\ we set to zero any cells with intensities below this threshold, and set to 1 any cells with intensities above the threshold). After doing this, we are left with some number of distinct regions that remain non-zero, each of which is bounded by an isosurface corresponding to the selected threshold intensity. Each of the distinct surfaces constructed in this way is identified by the local maxima that it contains and corresponds to a point in the dendrogram. 

Now consider what happens if we change our intensity threshold. Over some range of threshold values, there will be no essential change: we will recover the same number of distinct regions (albeit with slightly different shapes), containing the same local maxima. This behaviour yields a set of vertical {\em branches} in our dendrogram (one per distinct isosurface), with a length corresponding to the range of contour levels over which we can vary our threshold without the number of distinct regions changing. If we vary our threshold enough, however, the set of distinct regions will change. For any pair of regions, there is some critical value of the intensity above which the regions are distinct, and below which they merge to form a single connected volume. This critical value is known as the merge level for that pair of regions, and corresponds to a horizontal line in our dendrogram connecting two branches. If we now further decrease our intensity threshold, the process repeats: we have some range of values for which our new structure remains distinct, which yields a new vertical branch in our dendrogram, but eventually this structure too will merge with some other structure. This procedure only terminates when we are left with only one single connected isosurface, corresponding to the {\em root} of the dendrogram. 

If we move through the dendrogram in the other direction, starting at the root with a low threshold and then systematically increasing the threshold, we will find that the root will eventually split into two branches, each of which may then in turn split when the threshold is increased further, etc. Branches that do not split no matter how high we place the threshold are known as the {\em leaves} of the dendrogram.

\begin{figure*}
    \centering
    \includegraphics[width=0.8\textwidth]{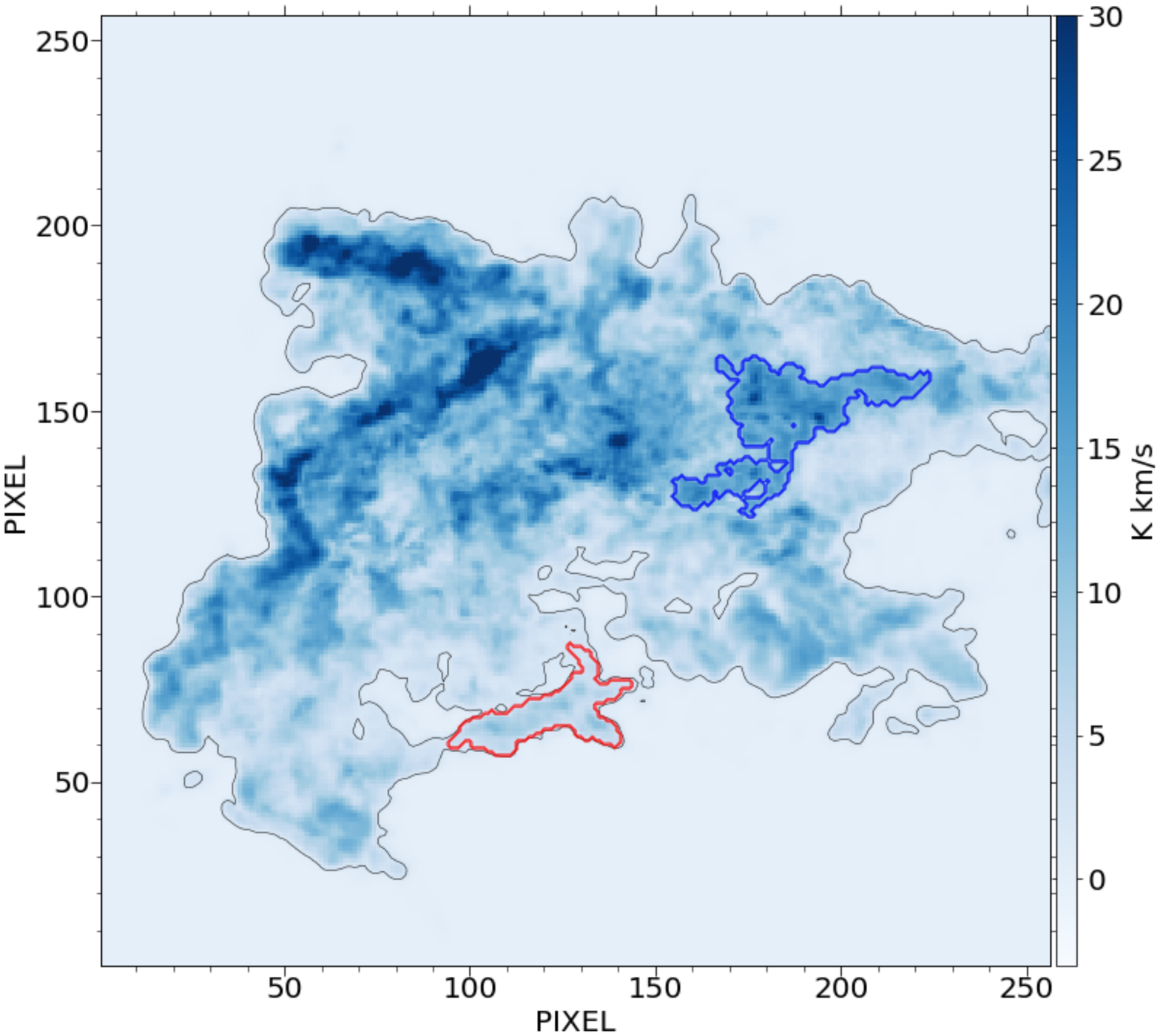}
    \includegraphics[width=0.8\textwidth]{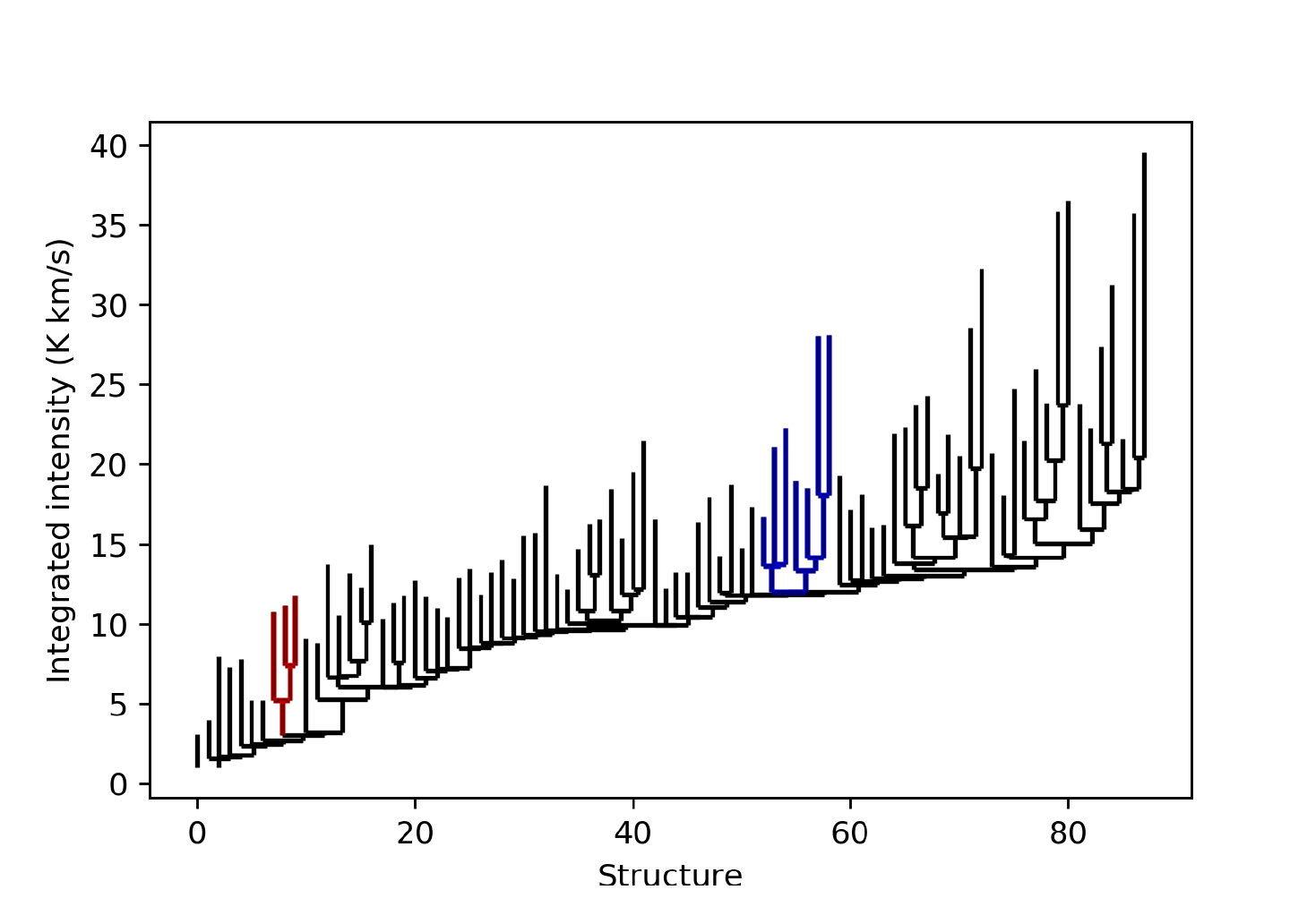}
    \caption{{\it Upper panel}: synthetic $^{12}$CO $J=1-0$ emission map, computed for one of the solar metallicity molecular clouds simulated by \citet{2012MNRAS.426..377G}. The map was produced by post-processing the data from the simulation using {\sc radmc-3d}, as described in more detail in the paper, and shows the velocity-integrated intensity of the line. The physical size of the displayed region is $16.2 \times 16.2$~pc. 
    To make it easier to pick out the fainter structures in the map by eye, we have limited the maximum integrated intensity displayed here to $30 \, {\rm K \, km \, s^{-1}}$. {\it Lower panel}: a dendrogram computed for the synthetic emission map shown in the upper panel. The dendrogram was constructed using the {\sc Astrodendro} Python package (see Section~\ref{useful}). The minimum value is set to $2 \, {\rm K \, km \, s^{-1}}$ and we retain only leaves that are separated from the merge level by $2 \, {\rm K \, km \, s^{-1}}$ and that include at least 20 pixels. We have highlighted two of the branches in the dendrogram and the corresponding regions in the emission map to show the connection between them.}
    \label{fig:dendro}
\end{figure*}

An example of the results of this process is shown in Figure~\ref{fig:dendro}. The dendrogram shown here was computed by applying the method described above to the synthetic CO emission map that is also shown in the Figure. We have highlighted several branches in the dendrogram and the corresponding regions in the emission map to show the connection between them. 

When applied to real data, it is necessary to modify the dendrogram method slightly to account for the effects of noise, which would otherwise create a large number of leaves that correspond only to noise fluctuations, rather than to real structures in the data. This is typically done by creating leaves only for regions that have a size (in pixels or velocity channels) greater than some minimum threshold, and that have a local maximum intensity that is separated from the merge level by several times the noise \citep[e.g.][consider only structures that are separated from the merge level by at least 4 sigma to be real]{2008ApJ...679.1338R}.

Dendrograms are useful as a means of identifying and characterising hierarchical structure in real and synthetic observations, and also provide a means of directly comparing real and synthetic data. One way in which this can be done is by associating physical properties (e.g.\ sizes, integrated intensities, velocity dispersions) with each leaf and branch in the dendrogram and then comparing the distributions of these properties. Examples of this approach can be found in \citet{2008ApJ...679.1338R}, in \citet{2009Natur.457...63G} and in \citet{2012A&A...539A.116W}. Alternatively, one can compare statistical properties of the dendrograms themselves (e.g.\ the number of structures, the distribution of lengths of the leafs and branches), as in \citet{2013ApJ...770..141B} and \citet{2017MNRAS.471.1506K}. The latter approach is more abstract, but has the advantage that it allows one to avoid the ambiguities involved in relating distinct physical properties to the leaves and branches of the dendrogram \citep[see e.g.\ the detailed discussion of this point in][]{2008ApJ...679.1338R}.

\subsubsection{Line ratios}
\label{sec:linerat}
Measuring the intensity of individual emission lines from atoms or molecules provides us with some information on the physical properties of the emitting region, but the information provided by any single line is often highly degenerate. For example, consider the case of the [CII] 158~$\mu$m fine structure line emission produced by a cold cloud of atomic hydrogen. If the density of the cloud is less than the critical density of the [CII] line, then the strength of the line we observe will depend on the density of the gas, its temperature, and the abundance of singly-ionized carbon (C$^{+}$) in the cloud. In addition, if the line is optically thick, its integrated intensity will also depend on the velocity dispersion of the C$^{+}$ ions. Measuring the intensity of the line may allow us to put upper or lower limits on some of these quantities, but cannot provide unambiguous information on all of them. 

The situation is very different, however, if we combine information from different transitions by examining their ratios. By comparing the strengths of transitions with similar excitation temperatures but very different critical densities, we can break the degeneracy between density and temperature. Similarly, by comparing the strengths of different transitions from the same molecule (e.g.\ CO), we can derive information on the temperature and density of the gas in a way that is independent of the chemical abundance of our tracer molecule, provided that the transitions are optically thin. 

Line ratios are therefore a potentially powerful diagnostic of the physical conditions in the ISM, and offer a convenient way of comparing simulations with observations. They clearly are most useful when there are a large variety of different lines to observe, and hence are primarily used to study moderately energetic regions such as PDRs or HII regions. 
Grids of PDR models with a simple 1D slab geometry have been computed by several different groups
(see e.g.\ \citealt{1999ApJ...527..795K}; \citealt{2006ApJ...644..283K}; 
\citealt{2005ApJS..161...65A}; \citealt{2012A&A...541A..76L}), and the values of many of the resulting line ratios have been tabulated as a function of the strength of the interstellar radiation field ($G_{0}$\footnote{$G_0$ is the Habing unit, which is the integral of photon flux in the range 912--2400~\AA$\,$ and is equal to $1.6 \times 10^{-3}$\,erg\,cm$^{-2}$\,s$^{-1}$}.) and the number density ($n$), allowing one to determine in an automated fashion the values of $G_{0}$ and $n$ that best fit a given set of spectral line data (see e.g.\ the Photo Dissociation Region Toolbox\footnote{\url{http://dustem.astro.umd.edu/pdrt/index.html}}, described in \citealt{2008ASPC..394..654P}, or the Interstellar Medium Database\footnote{\url{https://ism.obspm.fr/?page_id=403}}, which is based on the \citealt{2012A&A...541A..76L} PDR models).

For HII regions, a quantity of interest is often the hardness of the radiation field, and an extremely widely used diagnostic of this is the so-called BPT diagram (named after Baldwin, Phillips, and Terlevich; see \citealt{1981PASP...93....5B}). The most famous version of this diagram is a plot which compares the intensity ratio of the [NII] 6584~\AA$\,$ and H$\alpha$ lines against the intensity ratio of the [OIII] 5007~\AA$\,$ and H$\beta$ lines. Higher energy UV photons are required to produce doubly-ionised oxygen (O$^{++}$) than are needed to produce singly-ionised nitrogen (N$^{+}$), and so the relative strengths of the [OIII] and [NII] transitions can be used as a diagnostic of the hardness of the ionising radiation field. The normalisation of both lines by the intensity of hydrogen recombination lines lying close to them in wavelength allows one to correct for the effect of extinction, which otherwise would artificially depress the strength of the [OIII] line relative to the [NII]. Other similar sets of line ratios, using transitions from S$^{+}$ or neutral oxygen in place of N$^{+}$ are also in common usage \citep[see e.g.][]{2006MNRAS.372..961K}. BPT diagrams are most widely used in the extragalactic community, e.g.\ for distinguishing between galaxies dominated by star formation or by AGN, but the advent of wide-field integral-field spectrographs such as MUSE\footnote{\url{https://www.eso.org/sci/facilities/develop/instruments/muse.html}} has lead to the use of this kind of diagnostic becoming more common in the study of Galactic H$\,${\sc ii} regions \citep[see e.g.][]{2015MNRAS.450.1057M,2016MNRAS.455.4057M}.

One problem with line ratio methods, however, is that their interpretation is not straightforward if the resolution of the observations is not high enough to resolve relevant changes in the physical conditions in the source region (e.g.\ density substructure or fluctuations in the radiation field strength). For example, consider the simple case of a small, dense clump embedded in an extended, lower density cloud. In general, the line ratios produced by the gas in the clump will differ from those in the extended cloud. However, if our observations cannot resolve the clump, then the line ratios that we will actually measure will be some average of those in the clump and the cloud, with the details of the averaging depending on the relative strength of the two sets of lines and the angular size of the clump relative to the size of the beam. If we are unlucky, the physical conditions that we then infer based on the measured line ratios may not be a good representation of the conditions in either the clump or the extended cloud. Synthetic observations provide a powerful method for investigating the sensitivity of common line ratio diagnostics to this kind of effect. Examples of work looking at the reliability of molecular cloud and PDR diagnostics include the work on CO line ratios by \citeauthor{2017MNRAS.465.2277P}~(2017a,b)\nocite{Penaloza2017b} and on unresolved weak shocks in PDRs by \citet{2012ApJ...748...25P}. {An example of a similar study applied to H$\,${\sc ii} diagnostics is the \citet{2012MNRAS.420..141E} paper discussed in more detail in Section~\ref{sec:pillars}}. \textbf{Another limitation of line ratio diagnostics is that they are only useful if the emission is originating from the same volume (emission from the same system might originate from different parts of the volume along the line of sight).}

\subsubsection{Other methods} 
The techniques discussed above are only some of the many that have been used for comparing synthetic and real observations, albeit some of the more popular and widely used examples. Other techniques include the genus statistic 
\citep{2007ApJ...658..423K,2008ApJ...688.1021C}, the Modified Velocity Centroid (MVC) method \citep{2003ApJ...592L..37L}, the closely-related Velocity Coordinate Spectrum (VCS) and Velocity Channel Analysis (VCA) methods \citep{2000ApJ...537..720L,2004ApJ...616..943L,2006ApJ...652.1348L,2009ApJ...693.1074C}, the Spectral Correlation Function (SCF; 
\citealt{1999ApJ...524..887R,2001ApJ...547..862P,2003ApJ...588..881P}, the Histogram of Oriented Gradients (HOG; \citealt{2013ApJ...774..128S}), the Cramer statistic \citep{BF2004}, and the $Q^{+}$ algorithm \citep{2017MNRAS.466.1082J}. A comprehensive overview of the strengths and weaknesses of all of these methods is beyond the scope of this review, and would inevitably be incomplete, since new methods are being introduced all the time (see e.g.\ \citealt{2016A&A...585A..98A} for a recent example). However, there have been several recent efforts to determine which statistical tools are the most useful in which circumstances, as we discuss in more detail in the next section.

\subsubsection{Selecting the best method}
\label{sec:bestmethod}
As we have seen, there are many different statistical tools that we can potentially use for comparing synthetic and real observations. How do we determine which of these tools we should actually use in any given case? This is a difficult question to answer, and there has as yet been only a little work done on this topic. 

For a statistic to be useful, it must allow us to distinguish between different models despite the presence of observational limitations such as noise, or the limited dynamical range of many observed tracers. In addition, we should ideally be able to relate changes in the statistic in a relatively simple way to changes in the basic physical parameters of the model. An example of a statistic that often fails on this score is the column density PDF. We know from simulations that turbulent gas that is not self-gravitating develops a log-normal column density PDF (see Section~\ref{sec:pdfs}). The width of this PDF depends on the turbulent Mach number, the magnetic field strength, and also the balance between solenoidal and compressive modes in the turbulence 
\citep[see e.g.][]{2008ApJ...688L..79F,2012MNRAS.423.2680M}. In addition, it can also be artificially reduced if we are observing it with a tracer with a limited dynamical range (e.g.\ $^{13}$CO emission; \citealt{2009ApJ...692...91G}), or if our observational resolution is too poor to pick out small-scale fluctuations in the column density. What then can we conclude from the fact that an observed cloud and a simulated cloud have a log-normal column density PDF with the same width? Unfortunately, not very much, because the correspondence is not unique: if we changed both the Mach number {\em and} the magnetic field strength, we could probably find a whole range of other simulated clouds with the same PDF. The fact that our simulation produces the same PDF as we observe is a necessary condition for our simulation to be a good description of the real cloud, but not a sufficient condition. 

One obvious way in which we can try to overcome this problem is to examine multiple different statistics when comparing our synthetic observations with real observations. For example, \citet{2013ApJ...777..173B} consider nine different diagnostics when trying to determine which of two simulations -- one from \citeauthor{2011MNRAS.415.3253S}~(2011b) and one carried out specifically for the paper, using the same approach as in \citet{2013ApJ...770...49O} -- is a better match for the Perseus molecular cloud. They find that neither of the simulations does a good job of reproducing all of the observed properties of Perseus, even though some individual properties are reproduced very well, demonstrating the importance of considering multiple diagnostics \citep[see also][for additional discussion of this point]{2011IAUS..270..511G}.

Another important approach, which has seen increasing attention over the past few years, is to use carefully designed sets of simulations to establish which statistical techniques are sensitive to which physical conditions in the simulations. For example, \citet{2014ApJ...783...93Y} carried out simulations of self-gravitating isothermal turbulence in which they varied the turbulent Mach number, the initial magnetic field strength, the shape of the turbulent velocity spectrum and the temperature of the gas. They then analyzed PPV cubes of $^{13}$CO emission derived from these simulations using three different statistical techniques: PCA, the spectral correlation function, and the Cramer statistic \citep{BF2004}. They showed that all three statistics were sensitive to changes in the Mach number and the adopted temperature, although the relative sensitivity differed between statistics and also varied over time in the simulations. However, none of the statistics showed a meaningful sensitivity to changes in the initial magnetic field or the shape of the velocity spectrum. Another important point made by \citet{2014ApJ...783...93Y} is that when carrying out this kind of study, it is important to vary multiple parameters at a time in the simulations in order to be able to detect interactions between parameters. The results of ``one factor at a time'' studies, in which only one parameter is varied while the others are held constant, can be misleading if significant interactions exist between parameters. 

More recently, a similar study has been carried out by \citet{2017MNRAS.471.1506K}. They also consider simulations of self-gravitating isothermal turbulence, but vary five different parameters (the Mach number, plasma $\beta$ parameter, virial parameter, driving scale and solenoidal driving fraction), and consider 18 different statistical techniques. They find that many statistics are sensitive to changes in the driving scale or Mach number, but that far fewer are sensitive to changes in $\beta$, with the best technique in this case being the Velocity Coordinate Spectrum (VCS) statistic of \citet{2006ApJ...652.1348L}. Many of the statistics are also poor at detecting interactions between parameters, although the bispectrum (i.e.\ the Fourier transform of the three-point correlation function; see e.g.\
\citealt{2009ApJ...693..250B}) and the spectral correlation function perform particularly well here. 

Finally, the same approach has also been used by \citet{2016ApJ...833..233B} to study MHD simulations of wind-driven turbulence carried out by \citet{2015ApJ...811..146O}. Again, they apply a large number of different statistical tools to a set of simulations in which several key parameters are varied: in this case, the initial magnetic field strength, total stellar mass-loss rate and the simulation run time. They find that many different statistics are sensitive to the stellar mass loss rate, but few are sensitive to the magnetic field strength, with the exception of the Cramer statistic. 

Together, these three studies are the best guide that we currently have as to which statistical tools should be used in which situation when comparing simulations with observations. However, it is important to mention their limitations. First, these studies used what is known as {\em full factorial design}, where simulations are run for all of the possible combinations of the input parameters that are being varied. This quickly becomes very costly: even if each parameter has only two possible values, an experiment in which five parameters are varied (as in \citealt{2017MNRAS.471.1506K}) requires a total of $2^5 = 32$ simulations. Consequently, the simulations have to be fairly cheap, and so most of the studies carried out so far have used fairly low resolution simulations and have omitted effects (e.g.\ non-equilibrium chemistry) that may have an important impact on the results. Improving on this by carrying out higher resolution simulations with more physics that vary a wider range of input parameters is a clear priority for future research, but will likely require the use of a more sophisticated fractional factorial design in place of full factorial design (E.\ Koch, private communication).

Second, these studies each consider only a single observational tracer: $^{13}$CO $J=1-0$ line emission in the case of \citet{2014ApJ...783...93Y} and \citet{2017MNRAS.471.1506K}, and $^{12}$CO $J=1-0$ line emission in the case of \citet{2016ApJ...833..233B}. However, an increasing number of numerical studies have shown that the choice of tracer, and whether or not it is optically thick, can have a large impact on the statistical properties of the emission 
(see e.g.\ \citeauthor{2013ApJ...771..122B}~2013a,b; \citealt{2014MNRAS.440..465B};
\citeauthor{2015MNRAS.446.3777B}~2015a,b; \nocite{2015MNRAS.451..196B} \citealt{2015ApJ...811L..28B, 2015ApJ...799..235G, 2016ApJ...818..118C}). Ideally, therefore, one would like to repeat this kind of study for a range of different possible tracers.

Finally, the studies discussed here focused on only two different physical situations: isothermal, self-gravitating turbulence, and wind-driven turbulence, both in the context of gas within a giant molecular cloud. There is a clear and urgent need to carry out a similar type of statistical comparison for many of the other astrophysical scenarios in which synthetic observations are used (see Section~\ref{sec:apps}). 

It is important to stress that we mention these limitations not to criticize the work by \citet{2014ApJ...783...93Y}, \citet{2016ApJ...833..233B} and \citet{2017MNRAS.471.1506K}, but rather in an effort to inspire other researchers to carry out similar studies along these lines. More work of this type is desperately needed if we are to realize the full potential of the field of synthetic observations. We include this as a key technical development in section \ref{sec:technicaldevs}.

\subsubsection{Some useful resources for interpreting and manipulating synthetic observations}
\label{useful}

Here we include a list of some useful resources for interpreting and manipulating synthetic (and real) observations. Note that in this section we are not intending to summarise radiation transport or astrochemical codes, or tools for implementing instrumentational effects which are themselves discussed throughout this review. The tools highlighted here are applied to the completed synthetic images. 

\begin{itemize}

    \item \textsc{astrodendro}\footnote{\url{http://dendrograms.org/}} is a Python package for computing dendrograms (discussed in section \ref{sec:dendrograms}) from real or simulated data. 
    
    \item \textsc{dendrofind}\footnote{\url{http://galaxy.asu.cas.cz/~richard/dendrofind/}} is a Python algorithm for clump finding and for  making dendrograms. It operates on a position-position-velocity datacube of the brightness temperature \citep{2012A&A...539A.116W}.
    
    \item \textsc{turbustat}\footnote{\url{http://turbustat.readthedocs.io/en/latest/}} is a Python package that offers a range of tools for analysing the statistical properties of synthetic data cubes of (in particular turbulent) astrophysical media \citep{2017MNRAS.471.1506K}. These tools include structure analysis (e.g.\ dendrograms), distribution analysis (intensity PDFs, high order moment maps and more) and turbulence analysis (power spectra, velocity centroids, $\Delta$-variance, etc.).
    
    \item \textsc{glue}\footnote{\url{http://www.glueviz.org/en/stable/}} provides a framework for intuitive linking between different visualisations of the same (or related) datasets \citep{2015ASPC..495..101B}. As a hypothetical example, consider as two datasets an image of a star-forming region and a histogram of stellar ages from that region. Highlighting a feature in the plot of the stellar ages (e.g.\ a group of much younger objects) simultaneously highlights those objects on the image of the region. In the event that these young objects were at the periphery of the star-forming region, one might therefore look into the possibility of sequential or triggered star formation.  This interactive approach is a powerful way of exploring and understanding  data. 
    
       \item \textsc{galario}\footnote{\url{https://github.com/mtazzari/galario}} is a GPU-accelerated library for studying interferometric data directly in the Fourier plane (the domain in which the observations are made). This approach is advantageous compared to the more intuitive image plane analysis since the uncertainties are better understood and the overall evaluation of a model likelihood is much faster compared to the computation of a synthesized image from the visibilities. \textsc{galario} computes visibilities sufficiently quickly that it can be used to calculate a large number of likelihoods over a possible parameter space of input models using techniques such as Markov Chain Monte Carlo or $\chi^2$ optimisation. \textsc{galario} is also easy to use and to integrate with existing codes/scripts \citep{2017arXiv170906999T}.

    \item A number of different filament finding programs exist, such as \textsc{disperse} \citep{2013arXiv1302.6221S} or \textsc{getfilaments} \citep{2013A&A...560A..63M}. Similarly, widely used programs for identifying clumps in images include \textsc{clumpfind} \citep{1994ApJ...428..693W} and \textsc{gaussclumps} \citep{1990ApJ...356..513S}.
    
    \item{Software for visualising data includes \textsc{casa}, \textsc{gaia}, \textsc{iraf}, \textsc{ds9} {and the python packages \textsc{kapteyn} and \textsc{astropy}}.}
    
    \item{{\textsc{gildas}\footnote{\url{http://www.iram.fr/IRAMFR/GILDAS}} is a suite of packages for observation planning and image processing of sub-mm and radio observation applications with single dishes or interferometric instruments. }}
    
    \item{{\textsc{montage}\footnote{http://montage.ipac.caltech.edu/} is a resource for assembling collections of fits datafiles into mosaics. }}
    
\item \textsc{scouse}\footnote{\url{https://github.com/jdhenshaw/SCOUSE}} performs spectral line fitting, and can do so on a region-by-region basis. Although it is only a semi-automated method, \textsc{scouse} is capable of fitting large spatial regions of PPV datacubes in a quick and systematic way \citep{2016MNRAS.457.2675H}.     
\end{itemize}

Finally we note that much of the above software is reliant upon, or part of, the community-developed core Python package for Astronomy, \textsc{astropy}\footnote{\url{http://www.astropy.org/}} \citep{2013A&A...558A..33A}. 

\subsection{Caveats regarding discretisation/resolution}
\label{sec:pitfalls}

Throughout this entire part of the review we have discussed the methods and tools involved in the production and manipulation of synthetic observations. Where appropriate we have also highlighted various issues and, where known, the ways of overcoming them. A key problem is the resolution/discretisation of a model, which is sufficiently important that it warrants some further dedicated comments.

One of the main caveats involves how one goes about mapping a simulation output onto a domain from which the synthetic observation will be computed. This is particularly important if the discretisation schemes of the two codes are different, for example when importing the output of a Lagrangian SPH simulation into an Eulerian Monte Carlo code. This issue is explored in great detail in \citet{2017ApJS..233....1K} and \citet{2017arXiv171007108P}. However, it is still potentially an issue even if no re-gridding occurs. For example a dynamically captured shock may lie well away from a coarsely resolved, but observationally important ionisation front or H-H$_2$ transition. As another example, an extremely diffuse region that is poorly sampled by SPH particles is difficult to reliably map onto a grid, but such regions can still be important emitters. In short, a dynamically resolved calculation is not necessarily ``microphysically'' (or ``observational characteristcally'') resolved.

In line transfer calculations, a related effect involves the discretisation of velocities. If the velocity bins used for the line transfer calculation are narrower than the typical cell-to-cell difference in velocity in the simulation, then a naive treatment of the problem can lead to one missing emission in the intermediate velocity bins, since there is technically no gas at these velocities in the simulation (even though there clearly would be in the real cloud). This problem can be addressed by using the ``Doppler catching'' method discussed in \citet{2009ApJ...704.1482P} and implemented in e.g.\ {\sc radmc-3d}.

Finally, a further complication that one faces when modelling emission from point sources (usually stars) is that it may be the case that the dynamical simulation does not sufficiently resolve the gas distribution close to the source (e.g.\ discs or the innermost parts of accretion flows) which may be responsible for much of the emission, or for reprocessing the radiation field. In such cases, it may be necessary to introduce a sub-grid model to artificially increase the resolution of the dynamical calculation in the emitting region \citep[e.g.][]{2004MNRAS.351.1134K, 2017ApJS..233....1K}.

\section{\Large Part 3: Applications to star formation and the ISM}
\label{sec:apps}

We now review applications of synthetic observations in the literature to date. The sheer scope of application illustrates the power and growing popularity of synthetic observations. It also makes capturing the entirety of the literature very difficult. This review includes a substantial \total{citnum}\ references, nevertheless there will certainly be some omissions. We have attempted to group this part of the review by astrophysical application rather than observational/microphysical regime. Note that this should not be considered a review of all the respective fields included in this section, rather of their application of synthetic observations.  \\

\subsection{Massive young stellar objects}
Massive star formation continues to pose a considerable challenge, in no small part because observations of massive young stars are incredibly difficult. Massive stars are short lived and form from geometrically complex, dense clouds. They are effectively viewed from only a single point in time of their evolution. Furthermore, once a massive star forms, feedback (ionizing radiation, winds and supernovae) will act to clear the surrounding medium and thus potentially erase signatures of the star formation process. This same feedback has historically led to the issue of how a massive star can continue accreting once it grows to a certain luminosity. Lastly, candidate massive young stars in the process of forming are all at $>$kpc distances, making them difficult to resolve and hence interpret. So, we are limited in our observed information on massive star formation and it isn't obvious whether it proceeds as a scaled up version of low mass star formation, or by some other mechanism. In particular, how a massive star cluster is formed is a challenging problem as feedback could rapidly disperse star forming gas. Being such a challenging problem, synthetic observations will be crucial for its resolution.

\subsubsection{Massive young stellar object discs and outflows}

Radiation pressure is expected to drive away material once a star reaches 20--40$M_\odot$ \citep[][]{1974A&A....37..149K, 1987ApJ...319..850W}, which would halt further mass accumulation. Since stars are observed with masses greater than this threshold \citep[possibly over $100\,M_\odot$, e.g.][]{2010MNRAS.408..731C, 2013A&A...558A.134D}, some mechanism must continue to permit accretion. A strong candidate for this is accretion through a disc \citep[e.g.][]{2010ApJ...722.1556K, 2016MNRAS.463.2553R, 2016ApJ...832...40K, 2016ApJ...823...28K,  2017arXiv170604017H}. Such discs have been difficult to detect to date, though in recent years there have been some candidates around stars up to $30-60$~M$_{\odot}$ \citep[e.g.][]{2015ApJ...813L..19J,2016ApJ...823..125C, 2016MNRAS.462.4386I,2017A&A...602A..59C}. However these observations do not resolve the discs well owing to their being at $\sim$kpc distances. Centroid analysis therefore typically has to be employed, where the spatial location of the peak intensity of emission in each spectral channel is fitted and used to try and infer the presence of a disc. This kind of analysis is difficult to interpret. The nature of circumstellar discs about massive young stars is hence still uncertain. 

Synthetic observations of discs around massive young stars were computed by \cite{2007ApJ...665..478K}, with an aim of predicting the ability of the Expanded Very Large Array (EVLA) and ALMA to detect discs around massive young stellar objects (YSOs) and substructure within them. They post-processed radiation hydrodynamic calculations of the collapse and fragmentation of massive protostellar cores from \cite{2007ApJ...656..959K} and generated synthetic radio observations of ammonia (NH$_{3}$) inversion transitions at frequencies of $\sim25$\,GHz and sub-mm observations of rotational transitions of methyl cyanide (CH$_3$CN) at 
$\sim220$\,GHz. They found that the inner 200\,AU of their disc could not be kinematically detected in CH$_3$CN emission, owing to the large optical depth of the rotational transitions there, but that the longer wavelength NH$_{3}$ transitions did probe this region well. They also found that CH$_3$CN became a much better tracer of disk kinematics at radii of $\sim$200--2000\,AU. Overall, they predicted that discs around massive young stars should be at least marginally resolved, which is essentially the case given the evidence for discs in recent years mentioned above. The spirit of this comparatively early work is similar to many current models/synthetic observations and their comparison with real observations. However, it suffers from some significant limitations compared to more recent work. For example, convolution of their images only took place using a Gaussian beam, whereas in practice interferometric effects can be much more serious (see e.g.\ Section \ref{sec:instrumentation}). Furthermore, they assumed a constant CH$_3$CN abundance, the variation of which is now better understood (though still very limited). 

\cite{2017arXiv170604017H} ran radiation hydrodynamic calculations of the collapse of a core to form a massive (25\,M$_{\odot}$) young star. Collapse results in the formation of a disc which is resilient enough to radiation pressure that it permits continued accretion once  bipolar cavities are blown. These calculations were carried out with the \textsc{torus} radiative transfer and dynamics code, so synthetic observations of the resulting models were also easily obtained. They computed the time evolution of the SED, finding that initially strong 10\,${\mu}$m absorption, due to the embedded nature of the system, gets weaker over time as the ambient material is dispersed. They also computed \textit{Herschel} and ALMA continuum images, the latter of which were processed using \textsc{casa}. They compared these scattered light continuum images with observations by \cite{2004A&A...427..505A}, finding that typically their conical nebulae (the scattered light signatures of bipolar outflows) were a factor 4 larger in spatial extent than in the aforementioned observations, assuming kpc distances. They also found that spiral features in their discs should be observable in ALMA 1\,mm continuum images. \cite{2017arXiv170604017H} additionally produced synthetic CH$_3$CN observations of their model, but without accounting for interferometric effects. They found similar conclusions to \cite{2007ApJ...656..959K}, but further commented on the fact that the kinematic asymmetry of the disc is dependent upon the observer viewing angle. For example, depending on this the line profile can  be symmetric, or red or blue-asymmetric.

\begin{figure*}
    \centering
    \includegraphics{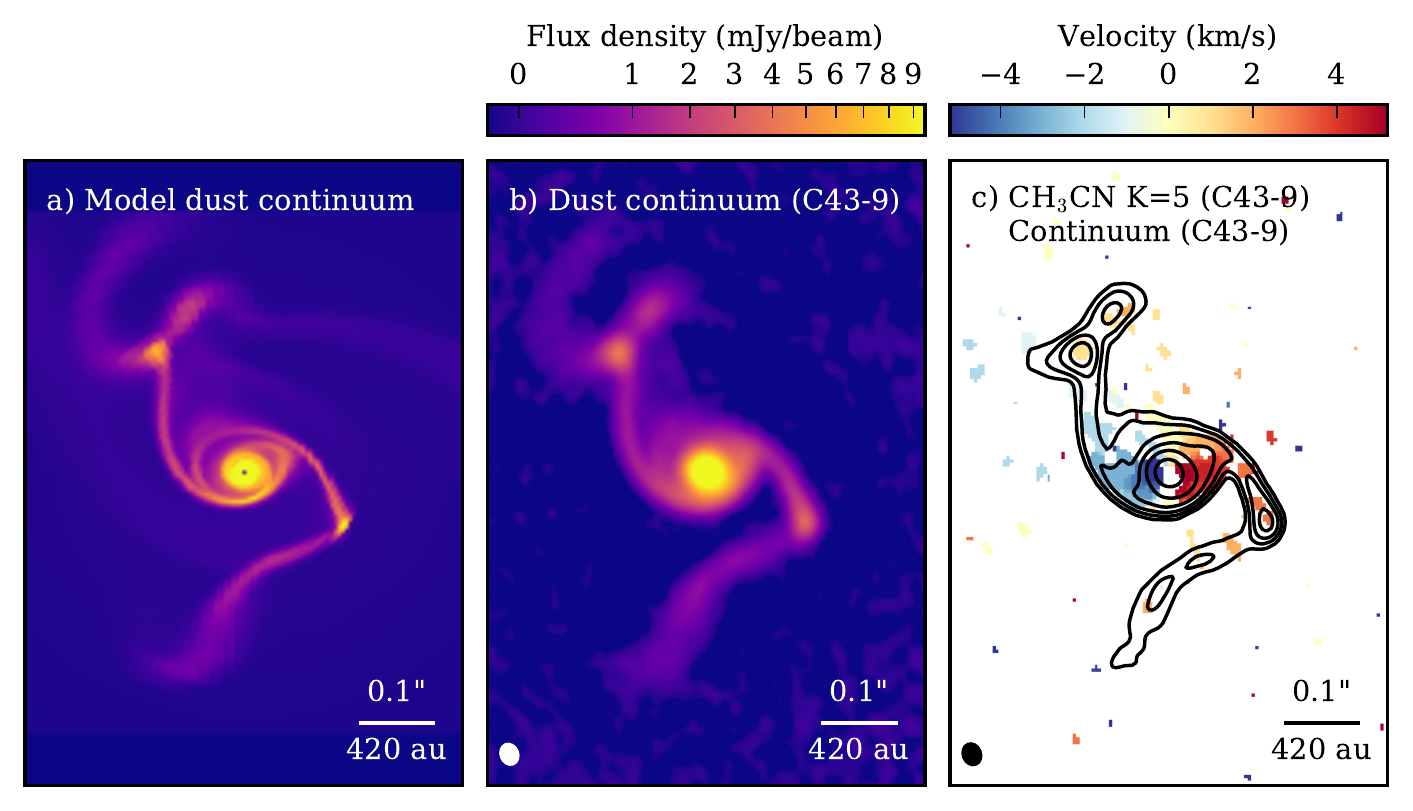}
    \caption{(a) 1.2 mm dust continuum image generated with the radiative transfer code {\sc radmc-3d} \citep{2012ascl.soft02015D} of a simulation of an accretion disk around a young high-mass star \citep{2017arXiv171001162M}, showing two gaseous clumps forming in the fragmenting disk. (b) Same as (a), but processed using the {\sc CASA}  ``simobserve'' command to show the disk as it would appear if observed by ALMA (configuration C43-9, 2.1 hr on source and 1.8 GHz bandwidth).
    (c) The first moment map of the CH$_3$CN $J = 13–12, K = 5$ line with the 1.2 mm continuum emission over-plotted in contours. The simulations recover emission and kinematics similar to those observed towards AFGL 4176 by ALMA in Cycle 1 \citep{2015ApJ...813L..19J}. This Figure is used Courtesy of Katharine Johnston.}
    \label{fig:MeyerJohnston}
\end{figure*}

\cite{2017arXiv171001162M} recently studied the fragmentation of massive YSO discs to form the precursors of spectroscopic massive binaries. They included synthetic ALMA mm continuum observations, predicting that spiral structure and even fragmented cores should be detectable from a 2000~AU, 30\,M$_\odot$ disc at distances of $\sim4$\,kpc in 3 hours using ALMA configuration 4.9. That this work and \cite{2017arXiv170604017H} both predict mm continuum observability of spirals, but none have been detected to date, is perhaps suggestive of the fact that the dust disc may be of smaller radial extent than the gas, which has been observed to be the case in lower mass systems \citep[e.g.][]{2013A&A...557A.133D}. Decoupled dust-gas dynamics may therefore be important in this scenario. An illustration of continuum and molecular line synthetic observations for this model is given in Figure \ref{fig:MeyerJohnston}. 

\cite{2018arXiv181011398J} computed CH$_3$CN synthetic observations for lines that vary more significantly in the temperatures that they trace, using synthetic observations computed from semi-analytic density, temperature and velocity models \citep[e.g.][]{2016MNRAS.458..306H}. These included full interferomentric processesing using \textsc{casa}. They also included a prescription for molecular freeze out and thermal dissociation. Although these calcualtions lack the full dynamical complexity of the \cite{2007ApJ...665..478K} and \citet{2017arXiv170604017H} models, their comparatively low computational cost meant that a large parameter space could be computed, probing disc structure, viewing angle and different lines. They found that detecting discs around massive stars (and in particular detecting substructure in such discs) at $\sim$\,kpc distance is difficult. However, they were able to propose optimal lines and ALMA configurations for detecting discs/substructure, and also tested processing techniques such as Gaussian and Keplerian subtraction to make spiral features/other substructure more detectable.

\cite{2014ApJ...788...14P} performed 3D radiation hydrodynamic calculations of massive YSOs with a sub-grid model for launching an outflow. Unlike the other models discussed in this section, this work considered multiple stars in a given model, finding that stars forming from the same accretion flow can have aligned outflow axes, resulting in a collective outflow. From these models they used \textsc{radmc-3d} to compute synthetic $H_2$ and CO observations. \cite{1999A&A...343..953V} had previously computed H$_2$ S(1) 1-0 line synthetic observations from models of outflows where the outflow is both precessing and changing its outflow velocity sinusoidally. They found that the (shock tracing) H$_2$ S(1) 1-0 line traces the edge of the outflow. \cite{2014ApJ...788...14P} self-consistently treat the precession and outflow velocity based on the sink particles in their models, meaning that their outflows can achieve higher velocities and hence shock excitation and H$_2$ S(1) 1-0 line emission throughout a larger extent of the outflow (i.e. not just at the edges). This self consistent launching, combined with the fact that there are multiple sources in their models mean that their collective outflows are also much more structurally complex than those of \cite{1999A&A...343..953V}.

\cite{2015ApJ...799...53K} tested the classification of high mass YSOs in centre  of the Milky Way Galaxy (the central molecular zone). They set up a grid of geometrically simple models, from which they used the radiative transfer code \textsc{Hyperion} \citep{2011A&A...536A..79R} and \textsc{FluxCompensator} \citep{2017ApJ...849....3K} to model the 8, 24 and 70 micron continuum observations. They demonstrated that previously classified young high mass YSOs in that region were older than previously inferred. The star formation rate \citep{Yusef-Zadeh:2009} extracted directly from the 24-micron continuum point sources in that region could therefore be much lower, which would agree better with indirect estimates of the star formation rate using free-free emission \citep{2013MNRAS.429..987L}.

\subsubsection{Ultracompact HII regions}
\label{sec:uchiireg}

Before blowing out into the larger extended H\,\textsc{ii} regions, photoionised bubbles about massive YSOs are comparatively very compact and come in a variety of morphologies which \cite{1989ApJS...69..831W} classify as spherical, cometary, irregular, core-halo or ``shell''. This class of system (specifically, ionised regions with extent $<0.1$\,pc and number density $>10^4$\,cm$^{-3}$) are referred to as ultracompact H\,\textsc{ii} regions \citep[for reviews see][]{2002ARA&A..40...27C, 2007prpl.conf..181H}. Some of the main drivers of research into ultracompact H\,\textsc{ii} regions are to understand the different morphologies and their link to the massive star formation process. A major difficulty in the study of this early phase of massive star formation is that the objects are usually deeply embedded. Being more optically thin, radio emission from ultracompact H\,\textsc{ii} regions is hence one of the earliest possible \textit{direct} probes and hence complementary pan-chromatic direct tracers are not available. However, indirectly much of the stellar emission is reprocessed by dust and re-radiated in the sub-mm and infrared. Synthetic observations to interpret the different types (e.g.\ morphologies) of radio emission detected, and to determine what the reprocessed emission might tell us, are hence essential. 

\cite{2006ApJS..165..283A} calculated the radiation hydrodynamic evolution of the medium about both static (in a non-uniform medium) and propagating O stars. Their calculations resulted in compact (so not quite ultracompact, but still small, $\sim 0.1-0.3$\,pc) cometary H\,\textsc{ii} regions. From these models they also derived emission maps and long-slit spectra, including convolution with a $1''$ Gaussian beam. They found that a wind can result in limb brightening of the morphology because the wind bubble diverts the photoevaporative flow. However, the strength of this limb brightening is a function of the initial density gradient. Permitting the star to move (rather than holding it stationary) pushes the peak of emission closer to the star because the increased ram pressure makes the wind-blown region smaller (and hence the photoionised region is closer to the star). They also found that the velocity of the shell bounding the photoionised gas is more sensitive to the density of the medium than the propagation velocity of the star. Lastly, they found that the speed of the star affects the location of the largest line widths, which for moving stars is located closer to the star. 

\begin{figure*}
     \centering
     \includegraphics[width=18cm]{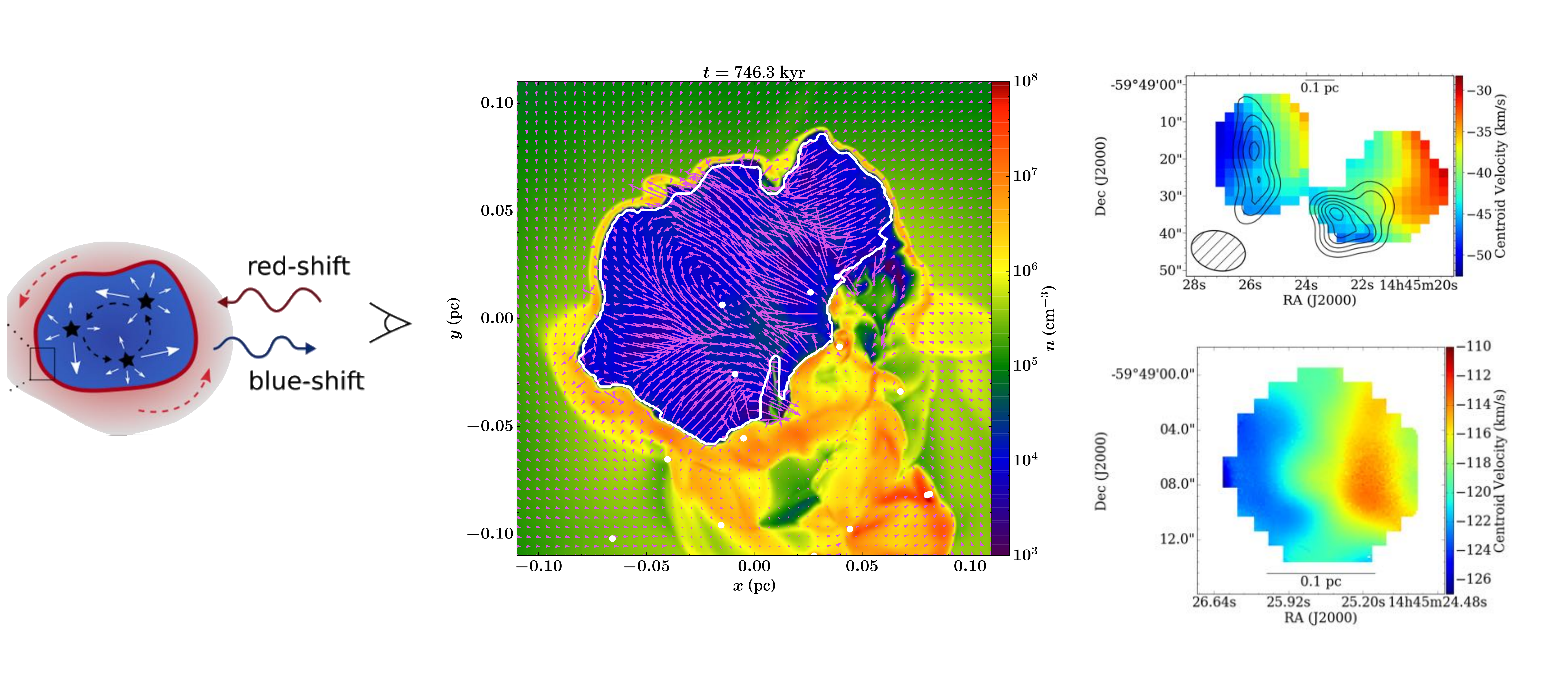}
     \caption{Illustrations from the work of \cite{2018arXiv180500479D}. The centroid velocity maps of real UCH\,\textsc{ii} regions show large velocity gradients (upper right panel). Synthetic observations of radiation hydrodynamical models of UCH\,\textsc{ii} region evolution (lower right and middle panel respectively) give similar features, which are explained by circulation in the ionised gas driven by angular momentum inherited from the initial molecular cloud, as well possible driving by the massive YSOs in the ionised gas. }
     \label{fig:dalgleish}
 \end{figure*}

\citeauthor{2010ApJ...711.1017P}~(2010a,b) \nocite{2010ApJ...719..831P} and \citet{2011ApJ...729...72P} simulated the formation of O-type stars in rotating, flattened molecular clouds, and followed the growth of their \hii regions. They then produced synthetic images of these \hii regions in the radio, and in dust continuum emission, and showed that the interaction of the ionised gas with neutral gas flow -- particularly accretion flows -- gave rise in a single model (but perhaps at different epochs) to all the major \hii region morphologies identified in the seminal work of \cite{1989ApJS...69..831W}. They also found (as did \citealt{2005MNRAS.358..291D}) quenching of the \hii regions by neutral gas caused them to flicker on timescales short enough to be observable. \citet{2011MNRAS.416.1033G} further analyzed these synthetic observations and made detailed statistical predictions for the time variability of the radio emission. Variations in the radio continuum emission from ultra-compact \hii regions in Sgr B2 with durations and magnitudes consistent with this theoretical picture have subsequently been reported by \citet{2014ApJ...781L..36D,2015ApJ...815..123D}. It therefore seems promising that the variety of radio continuum morphologies might be a time-dependent interplay between the stellar feedback and (accretion-)flow of the ambient medium. {Another recent development is that of \cite{2018arXiv180500479D} who predict the kinematic signatures of H\,\textsc{ii} region gas that has rotation both due to the initial angular momentum of the collapsing cloud, as well as due to being stirred by young massive stars. They found that this readily explains the large velocity gradients in ionised gas observed towards G316.81–0.06. Figure \ref{fig:dalgleish} shows a cartoon of the end result of this scenario (left panel). It also shows a snapshot from a radiation hydrodynamic calculation of massive YSOs forming from a collapsing cloud and producing an UCH\,\textsc{ii} region (middle panel), a synthetic centroid velocity map of this model (lower right panel) and a centroid velocity map towards a real UCH\,\textsc{ii} region (upper right panel). The large velocity gradients observed do appear to be consistent with the numerical models. Of future interest in this regard is whether the initial angular momentum of the cloud or propagation of the massive YSOs dominates the stirring, as well as understanding for how long such a signature survives.  }
 
\cite{2016ApJ...818...52T}, \cite{2017ApJ...849..133T} studied the breakout and evolution of UCH\,\textsc{ii} regions as the cavities blown by the young massive stellar object become optically thin enough to the UV to be photoionised. Early on it is the parts of the cavity most offset from the poles that have the lowest optical depth, which result in a cross-like morphology (dependent upon the viewing angle). However after less than 10\,kyr from the H\,\textsc{ii} region onset  in the cavity, the whole region becomes photoionised. They produced synthetic radio free-free continuum and LTE H\,$\alpha$ recombination line observations from their models, assuming a 1\,kpc distance. In addition to studying the morphology, with the latter they predicted line widths of $\sim100\,$km/s.  From the SED (which includes both dust and free-free continuum emission) they find that free-free dominates over the dust at <100\,GHz and is insensitive to the viewing angle (whereas the dust dominated component is dependent on viewing angle). In \cite{2017ApJ...849..133T} accretion bursts were shown to be able to modify the free-free emission in the cavity by roughly an order of magnitude on timescales of tens to hundreds of years.

\subsection{Evolved HII regions and wind-blown bubbles}
Once the natal envelope of a massive star or a group of such stars is dispersed, their winds and ionising radiation heat up and displace the surrounding ISM -- a process known as feedback \citep[see][for a recent review]{Dale:Review:2015}. Given that stars form at similar times in clusters, this heating and displacement of star-forming material most likely shuts off the bulk of the subsequent star formation potential in that region (at least in the current star formation epoch). However, the action of feedback may also induce new (so-called ``triggered'') star formation, the importance of which is still debated, owing to a lack of agreement over how to distinguish triggered star formation from star formation that would have occurred regardless of the action of the feedback \citep[e.g.][]{2015MNRAS.450.1199D}. 

In small H\,\textsc{ii} regions, powered by only a few massive stars, photoionisation is the dominant feedback process  with radiation pressure acting at only the ten per cent level or less \citep[e.g.][]{2014MNRAS.439.2990S, 2015MNRAS.453.2277H} and winds clearing only a small interior zone \citep[see e.g.]{1984ApJ...278L.115M}. The majority of the H\,\textsc{ii} in the Galaxy are of this type. However, since the slope of the star cluster mass function is shallower than $-2$, we know that most of the massive stars themselves form in massive clusters, in which winds and radiation pressure play a dominant role \citep{2009ApJ...703.1352K, 2015RMxAA..51...27R, 2016MNRAS.462.4532G, 2017MNRAS.470.4453R}. Examples of clusters of this type include the Arches cluster close to the Galactic Centre, the 30 Doradus cluster in the Large Magellanic Cloud, or the super star clusters that one observes in starburst galaxies.

The region directly impacted by a wind is heated to very high temperatures, and can emit strongly in the X-rays. However, often wind-blown cavities are so diffuse that they are still incredibly difficult to detect. Indirect methods of detecting wind-blown bubbles, for example using infrared arcs, are discussed in Section \ref{sec:movingOBstars}. Elsewhere in this section, we will generally focus on photoionisation-dominated regions.

The H\,\textsc{ii} region can be tens of parsecs in diameter, with an evolving morphology determined at least in part by the ambient (pre-feedback) ISM. In a region dominated by an isolated massive star, or one dominant massive star, the effects of stellar motion can also become important (see Section \ref{sec:movingOBstars}). Furthermore, pre-feedback structures such as filaments are sculpted into pillars and bright-rimmed clouds as feedback operates, which are themselves objects of interest but also may be sites of triggered star formation. The energy and momentum injected through feedback may also contribute significant energy input and driving of turbulence in the ISM, though whether the latter is dominant over gravitational instability is uncertain \citep[e.g.][]{2008PhST..132a4026P,2009ApJ...694L..26G, 2016MNRAS.458.1671K}. Furthermore, sculpting of the ISM provides diffuse channels through which supernova energy and momentum can escape more efficiently into the wider ISM \citep[e.g.][]{2013MNRAS.431.1337R}. Feedback on the scale of a star-forming region can hence have an effect on the evolution of the host galaxy. Here we discuss synthetic observations applied to H\,\textsc{ii} regions, structures within them such as pillars, and the medium surrounding moving OB stars. 

\subsubsection{Feedback and the global structure of HII regions}
Synthetic observation studies focusing on the larger scale structure of H\,\textsc{ii} regions are typically aiming to compare the H\,\textsc{ii} region morphology with observed systems, or to interpret the pre-feedback nature of the ISM and star-forming material. There are also many observational diagnostics of H\,\textsc{ii} regions, for example using line ratios to infer temperature/electron density and interpreting the structure of the ionised gas distribution, that can be tested in complex scenarios with synthetic datasets. 

\begin{figure*}
  \centering
  \includegraphics[width=\linewidth]{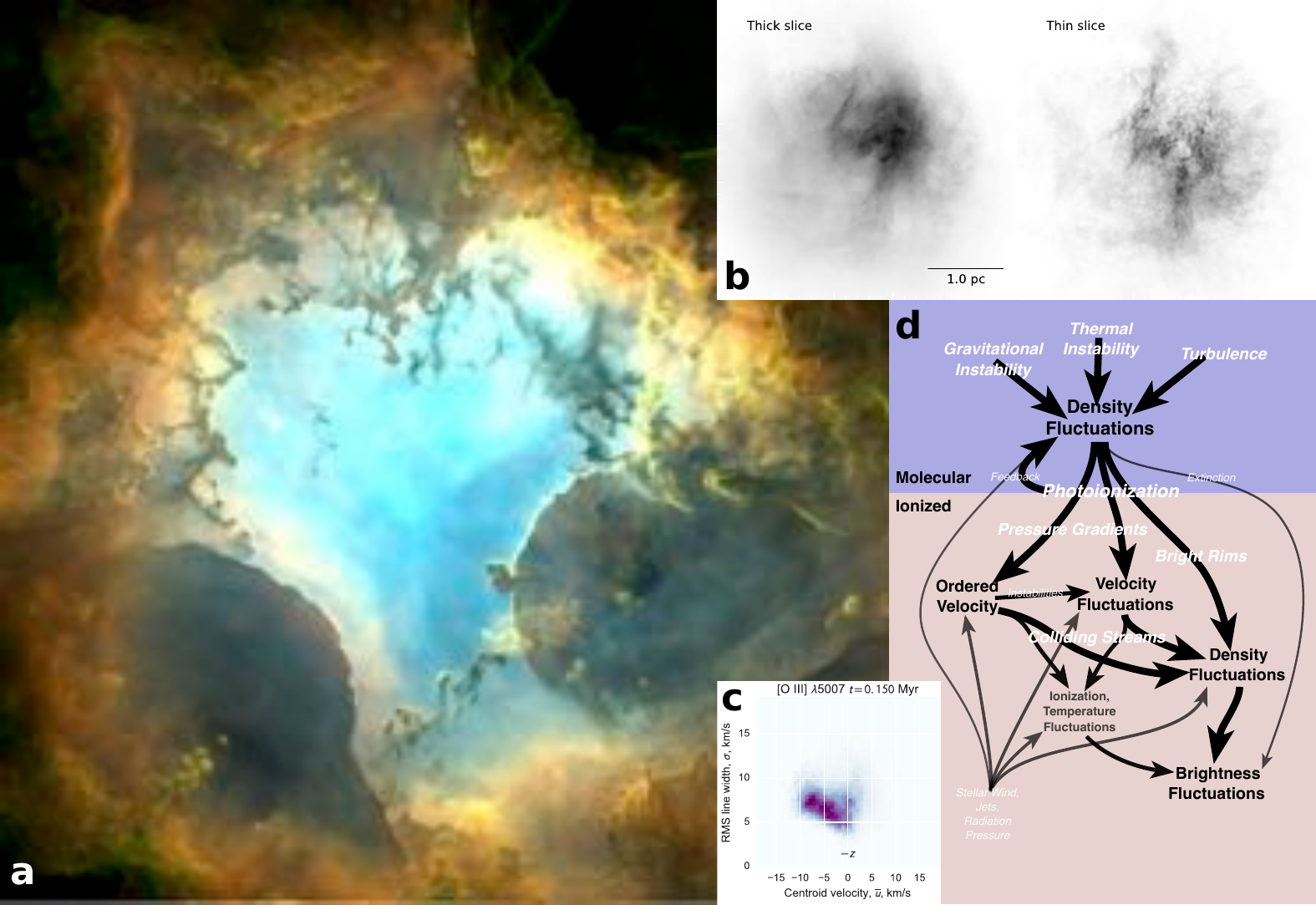}
  \caption{Turbulence in simulated H\,\textsc{ii} regions. (a) Simulated optical
    emission line image of H\,\textsc{ii} region at age of 250,000~years from
    \citet{2014MNRAS.445.1797M}, building on earlier work of
    \citet{2006ApJ...647..397M, 2011MNRAS.414.1747A}. (b) Comparison of synthetic
    spectral maps using thick (left) versus thin (right) velocity
    slices.  The thick slice is sensitive only to emissivity
    fluctuations, whereas the thin slice is also sensitive to velocity
    fluctuations and therefore shows more fine-scale structure.  (c)
    Predicted joint distribution (PDF) of linewidth and centroid
    velocity for a simulated H\,\textsc{ii} region that shows a champagne flow
    towards the observer \citep{2016MNRAS.463.2864A}.  (d) Causal
    relationships between different types of fluctuations in molecular
    clouds and H\,\textsc{ii} regions, as deduced from comparison between 
    synthetic observations and real observations of the Orion Nebula.
    Turbulence and ordered photoevaporation flows are found to
    contribute roughly equally to the observed density fluctuations. Figure courtesy of William Henney and Jane Arthur. }
    \label{fig:ArthurHenneyFig}
\end{figure*}

\cite{2006ApJ...647..397M}, \cite{2011MNRAS.414.1747A} and \cite{2014MNRAS.445.1797M} model the photoionisation of turbulent boxes using the radiation-hydrodynamics code described by \cite{2006NewA...11..374M}. Two-level models assuming that the ionisation fractions of heavier elements are constant fractions of the ionisation fraction of hydrogen were used to create brightness maps in the H$\alpha$, [O III], [S II] and [N II] lines, rendered using the usual HST colour map (see Figure \ref{fig:ArthurHenneyFig}, panel A). \cite{2006ApJ...647..397M} and \cite{2011MNRAS.414.1747A} use their results mostly for morphological comparisons regarding, for example, the apparent smoothness of the bubble inner edge and the prevalence of pillars, particularly when comparing the case of a bright O-star and a more common but fainter B-star. \cite{2014MNRAS.445.1797M} went further and generated synthetic second-order structure functions of velocity centroids, and velocity channel maps with which to analyse the turbulent velocity field (Figure \ref{fig:ArthurHenneyFig}, panels B and C). They find that projection effects (smoothing, or anisotropic bulk motions viewed from certain angles) seriously hamper the accurate recovery of the structure function slopes, but that velocity channel analysis (VCA) yields much more robust results. They employ these results in the analysis of genuine data from the Orion region in \cite{2016MNRAS.463.2864A}. They test diagnostics of velocity fluctuations in the ionised gas, finding that of structure functions, velocity channel analysis, linewidths and PDFs, VCA is the most reliable method for determining the spectrum. For Orion they suggest that a global champagne flow and smaller-scale turbulence contribute equally to the total velocity dispersion, and furthermore that the photoionised gas is based in a thick shell, with a possible wind-blown inner cavity. Overall a large number of inter-related processes are responsible for brightness fluctuations in H\,\textsc{ii} regions such as pressure fluctuations, inflowing material from the irradiated globules and the rims of the region and stellar winds/outflows. These processes and their relationships are illustrated in panel D of Figure \ref{fig:ArthurHenneyFig} \citep{2016MNRAS.463.2864A}.

\citet{2012MNRAS.427..625W,2013MNRAS.435..917W,2015MNRAS.452.2794W} performed high-resolution SPH simulations of GMCs with different fractal dimensions ($d_f\simeq2.0-2.8$) in an attempt to understand the importance of triggered star formation and the overall structure of H$\,${\sc ii} regions. They found that the fractal dimension is the key parameter that determines whether the geometry of the H$\,${\sc ii} region is ``shell-dominated''  (e.g.\ if $d_f\lesssim2.2$, large-scale structures are commonly found) or pillar-dominated (e.g.\ if $d_f\gtrsim2.6$, the shell breaks up in a number of smaller individual knots). Synthetic observations from these models were generated using {\sc radmc-3d} to estimate the dust $870\,\mu$m emission, whiched showed morphological agreement with the global structure of the RCW~120 galactic H$\,${\sc ii} region. Note that RCW~120 has also been interpreted in other ways, including as a moving O star (see Section~\ref{sec:movingOBstars}) and or a cloud-cloud collision (see Section~\ref{sec:ccc}). 

Obtaining dynamical information from radio observations of \hii regions requires modelling of radio recombination lines (RRLs) which itself requires consideration of the line profile function and departures from local thermodynamic equilibrium (LTE). \cite{2012MNRAS.425.2352P} created synthetic observations of spherically-symmetric \hii regions for comparison with ALMA and EVLA data. They used a version of {\sc radmc-3d} already modified to include free-free emission by \citeauthor{2010ApJ...719..831P}~(2010b), that they further improved to sample the Voigt line profile.
Departures from LTE were computed using a pre-computed table in temperature-electron number density space of the departure coefficients. They show that non-LTE effects give rise to asymmetric line profiles even in symmetric \hii regions. They then use {\sc casa} to create synthetic visibilities using various ALMA and EVLA array configurations, integration times and, in the case of ALMA, different levels of water-vapour-induced noise. They find that the results depend very strongly on whether the observations have sufficient angular resolution and sensitivity to probe the optically-thick core of the H\,\textsc{ii} region. The ability to discern line asymmetries also depends on angular resolution, since poor resolution blends the optically-thin emission (which has symmetric line profiles) from the outer parts of the H$\,$\textsc{ii} region with the optically-thick central emission. However, increasing the angular resolution too much leaves the observations too insensitive to detect RRLs at all. They use these outcomes to compute under what conditions asymmetric line profiles \textit{could} be observed.

In the paper series of \cite{2017ApJS..233....1K,2017ApJ...849....1K,2017ApJ...849....2K}, IRAC, MIPS, PACS and SPIRE synthetic continuum observations were produced using hydrodynamical simulations from \cite{DaleBoth:2014} and the radiative transfer code \textsc{Hyperion} \citep{2011A&A...536A..79R}. In the first paper in the series, \cite{2017ApJS..233....1K} describe in detail how they produced these synthetic observations and how to overcome the obstacle of mapping a particle-based simulation onto a Voronoi tessellation before being able to solve the radiative transfer problem. They also describe their tests on the importance of the radiation originating from regions beyond the resolution limit of the adopted SPH simulation. Observational limitations characteristic of real observations were accounted for using the \textsc{FluxCompensator} package \citep{2017ApJ...849....3K}. 
They present about 6000 synthetic continuum observations at multiple wavelengths for models of different evolutionary stages, distances and orientations. These resulting synthetic observations are all freely available to the community \citep{2017ApJS..233....1K}. In \cite{2017ApJ...849....1K} They used these synthetic observations to test mass estimates using the commonly-applied modified black-body fitting technique. The recovered gas mass, dust density, and dust temperature show large errors, especially on a pixel-by-pixel basis in the vicinity of young stars, in agreement with the findings of \cite{2015ApJ...809...17W}, where they tested the same technique for cores. They then went on to propose that these pixels can be cut away through $\chi^2$-clipping to improve observational mass estimates. \cite{2017ApJ...849....2K} also showed that the star-formation rate estimated from diffuse infrared continuum emission is erroneous if this method is applied to a single star-forming region, but that there was a promising relation between the counted YSOs in the continuum and the instantaneous star-formation rate.

\subsubsection{Pillars and bright-rimmed clouds}
\label{sec:pillars}
Pillars and bright-rimmed clouds at the boundaries of H\,\textsc{ii} regions have been the subject of multiple synthetic observation efforts. Famous examples of such objects are the ``Pillars of Creation'' in M16 and the Horsehead nebula in Orion. Generally these features are thought to have once been overdensities in the turbulent star-forming medium, which are more resilient to the action of feedback from massive stars than the lower density gas surrounding them \citep[e.g.][]{1989ApJ...346..735B,2006MNRAS.369..143M, 2010ApJ...723..971G,2011ApJ...736..142B}. The lower density regions are excavated by feedback, leaving behind the pillars and rims that photoevaporate more slowly. However, pillars can also be the result of instabilities acting even in initially very smooth gas \citep[e.g.][]{2002MNRAS.331..693W, 2006ApJ...647.1151M, 2008ApJ...672..287W} or of perturbations in the shape of the ionisation front \citep[e.g.][]{2012A&A...546A..33T}.

Seminal work on bright-rimmed clouds and cometary globules was the early radiation hydrodynamic calculations of \cite{1994A&A...289..559L}.  In this, their first in a series of three papers, they compared a measure of the optical emission from their models with the morphological characteristics of real systems, showing that a clump being compressed over time can reproduce all observed morphologies depending on the time in its evolution. In \cite{1995A&A...301..522L}, they used their models to compute approximate position-velocity diagrams for an arbitrary optically thin transition with a critical density of 10$^3$\,cm$^{-3}$. They compared this with CO line observations towards the globule CG7S, finding a similar kinematic profile and velocity gradients. Finally in \cite{1997A&A...324..249L} they used a procedure developed from their models to diagnose radiatively driven implosion and triggered star formation in the bright-rimmed cloud IC 1848, for example comparing the pressure in the ionised layer bounding the cloud with the internal pressure to determine whether the neutral cloud should be being compressed by the ambient medium.

\begin{figure*}
    \centering
    \vspace{-2.5cm}
    \includegraphics[width=16cm]{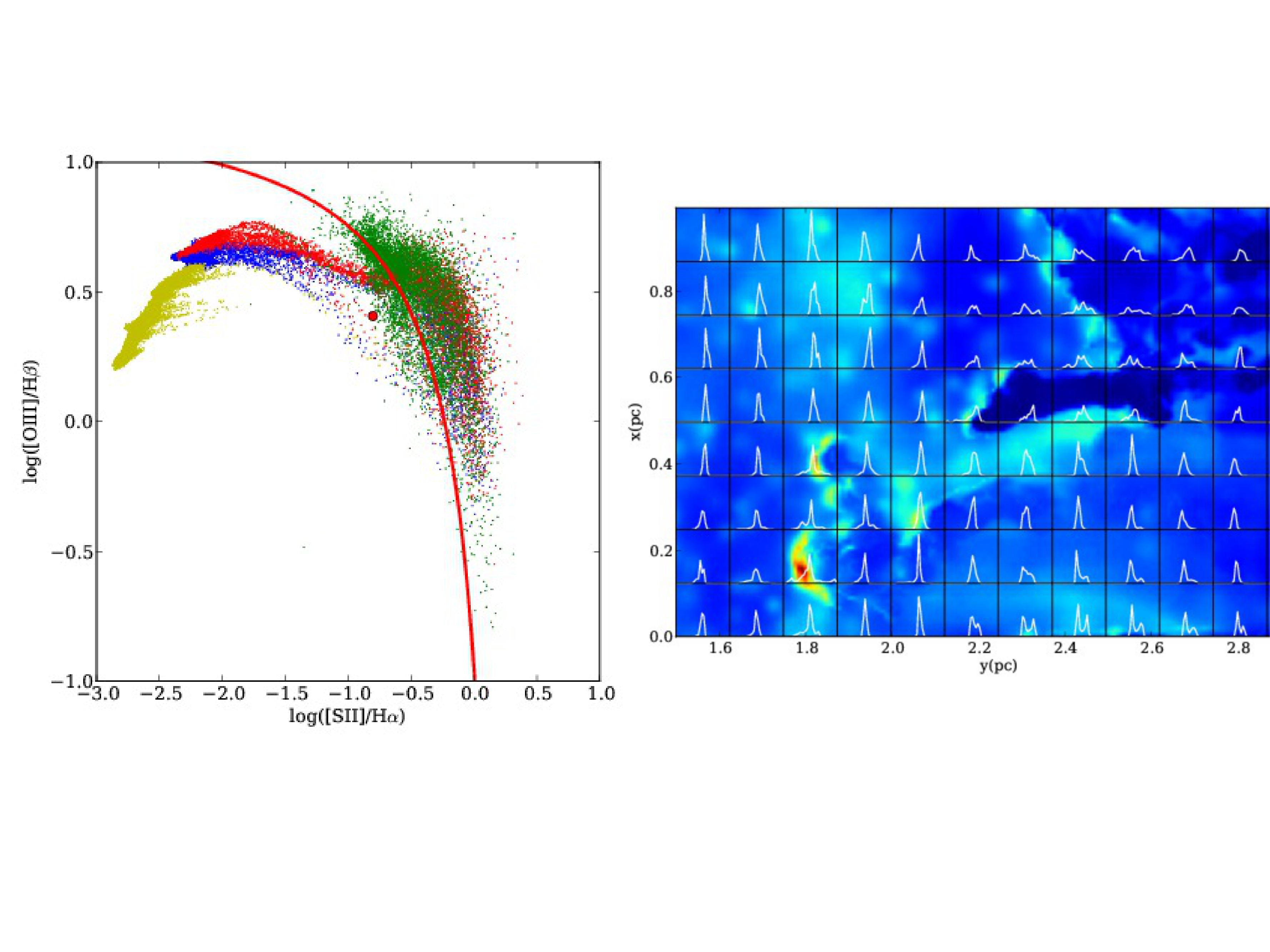}
    \vspace{-2.8cm}
    \caption{Figures adapted from \citet{2012MNRAS.420..141E} with permission from the authors. The right-hand panel is a synthetic [O III] observation of the radiation hydrodynamic calculations of \citet{2010ApJ...723..971G}, including spatially averaged line profiles. The left-hand panel is a BPT diagram from such a model. The red line separates the boundary between photoionised and shock-dominated systems, with the latter lying below the line. The averaged point over the whole simulation is the filled red circle, but when spatially decomposed, the BPT diagnostic breaks down, incorrectly identifying some components as predominantly shock-excited when no such conditions exist in the model.}
    \label{fig:ercolano}
\end{figure*}

 \cite{2009MNRAS.398..157H} ran radiation-magnetohydrodynamic calculations of the photoionisation of magnetised globules. These included three-colour synthetic images of  [N II] (6584\,\AA, red), H$\alpha$ (6563\,\AA, green) and [O III] (5007\,\AA, blue), though since these species were not directly included in their calculations their ionisation state was approximated as a function of the hydrogen ionisation fraction. Although these calculations included magnetic fields, they did not produce synthetic polarization images.  \cite{2011MNRAS.414.1747A} and \cite{2011MNRAS.412.2079M} computed projected Stokes Q and U parameters for a larger scale star-forming region and for magnetically threaded pillars/globules respectively, but to our knowledge there have been no synthetic images of the polarisation state of such clouds produced using full radiative transfer. 

\cite{2010MNRAS.403..714M} produced H$\alpha$ emission maps from their radiation hydrodynamic calculations of irradiated clumps. They found that the clumpiness of their resulting trunks (which is observed in molecular line, i.e. sub-mm, emission) is not observed in the optical, in agreement with HST observations of real pillars. Although not strictly a synthetic observation, they also collapsed their 3D simulations to produce position-velocity diagrams, finding that initial conditions that are already pillar-like are kinematically distinct from pillars that result from more clump-like initial conditions. 

\cite{2010ApJ...723..971G} ran radiation hydrodynamic simulations of turbulent boxes that are irradiated from one side by an ionising radiation field. As mentioned above, the low density components are quickly evaporated, leaving a series of higher density remnant pillars. \cite{2012MNRAS.420..141E} post-processed these calculations to compute the detailed photoionisation structure using the Monte Carlo photoionisation code \textsc{mocassin}. From this they generated synthetic images of recombination and forbidden lines. They also computed line profiles (right hand panel of Figure \ref{fig:ercolano}) which they compared with observations, though since they did not include stellar winds in the dynamical calculations only the components of the profile within a factor of a few of the ionised gas sound speed were reproduced. They also computed BPT diagrams for the simulated H\,\textsc{ii} regions (see Section~\ref{sec:linerat}). These give an excellent example of the point that we made at the end of Section~\ref{sec:linerat}, namely that the complex 3D distribution of gas in real H\,\textsc{ii} regions or PDRs can potentially lead to one drawing incorrect conclusions from line ratio studies. \citet{2012MNRAS.420..141E} show that while the BPT diagram correctly identifies some components of the emission as being photoionisation-dominated, it incorrectly finds that others are shock-dominated, even though the original calculation did not include the effects of stellar winds (see Figure~\ref{fig:ercolano}). \cite{2013ApJ...769...11Y} similarly illustrated the sensitivity of BPT diagrams to substructure in resolved H\,\textsc{ii} regions.  

\cite{2015MNRAS.450.1057M} used \textsc{mocassin} synthetic observations similar to those discussed in the previous paragraph to help interpret new MUSE data towards the famous pillars of creation. Similar to the above, they computed the ionisation state of the medium external to the pillars and used the resulting line profiles to estimate the 3D structure of the pillars.

\cite{Haworth:2012} post-processed radiation hydrodynamic simulations of bright-rimmed cloud (BRC) formation from \cite{2012MNRAS.420..562H} to produce synthetic recombination line, forbidden line and radio continuum observations, as well as SEDs. They tested observational diagnostics such as mass/temperature estimates of the clumps and the comparison of internal/external pressures of the clump to determine whether they are being compressed. \cite{2013MNRAS.431.3470H} processed the same models to compute synthetic molecular line observations. Again they tested mass/temperature diagnostics from these lines, but also studied kinematic signatures. They found that the kinematic profile of a clump being compressed can take on a variety of forms depending on the viewing angle. This includes having a single peak with wing features, as well as double or even triple peaked profiles. In all cases there is a peak from the central part of the cloud. Secondary and tertiary features along the line of sight depend on what parts of the swept up shell of material that is driving into the cloud are observed. For example, a shell propagating away from the observer into a clump produces a double peaked profile, whereas a shell wrapped around a clump might produce a triple peaked profile (one component from the central cloud, one driving away from the observer into the cloud and a third driving into the cloud from the far side, towards the observer). This variety of line profiles has been observed \citep[e.g.][]{2013A&A...560A..19T}.

\subsubsection{Moving OB stars}
\label{sec:movingOBstars}
Isolated massive stars propagating through the ISM create aspherical H\,\textsc{ii} regions and wind-blown bubbles. Stars propagating supersonically will create bow shocks \citep[e.g.][]{1988ApJ...329L..93V}, but \cite{2015A&A...573A..10M} showed that even low velocity stars can produce very non-symmetric ambient media. The idea of a perfectly spherical H\,\textsc{ii} region or wind-blown bubble is hence not expected in practice, nor is it observed, perhaps with the exception of the bubble nebula \citep[e.g.][]{2002AJ....124.3313M}, which could just be a result of the viewing angle (Green et al., in prep.). 

\begin{figure*}
    \includegraphics[width=16cm]{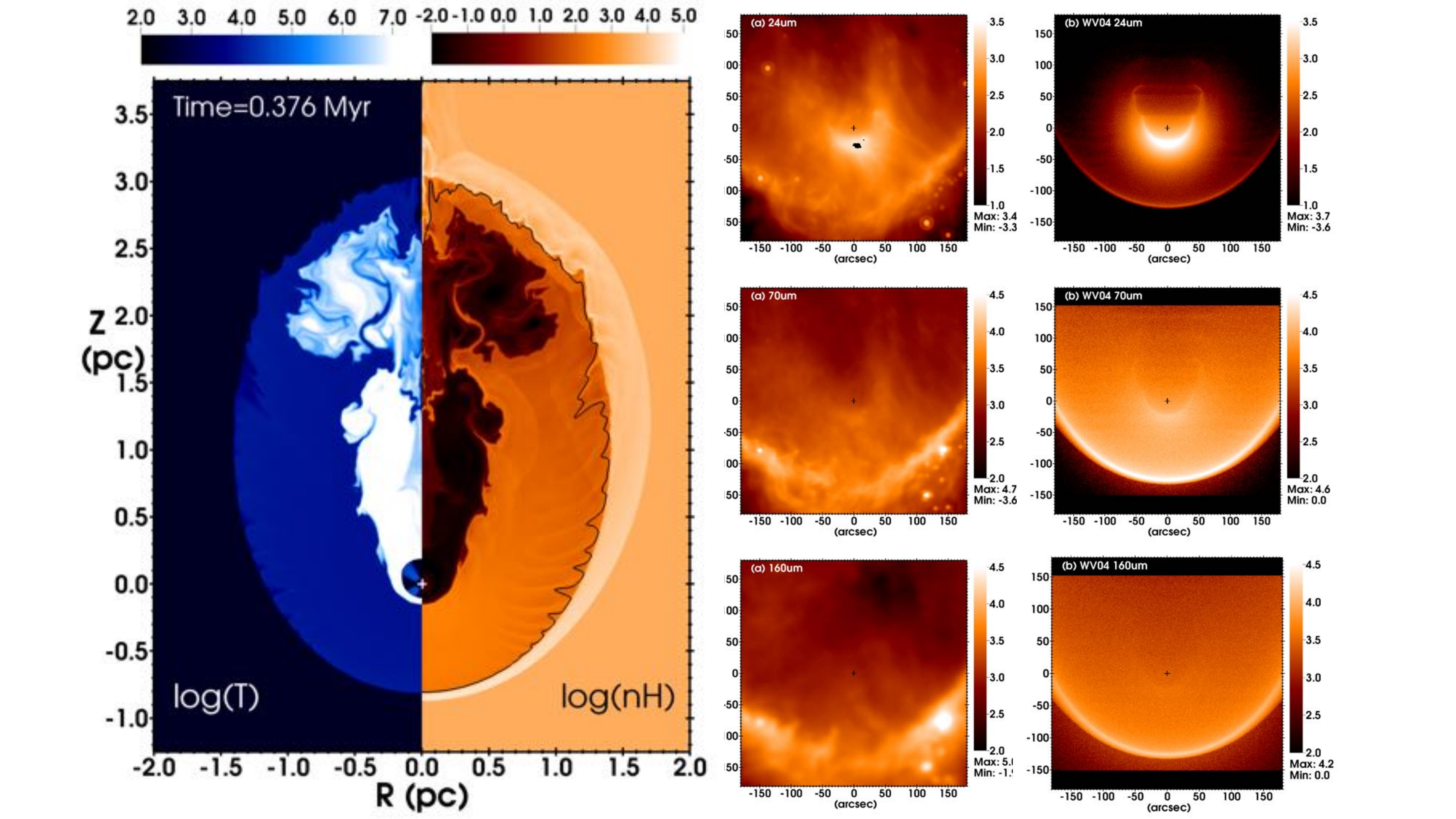}
    \caption{\textit{Herschel} (middle column) and synthetic (right column) continuum images of RCW120 based on a radiation hydrodynamic simulation with ionising radiation and winds (left). The infrared arc observed at 24\,$\mu$m (top panel) that gradually disappears at longer wavelengths (70 and 160\,$\mu$m in the middle and bottom panels) traces the edge of the wind-blown bubble. Such bubbles are historically elusive, but infrared arcs (validated by synthetic observations) appear to be a valuable signpost of them. This Figure is adapted from calculations from \citet{2016A&A...586A.114M}, used with permission of Jonathan Mackey.}
    \label{fig:mackey}
\end{figure*}

The medium surrounding isolated propagating OB stars is an important topic of study. Massive stars are typically thought to form in larger clusters, so isolated OB stars are likely ejected as a result of dynamical encounters within the clusters. Isolated massive stars can hence provide information about the process of massive star formation.  Furthermore wind-blown bubbles from OB stars are difficult to detect owing to their extremely low density and, in clusters, their being a superposition of winds from many stars. In isolated propagating systems it is expected to be easier to probe and learn about the wind. Other problems include the stability of the ionisation/shock front, the ability of the bow shock to accumulate material and also the possibility of inducing further star formation. For example in the famous RCW 120 system there is low mass star formation coincident with the leading edge of the H\,\textsc{ii} region in the direction of propagation of the star \citep[e.g.][]{2009A&A...496..177D,  2010A&A...518L..81Z}, which may be a signature of triggered star formation.

Of further interest is that the observable characteristics of supernovae and their remnants can be affected by the structure of the circumstellar medium, either due to the presence of dense and massive circumstellar shells \citep{2014Natur.512..282M} or by strong asymmetries in the medium \citep{2015MNRAS.450.3080M}. For a recent review of the interaction of supernovae with circumstellar matter see \cite{2016arXiv161207459C}.

Wind-blown bubbles are notoriously difficult to detect since, although hot (at $>10^6$\,K) they are very low density and hence their emission measure is very low. \cite{2016A&A...586A.114M} studied infrared arcs in H\,\textsc{ii} regions, with the aim of determining whether these trace the edge of non-spherical wind-blown bubbles about propagating O stars. They did a case study of RCW 120 by post-processing dynamical models to produce synthetic observations. Good qualitative agreement between synthetic and real observations was found, with the continuum intensity of the arc getting more difficult to detect at 24, 70 and 160\,$\mu$m respectively. Furthermore the intensity of the infrared arc relative to the emission from the shell at the boundary of the H\textsc{ii} region decreases with wavelength in the same way as is observed.  An alternative explanation for such arcs is that dust dynamically decoupled from the gas is driven inwards from the edge of the H\,\textsc{ii} region and stalls at a radius at which it is in equilibrium with the stellar radiation pressure force, as explored in  \citet{2015MNRAS.449..440A} and \citet{2017MNRAS.469..630A}. These authors found it more difficult to reproduce the observed intensity profile at different wavelengths, but included a much more sophisticated treatment of the dust, including dynamical decoupling. \cite{2014A&A...566A..75O} also studied infrared arcs in terms of radiation pressure balance. 

\cite{2017MNRAS.466.1857G} found the first candidate wind-blown bubble around a main sequence B-type star. This interpretation again relied on comparison of the observed infrared fluxes with synthetic observations computed from radiation hydrodynamical models of a propagating B star.

\cite{2016MNRAS.456..136A} computed both radiation hydrodynamic models and multi-wavelength synthetic observations of the bow shock about a runaway ($v>40\,$km/s) O star. They found that instabilities in the bow shock lead to the production of warm clumps that contain dust, making them the dominant emitters in far infrared, H\,$\alpha$ and radio wavelengths. They also studied the spatial distribution of emission, finding that H\,$\alpha$ and dust continuum emission is concentrated near the forward shock, similar to the findings of models/emissions maps from \cite{1997RMxAA..33...73R}. Synthetic observations of the radio continuum also showed widespread emission, similar to the emission maps of \cite{1991ApJ...369..395M}, only with the addition of the bright clumps resulting from instability. These calculations didn't account for the dynamical decoupling of dust, which can segregate grains of different sizes and hence result in a different morphology at different wavelengths in the continuum \citep[see e.g.][]{2011ApJ...734L..26V}.

\cite{2012A&A...541A...1M} computed bespoke models of the bow shock of Betelgeuse using 3D SPH radiation calculations. From these they studied the continuum emissivity, as well as that from various lines. They found that the relatively circular morphology of the bow shock, combined with strong infrared fluxes, implies that the bow shock is young, since over time it become unstable in their models and more disordered. They also discussed that if Betelgeuse is propagating at high velocity then the bow shock heating will result in emission at short wavelengths, so observations in e.g.\ the UV might constrain the velocity. 

In a series of papers \cite{2014MNRAS.444.2754M,2015MNRAS.450.3080M, 2016MNRAS.459.1146M,2017MNRAS.464.3229M} studied the (radiation-magneto) hydrodynamical and emission properties of the medium about propagating OB stars and red supergiants, including H\,$\alpha$ and forbidden lines. They find that the magnetic field reduces the H\,$\alpha$ and forbidden line intensities significantly, as well as the dust continuum emission, which they propose contributes to the difficulty of observing the ionised bow shocks of such systems. They also studied the observability of propagating stars in different mass/velocity regimes using different tracers. They found that the infrared dust continuum is better for tracing relatively low velocity, high-mass systems ($\sim$20\,km/s, 40\,$M_\odot$) but that H\,$\alpha$ is more effective at tracing similar mass stars at higher velocities, such as 70\,km/s.

Finally we note that \cite{2006ApJS..165..283A} also computed radiation hydrodynamic models of ultracompact H\,\textsc{ii} regions about propagating OB stars, which we discussed in more detail in section \ref{sec:uchiireg}.

\subsection{The turbulent ISM and low mass star formation}

Since molecular cloud formation and star formation occur in the neutral medium, the importance of understanding and being able to explain its thermal and dynamical state cannot be overemphasised. The structure of the neutral medium can be studied using dust continuum observations, which may also be used to estimate the gas temperature if the two components can be considered well-coupled, on the assumption of a simple temperature structure along each line of sight. Dust emission is also one of the main tools for estimating star formation rates. Molecular observations reveal the structure and dynamics of the gas simultaneously. Both techniques are commonly used to compute column densities (and therefore gas masses, essential for measuring star formation efficiencies) and velocity fields derived from molecular emission are used to examine turbulence, and to infer the overall energy balance of clouds. This is obviously of crucial importance for understanding ongoing and future star formation activity.

\subsubsection{Turbulence}
\label{sec:turb}
Turbulence is one of the most important features of the ISM and swathes of observational and theoretical work has been dedicated to detecting and characterising turbulence (see e.g.\ \citealt{2004ARA&A..42..211E}, \citealt{2004ARA&A..42..275S} and \citealt{2012A&ARv..20...55H} for informative reviews). These efforts are strongly hampered by projection effects, arguably the biggest single stumbling block in the comparison of simulations and observations.  Observationally, the ambiguity in the third dimension is usually broken by by assuming a correspondence between PPP and PPV datacubes, and synthetic observations can be used to evaluate how well this assumption is likely to hold. Synthetic observations can also be used to explore the impacts of two other important effects: the fact that atomic or molecular tracers of the ISM are not chemically homogeneous, and hence provide a biased view of the underlying density and velocity fields; and the fact that these tracers can also become optically thick, which can have the effect of hiding small-scale fluctuations. In this section, in order to keep the size of our discussion manageable, we focus primarily on the use of synthetic observations to study these three effects, although we note that synthetic observations have also been used to validate and calibrate many of the statistical tools for characterising observations that are discussed in Section~\ref{sec:bestmethod}.

Early attempts at comparing detailed simulations of astrophysical turbulence with observations of real clouds were made by \citet{1994ApJ...436..728F}, \citet{1995ApJ...448..226D} and \citet{1999ApJ...524..887R}, but none of these studies created synthetic observations as such, in  the sense that they did not choose particular tracers, or perform any radiation transport calculations.
One of the first authors to create genuine synthetic observations of a simulated turbulent velocity field were \cite{1998ApJ...504..300P}. They used the Monte Carlo code of \cite{1997A&A...322..943J} to generate $^{12}$CO, $^{13}$CO and CS spectral maps of a periodic turbulent box. Their simulations did not follow the chemical evolution of the gas, and so they assumed constant abundances of CO and CS. They compared their results to observations by e.g.\ \cite{1990ApJ...359..344F}, and found concordance in several observed parameters, such as the ratio of line intensity to linewidth.

\cite{2001ApJ...547..862P} and \cite{2003ApJ...588..881P} took this work further by applying the SCF analysis of \cite{1999ApJ...524..887R} and concentrating on the $^{13}$CO(1--0) line as a compromise between large--scale maps and resolution. They fit power laws to the SCFs of the simulations, and to eleven observational datasets of various clouds, including Taurus, Perseus and the Rosette. Comparison of the normalisations and slopes of the fitted power laws were then used to rule out several otherwise plausible physical assumptions as being inconsistent with the observations, for example that the kinetic and magnetic turbulent energies are in equipartition.

\citet{2002ApJ...570..734B} performed a series of simulations of driven hydrodynamical and MHD turbulence in a periodic box using the {\sc zeus-3d} code. They then post-processed their results to generate maps of $^{13}$CO 1-0, $^{13}$CO 2-1 and $^{12}$CS 1-0 line emission. To approximately account for the effects of sub-thermal excitation at low densities, they assumed that the lines were excited only at densities above the critical density of the transition. They did not account for any effects due to chemical inhomogeneities in the gas. Using their synthetic emission maps, they examined how well structures seen in the position-position-velocity (PPV) datacubes matched up with real 3D structures (also referred to as position-position-position or PPP structures). They found that projection effects were severe for the $^{13}$CO observations, and that the inferred physical properties of clumps identified using the {\sc clumpfind} algorithm in PPV space show poor correspondence to real PPP structures. In the case of CS, which traces much higher density gas, projection effects are far less pronounced, although even in this case some mismatches between PPV and PPP structures remain. One important consequence of this is that projection effects artificially flatten the slope of the size-linewidth relationship measured using low density tracers compared to the underlying value in the turbulent gas \citep[see also][]{2010ApJ...712.1049S}.

\citet{2013ApJ...777..173B} took two simulations -- one using the {\sc orion} code and one run by \citeauthor{2011MNRAS.415.3253S}~(2011b) using {\sc zeus-mp} with simplified CO chemistry -- of turbulent boxes. Molecular line observations were then constructed using {\sc radmc-3d} for $^{12}$CO 1-0, $^{12}$CO 3-2 and $^{13}$CO 1-0 assuming LVG and with representative levels of noise in each cube (although the noise has very little influence on their results). Dendrograms were used to identify objects in PPP and PPV space which were then compared quantitatively. \citet{2013ApJ...777..173B} show that $^{13}$CO tracer gives a much more accurate representation of the density field than either $^{12}$CO line since, as in the work by \citet{2002ApJ...570..734B}, the more space-filling, low-density tracers are more likely to suffer line-of-sight confusion. While the masses, sizes and internal velocity dispersions of structures can all be recovered tolerably well when using $^{13}$CO, errors can often be as large as a factor of two. Consequently, the errors in parameters depending on a combination of these quantities can be even larger: for example, the virial parameters of individual clumps can be uncertain at the level of 0.3 dex or more. This is sufficient to make the boundedness of many such objects impossible to determine with any confidence based on CO observations.

There have also been various studies of how best to infer the statistical properties of the 3D gas distribution (e.g.\ the volume density PDF) based only on projected data. For example, \citet{2010MNRAS.403.1507B} present a method for inferring the 3D variance of a field from the statistical properties of the 2D projection of the field. The key to the method is the assumption of statistical isotropy, i.e.\ that the power spectrum of the 2D field is the same as that of the 3D field. When this assumption is valid, \citet{2010MNRAS.403.1507B} show that there is a simple relationship between the 2D and 3D variance, although care must be taken to properly account for the effects of finite beam size. They also use numerical simulations to demonstrate that the assumption of isotropy is reasonable for non-magnetised clouds dominated by supersonic turbulence. In the MHD case, isotropy also requires that the turbulent motions be super-Alfv\'enic (i.e.\ that the rms velocity of the turbulence is larger than the Alfv\'en velocity). 

In a follow-up paper, \citet{2010MNRAS.405L..56B} apply the method to the particular case of the density PDF. As previously noted in Section~\ref{sec:pdfs}, numerical simulations have demonstrated that there is a simple relationship between the 3D density PDF and the 3D density variance. However, the 3D density variance is not a quantity that can be directly measured from observations. However, \citet{2010MNRAS.405L..56B} use numerical simulations to show that if one measures the 2D density variance (i.e.\ the variance of the projected density field), applies the \citet{2010MNRAS.403.1507B} method to infer the 3D density variance, and then uses this to predict the form of the 3D density PDF, the resulting PDF agrees well with the real one, apart from some deviations in the extreme tails, provided that the turbulence is isotropic

An alternative method for inferring the 3D density PDF from 2D column density data is presented in \citet{2014Sci...344..183K}. They first use wavelet filtering to decompose the column density map into a set of structures with different spatial scales. Each of these structures is then modelled as a prolate spheroid oriented in the plane of the sky. Given the assumed orientation of this spheroid, the observations constrain its semi-major and semi-minor axes, and hence its volume. Since the mass is also known, the characteristic density of this structure can easily be inferred. \citet{2014Sci...344..183K} show using numerical simulations that although the density of any given structure is often quite inaccurate (owing to the unknown inclination of the real object on the sky), the final density PDF that one constructs by combining all of the structures is surprisingly accurate. The advantage of this method compared to the \citet{2010MNRAS.405L..56B} approach is that it does not restrict one to the case of isotropic turbulence. However, it is unclear how to generalize it to deal with quantities other than the density, whereas the \citet{2010MNRAS.405L..56B} method can in principle be applied to any field whose 2D variance can be measured.

Turning now to the effects of opacity, we note that several different groups have used synthetic observations to study the impact of opacity effects on the observed statistical properties of turbulent flows. One of the first studies along these lines was carried out by \citet{2002A&A...391..295O}. He post-processed simulations of turbulent clouds carried out by \citet{1999ApJ...524..169M}, \citet{2000ApJ...535..887K}, \citet{2001ApJ...547..280H}, and \citet{2001A&A...379.1005O} using his own {\sc SimLine} radiative transfer code to produce maps of $^{13}$CO line emission. As the simulations did not account for the chemical evolution of the gas, he assumed a constant fractional abundance of $^{13}$CO relative to H$_{2}$. Varying this ratio and producing maps for small, medium and large values provided a simple way to adjust the characteristic optical depth of the emission. \citet{2002A&A...391..295O} analyzed the resulting integrated intensity maps using the $\Delta$-variance statistic. He showed that prior to the onset of gravitational collapse, the $\Delta$-variance of the $^{13}$CO emission is dominated by large-scale modes and agrees well with the $\Delta$-variance of the total column density. However, after one free-fall time, the $\Delta$-variance of the $^{13}$CO emission is barely altered, while that of the column density is now dominated by small-scale modes. He demonstrates that this result is a consequence of line opacity: after one free-fall time, the column density map is dominated by a number of small, dense cores, but these cores are optically thick in $^{13}$CO and so do not make a prominent contribution to the $^{13}$CO integrated intensity map. 

\citet{2004ApJ...616..943L} studied the effects of line opacity on the VCA statistic of \citet{2000ApJ...537..720L}. This statistic was developed to deal with the problem that fluctuations in both the density field and the velocity field contribute to the power spectrum observed in a velocity slice from a PPV cube of line emission. VCA works by examining how the spectral index of the power spectrum changes as we change the width of the velocity slice, as in the optically thin case, this allows the degeneracy between velocity and density to be broken. \citet{2004ApJ...616..943L} show that in the optically thick case, the same results still hold for sufficiently thin slices (i.e.\ slices with optical depth $\tau < 1$). However, for very thick slices, the interpretation of the observed power-spectrum becomes more difficult, since the observed fluctuations in the intensity are no longer related to the fluctuations in the density in a simple fashion. \citet{2006ApJ...652.1348L} carried out a similar analysis of a closely related statistical tool, the VCS statistic. The VCA statistic was also tested by \citet{2006ApJ...653L.125P}, using synthetic $^{13}$CO observations of a high resolution turbulence simulation. They showed that the method successfully recovered the power spectrum of the underlying turbulent velocity field in their simulation. Finally, although most of these studies only considered the effects of line absorption, \citet{2017MNRAS.470.3103K} have recently studied the impact of dust absorption on both statistics.

\citeauthor{2013ApJ...771..123B}~(2013b) used synthetic observations of $^{13}$CO emission from turbulent HD and MHD boxes created with the {\sc SimLine} code to investigate the effects of optical depth on the power spectrum of the integrated intensity images. They find, in line with the predictions of \cite{2004ApJ...616..943L}, that the slope of the \textit{observed} power spectrum in an optically thick medium saturates at a value of -3, and that the intrinsic power spectrum of the density field therefore cannot be inferred.

\citeauthor{2013ApJ...771..122B}~(2013a) used a similar set of synthetic $^{13}$CO emission maps to study the impact of optical depth effects on the determination of the column density PDF. They show that although the PDF of $^{13}$CO integrated intensities is usually log-normal, it is not always a good match for the true column density PDF, being affected by sub-thermal excitation in low density gas at the low column density end, and line saturation due to high optical depths at the high column density end. They find that $^{13}$CO does the best job of tracing the true column density when its optical depth is close to 1.  

A further study along these lines was carried out by \citet{2016ApJ...818..118C}, who use synthetic $^{13}$CO maps based on fractional Brownian motion (fBm) models and 3D turbulence simulations to examine how well PCA performs in the case of optically thick emission. They show that in contrast to techniques such as VCA or VCS that can be significantly affected by high optical depths, PCA continues to perform well even in optically thick gas. They speculate that this is because the technique is sensitive to phase information present in the data, rather than just the magnitude of the different velocity modes, and this information is less easily obscured by high optical depths.

Next, we note that although some of the studies discussed above attempt to account in an approximate fashion for the chemically inhomogeneous nature of molecular clouds, e.g.\ by considering emission only from gas above some density threshold, the computational demands of performing high-resolution 3D simulations of turbulent clouds with time-dependent chemistry means that self-consistent studies of the effects of chemistry on turbulence statistics have only recently become possible.  

\citet{2014MNRAS.440..465B} carried out simulations of turbulent gas using {\sc zeus-mp} with a time-dependent chemical model based on \citet{2007ApJS..169..239G} and \citet{1999ApJ...524..923N}. They then post-processed the results of these simulations using {\sc radmc-3d} to produce synthetic $^{12}$CO and $^{13}$CO emission maps. They applied PCA to these maps and explored how well the results from this agree with what one obtains from applying PCA to either the raw density field or the H$_{2}$ and CO densities. They show that in low density models, the slope of the PCA pseudo-structure function derived from H$_{2}$ is somewhat steeper than that derived from total density, while in higher density models, there is very good agreement, since almost all of the hydrogen in these models becomes molecular. CO is too intermittently distributed in the low density models to allow a meaningful structure function slope to be derived, with PCA finding only a small number of components. In higher density models, PCA applied to $^{12}$CO density and $^{12}$CO emission produces pseudo-structure function slopes that are steeper than the true one, while there is good agreement between the results for $^{13}$CO and for total density. 

\citeauthor{2015MNRAS.446.3777B}~(2015a) used higher resolution versions of the same simulations to investigate the impact of chemical inhomogeneities on the structure function of line centroid velocity increments (CVIs). They found that there was a large difference between the slope of the structure function derived from the CVIs of the total density field and that derived from the CO density or CO intensity. In general, the slopes derived from the CO synthetic observations are substantially shallower than the true slopes derived from the raw simulation output. Moreover, they showed that this effect was primarily due to the chemical inhomogeneity of the gas, rather than being an effect of line opacity. 

\citeauthor{2015MNRAS.451..196B}~(2015b) used the same set of simulations to examine the sensitivity of the $\Delta$-variance method to the chemical inhomogeneity of the gas. They computed $\Delta$-variances for maps of total column density, $^{12}$CO 1-0 and $^{13}$CO 1-0 integrated intensity, and also applied the same statistic to maps of the velocity centroids for the total density, H$_{2}$ and CO densities, and $^{12}$CO and $^{13}$CO intensities. The $\Delta$-variance of the centroid-velocity maps can be easily converted to a size-linewidth relation and hence compared with similar information coming from the structure functions or power spectra. As in their earlier study, \citeauthor{2015MNRAS.451..196B}~(2015b) find that the slopes derived from the CO observations are significantly shallower than that of the underlying density/velocity field. Optical depth effects strongly influence the conclusions that can be drawn from the column--density and intensity maps. In model clouds with low mean densities, the CO emission is confined to dense filaments and is a very poor tracer of the overall cloud structure, with the result that species tracing different densities return $\Delta$--variance slopes which are not only different from each other, but are different (up to and including having the opposite sign) from the total--density or H$_{2}$--derived slopes. Above mean cloud number densities of 100 cm$^{-3}$, the $\Delta$--variance slopes of all tracers examined at least have the same sign, and \citeauthor{2015MNRAS.446.3777B}~(2015a) advocate this as a minimum threshold density above which CO reliably traces \textit{global} cloud structure.

\cite{2015ApJ...799..235G} model a turbulent box and compute emissivities using the {\sc 3d-pdr} code, with brightness maps constructed using {\sc radmc-3d}. The photochemical network is not solved self--consistently, but the elevated temperatures are likely not high enough to dominate the dynamics. They compute the emission from 16 species and analyse the results using the Spectral Correlation Function of \cite{1999ApJ...524..887R}. They find that they are robust against changes in viewing angle. The SCFs flatten out at larger length scales, with the critical length scale becoming larger for lower--density, higher volume--filling factor tracers. They partition the tracers into three groups depending on the slope of the SCF at length scales shorter than the critical length for each tracer. Tracers in groups are homologous with one another, in that they trace similar cloud length scales.

Finally, synthetic observations have also been used to study the role of stellar feedback processes in driving turbulence in molecular clouds. For example, \citet{2014ApJ...784...61O} perform simulations of turbulent molecular clouds in which they model the impact of protostellar outflows. They set the abundances of $^{12}$CO, $^{13}$CO and C$^{18}$O to constant values (zero for gas hotter than 900\,K) and use {\sc radmc-3d} to compute emission maps for these three tracers. The morphology of the emission shows good agreement with observed outflows, and $^{12}$CO is found to be by far the best tracer of the three. Similar calculations (although using {\sc flash}) were post-processed in a similar way by \citeauthor{2014MNRAS.437.2901P}~(2014a,b), \nocite{2014ApJ...788...14P} but the outputs were further processed using {\sc casa}, assuming source distances of 128\,pc and 3\,kpc respectively. In both cases, they were well able to recover known outflow masses, source luminosities and mass-loss rates.

\citet{2015ApJ...811..146O} used the {\sc orion} code to model the growth of stellar wind bubbles in the turbulent ISM typical of objects such as the Perseus giant molecular cloud. They assumed a CO abundance which is zero in gas above 800\,K, or more rarefied than $10$\,cm$^{-3}$, and in gas denser than $2\times10^{4}$\,cm$^{-3}$ (due to freeze-out) and constant otherwise. {\sc radmc-3d} with the LVG approximation was then used to compute $^{12}$CO 1-0 emission maps. \citet{2015ApJ...811..146O} then computed  turbulent power spectra based on the emission maps, but found no clear signature of the effects of winds. Although initially perhaps surprising, this finding is consistent with the current lack of observed features in the power spectra of wind-affected regions (e.g.\ NGC 1333). \citet{2015ApJ...811..146O} also used the CO emission maps to infer the amount of energy and momentum injected by the winds, finding that because much of the gas shocked by the winds lies in the density range from which $^{12}$CO 1-0 emission is observable, observations using this line effectively over-estimate the influence of winds on the cloud as a whole. However, the observed sizes and morphologies of the CO bubbles are consistent with those observed in Perseus. \cite{2015ApJ...811..146O} therefore infer \textit{post facto} that the wind terminal velocities and mass-loss rates they assumed are reasonable, and thus discount proposed wind models with much lower mass-loss rates.

\subsubsection{Filaments}
Astronomers have known since the beginning of the last century \citep[see e.g.][]{1919ApJ....49....1B} that many dark clouds display a pronounced filamentary structure. However, the importance of filaments and their connection to star formation was recently re-emphasized with the discovery by {\it Herschel} of a large number of parsec-scale filaments in nearby molecular clouds \citep{2011A&A...529L...6A,2014prpl.conf...27A}. This discovery, together with the claim that the filaments all have a characteristic width of around 0.1~pc, has prompted a renewal of interest in the properties and behaviour of interstellar filaments \citep[for a  recent review, see][]{2017arXiv171001030A}. Among the many papers written about filaments over the past few years are a number of studies making use of synthetic observations. 

These studies can be separated into two main types. The first type of study adopts a highly idealized filament model, e.g.\ a Plummer or Plummer-like profile \citep{1911MNRAS..71..460P}, which is often assumed to be static, and combines this with a detailed treatment of the microphysics of the gas and the dust. For example, \citet{2012A&A...542A..21Y} model filaments as cylinders with Plummer density profiles. They examine filaments with a range of different flattening radii and central densities, and carry out detailed multi-wavelength Monte Carlo radiative transfer simulations to compute the dust grain temperatures and the resulting thermal emission in a range of different {\it Planck} and {\it Herschel} bands. They use their computed emission maps to study whether the temperature structure of the dust along the line of sight (i.e.\ the fact that for an externally heated filament, the dust is warmer at the edge and cooler at the centre) can give rise to an anti-correlation between the observationally-derived dust temperature $T_{\rm dust, obs}$ and the dust emissivity spectral index $\beta$, as suggested by \citet{2009ApJ...696.2234S}. They show that instead they recover a weak correlation, and hence argue that the observed $\beta$--$T_{\rm dust, obs}$ anti-correlation must be indicative of an actual change in the grain optical properties with decreasing dust temperature.

Another example of this kind of study can be found in \citet{2016A&A...592A..90C}. They study three types of simple filament model: a constant density cylinder, a cylinder with a Plummer-like density profile, and a filament with a Plummer-like profile plus an embedded higher-density sphere (designed to represent a filament with an embedded prestellar core). They processes these model filaments using the \citet{2012ApJ...751...27H} Monte Carlo radiative transfer code, and use the resulting dust emission maps to explore whether the observationally-derived dust temperatures and column densities depend on the inclination angle of the filament with respect to the observer. They show that the recovered dust temperature is insensitive to the inclination angle, but that the column density varies significantly as a function of inclination angle. Consequently, accurate determination of the filament line mass (i.e.\ the mass per unit length, which plays a critical role in determining whether a filament is gravitationally stable) requires knowledge of the filament inclination, which cannot readily be determined purely from the dust emission map. 

The other type of study focusses more on the dynamical evolution of the filament and less on the detailed microphysics. For example, \citet{2012A&A...544A.141J} post-processed the high resolution MHD simulations of \citet{2011ApJ...730...40P} using the \citet{2012A&A...544A..52L} continuum radiative transfer code. They smoothed the resulting emission maps to account for the finite resolution of real observations, and also added white noise. The maps were then converted to column density images using standard analysis techniques. They selected and extracted a set of filaments from the column density images by eye, and fit the column density profiles of these filaments using a model consisting of a linear background plus a Plummer-like function. They then explored how well the properties of the filaments recovered in this way agreed with the true properties of the filaments in the 3D simulation. They showed that for observational parameters typical of \textit{Herschel} observations and an assumed filament distance of 100~pc, the recovered properties of the filaments agreed well with their true properties. However, they also found that the accuracy degraded quickly as the quality of the observations becomes worse: reducing the signal to noise by a factor of four or placing the filaments four times further away resulted in large errors in the recovered filament properties. To put these values into context, note that the filaments studied by \citet{2011A&A...529L...6A} in IC~5146 are at a distance of approximately 460~pc \citep{1999ApJ...512..250L}. 

Another important result of the \citet{2012A&A...544A.141J} study is their finding that not all of the filaments that they identify in 2D are actually connected structures in 3D. Rather, some of the structures are formed by the superposition along the line of sight of different 3D structures. This point was revisited by \citet{2014MNRAS.445.2900S}, who studied filament formation in turbulent molecular clouds using the {\sc arepo} hydrodynamical code \citep{SpringelVORO:2010}, coupled with a simplified non-equilibrium chemistry network and cooling function \citep{2012MNRAS.421..116G}. Unlike \citet{2012A&A...544A.141J}, they did not derive column densities from synthetic dust emission maps, but instead worked directly with column density projections of the simulation data. These maps are therefore not affected by any of the observational limitations or radiative transfer effects considered in \citet{2012A&A...544A.141J}. Despite this, \citet{2014MNRAS.445.2900S} show that there are significant difficulties involved in inferring the properties of real 3D filaments from 2D column density projections, even given perfect knowledge of the 2D distribution. In agreement with \citet{2012A&A...544A.141J}, they find that the filaments that they identify in 2D are typically not single structures in 3D, but are instead made up of networks of short ribbon-like sub-filaments, reminiscent of the velocity-coherent sub-filaments observed in C$^{18}$O emission in Taurus by \citet{2013A&A...554A..55H}. \citet{2014MNRAS.445.2900S} also show that fitting the filament profiles with Plummer-like functions is unwise, as the fits are degenerate: a range of comparably good fits can be produced by simultaneously varying the central density and the flattening radius in the fit.

\begin{figure*}
    \centering
    \includegraphics[width=15cm]{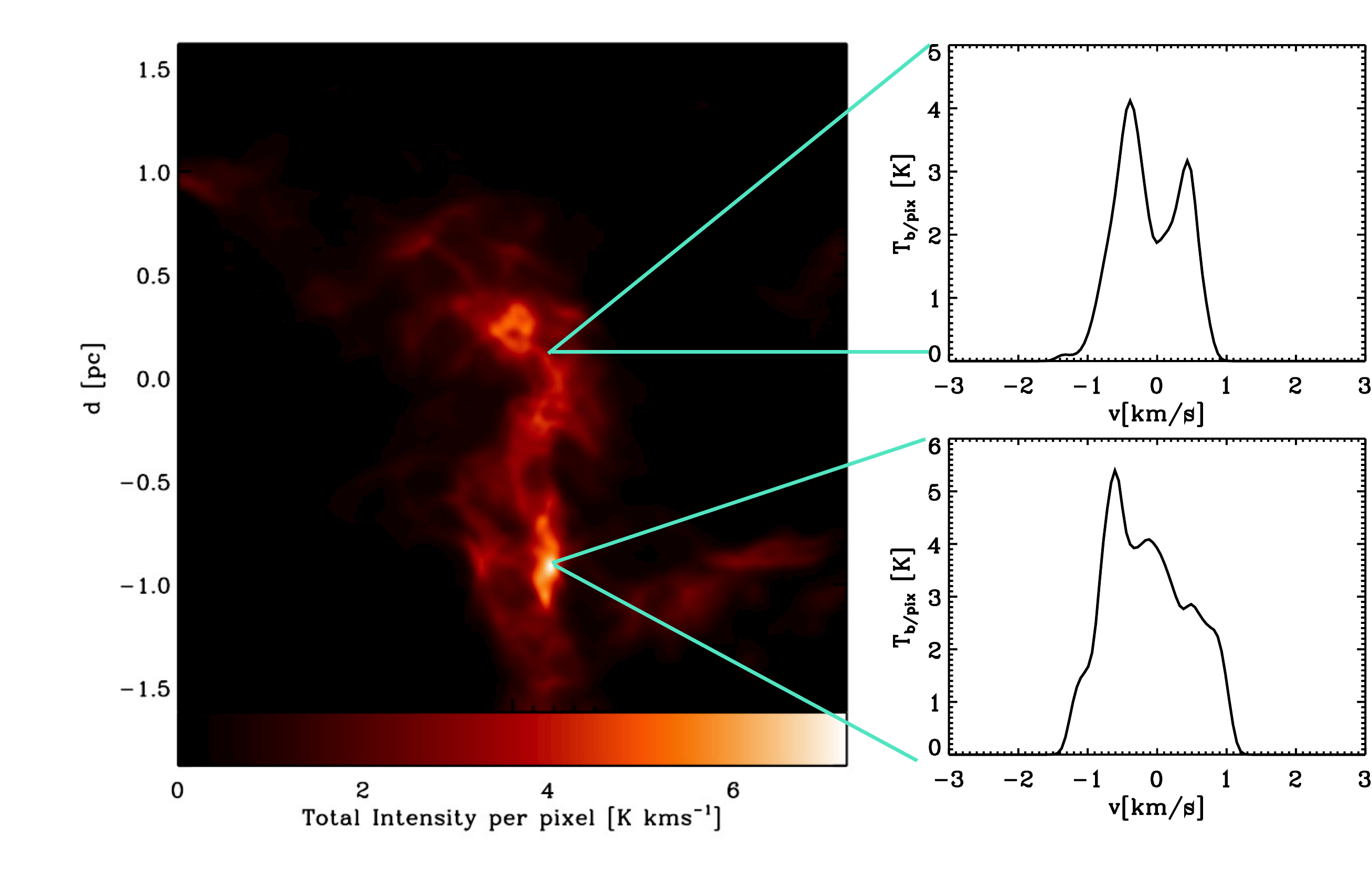}
    \caption{Synthetic emission map for the C$^{18}$O (1-0) line for one of the filaments formed in the simulations of \citet{2016MNRAS.455.3640S}. The panels on the right show the line profiles from the two highlighted regions. Multiple narrow velocity components are seen within the filament, in agreement with the observations of real filaments presented in \citet{2013A&A...554A..55H}. This figure is a copy of Figure~12 of \citet{2016MNRAS.455.3640S} and is used with permission of the authors.}
    \label{fig:Smithetal2016}
\end{figure*}

In a follow-up study, \citet{2016MNRAS.455.3640S} examined the velocity structure of the same set of simulated filaments. They also produced synthetic C$^{18}$O maps of the filaments using the {\sc radmc-3d} radiative transfer code \citet{2012ascl.soft02015D}. They demonstrate that many of their simulated filaments have multiple velocity-coherent sub-filaments that are visible in C$^{18}$O emission, in good agreement with the \citet{2013A&A...554A..55H} observations (see Figure~\ref{fig:Smithetal2016}). However, they show that the interpretation of these structures put forward by \citet{2015A&A...574A.104T} -- namely, that they are the result of fragmentation of the larger-scale filament -- is incorrect, at least for the simulated filaments. Rather than forming via top-down fragmentation, the sub-filaments form before the larger filament, and are then swept up by large-scale turbulent motions that gather them together, a process that \citet{2016MNRAS.455.3640S} term ``fray and gather''.

The formation of velocity-coherent sub-filaments was also investigated by \citet{2015ApJ...807...67M} using high-resolution AMR simulations of an isothermal, turbulent cloud. They post-processed their simulation results to yield synthetic C$^{18}$O emission maps using a highly approximate procedure: they assume that only gas with number densities $10^{3} < n < 10^{4.5} \: {\rm cm^{-3}}$ produces C$^{18}$O emission, and that this emission is optically thin and produced by molecules with a uniform excitation temperature, meaning that there is a linear relationship between C$^{18}$O column density and intensity. Despite these approximations, they find similar results to \citet{2016MNRAS.455.3640S}: velocity-coherent sub-filaments are naturally produced in their turbulent cloud, with properties consistent with the observed sub-filaments. They also show that some of the velocity structures showing up in spatially-connected filaments in projection are actually unassociated with them in 3D space, and are instead just chance projections (see also \citealt{2002ApJ...570..734B} and \citealt{2013ApJ...777..173B}, who find similar results for cores).  

In a recent study, \citet{2017MNRAS.467.4467S} also produced synthetic emission maps of filaments, both in dust continuum emission and in several transitions of $^{13}$CO and C$^{18}$O. Their synthetic maps were based on the simulations of collapsing magnetised filaments carried out by \citet{2016MNRAS.459L..11S}. These simulations were performed using the {\sc flash} AMR code \citep{2000ApJS..131..273F} and accounted for the chemical evolution of the gas using the {\tt react\_COthin} chemistry network from the KROME astrochemistry package \citep{2014MNRAS.439.2386G}. This network is one of the largest on-the-fly chemical networks used to date in a 3D simulation, and is an extended and updated version of the CO network presented in \citet{2010MNRAS.404....2G}. The results of the \citet{2016MNRAS.459L..11S} were post-processed using {\sc radmc-3d}, and the resulting emission maps were then analyzed to see how accurately the filament column densities and widths could be recovered from the synthetic observational data. \citet{2017MNRAS.467.4467S} show that for filaments with widths $> 0.1$~pc, the filament width can be recovered accurately using either line or dust emission, but that for narrower filaments, both tracers tended to over-estimate the filament width. They also show that although the dust emission provides a good estimate of the filament mass, the same is not true for the line emission: masses inferred from this can differ by up to a factor of 10 from the true mass. 

\subsubsection{Cores}
\label{sec:cores}
An important intermediate step between giant molecular clouds and stars comes in the form of cores, which are typically dense starless objects. Understanding the conversion of cores to stars is crucial for understanding star formation. Key questions include the efficiency with which cores are converted to stars, and whether this is independent of the core mass, the extent of core fragmentation, the number of stars produced per core and particularly whether there a link between the core mass function and the stellar initial mass function \citep[e.g.][]{2004MNRAS.347.1001H, 2007ARA&A..45..565M, 2008A&A...477..823G,  2010ApJ...725L..79C, 2011ApJ...732..101I,  2012MNRAS.419..760W, 2013MNRAS.432.3534H, 2014prpl.conf...27A, 2015MNRAS.450.4137G, 2016MNRAS.458.1242L}. Since the notion of cores has been around for a long time \citep[e.g.][]{1969MNRAS.145..271L} they have been the subject of a large number of observational studies. Furthermore, since they are \textit{comparatively} geometrically simple relative to, e.g. the turbulent ISM (being approximated in the first instance by a spherically symmetric model) they have also been the subject of a large number of synthetic observations. Indeed most observational studies of cores come with some form of chemical and/or radiative transfer model to assist in the interpretation \citep[e.g.][]{2001MNRAS.323.1025J}. Given the extremely large number of papers in this context we do not intend to review all of the literature, but rather to provide an overview of some of the general approaches and key results. 

Although there is still not a fully complete picture of the conversion of cores into stars, there is a clear distinction between different types of core. Starless cores are the broad class of overdense objects that, as the name suggests, contain no embedded sources. They may or may not eventually form stars. Pre-stellar cores are the gravitationally bound subset of starless cores that will eventually form stars. Gas within pre-stellar cores that collapses to very high densities and begins to collisionally dissociate H$_{2}$ passes through a brief phase where it is hydrostatically supported against collapse, and this hydrostatically supported gas is often referred to as the first hydrostatic core.
A core hosting embedded YSOs is referred to as a star-forming core. Finally, there are hot cores, which we discuss towards the end of this section. We now proceed roughly chronologically through some of the literature on cores that employ synthetic observations.

\cite{1995ApJ...439L..55B} computed synthetic SEDs of the first hydrostatic core (which they refer to as class $-$I objects) as a precursor to single or binary star formation.  They obtained their thermal and density structure using 3D spherical Eulerian radiation hydrodynamics models with a gray approximation for the radiative heating. At this stage, the core has a warm central region at $\sim200$\,K and a cooler outer envelope, and has an extent of only around 5\,AU \citep{1998ApJ...495..346M}. Furthermore the first hydrostatic core is relatively short-lived at only $\sim20$\,kyr. They were primarily interested in the detectability of first hydrostatic cores with existing (i.e.\ IRAS) and near future instrumentation at the time. Their synthetic observations predicted that the flux of such objects was below the sensitivity of IRAS, even in the nearest star-forming regions, but that they should be detectable by the infrared space observatory (ISO) and \textit{Spitzer} (at the time known as SIRTF). In reality, the complexity of core structure and fragmentation, coupled with the short lifetime and low flux from first hydrostatic cores made unambigous detection more difficult, though candidates have now been identified and studied in multiple wavelengths, such as B1-bN/B1-bS \citep{2012A&A...547A..54P}, Per-Bolo 58 \citep{2005A&A...440..151H, 2010ApJ...715.1344C, 2010ApJ...722L..33E} Chamaeleon-MMS1 \citep{2006A&A...454L..51B, 2014A&A...564A..99V}, CB17-MMS \citep{2012ApJ...751...89C} and L1451-mm \citep{0004-637X-838-1-60}.

\cite{2010ApJ...725L.239T} computed radiation hydrodynamic models of low-mass cores with only a small natal envelope. Such a core evolves differently from its more canonical counterparts because the accretion duration and accretion rate from the envelope is lower, meaning that the system evolves through angular momentum redistribution in the disc and is also longer lived. They termed these object Exposed Long-lifetime First cores (ELFs). Another key distinction with higher mass cores is that they are sufficiently optically thin for radiative cooling to become significant. They also computed synthetic SEDs for their models, finding that ELFs are fainter in the radio continuum but brighter in the mid-infrared than higher mass cores. The main obstacle for forming ELFs is that the probability of a low mass system collapsing gravitationally is small. 

 \cite{2011PASJ...63.1151T} produced non-LTE synthetic molecular line observations based on the radiation magnetohydrodynamic core collapse models of \cite{2010ApJ...714L..58T}. They discussed the many observational features of such classic cores, including a blue asymmetry in the line profile which is a signature of collapse (discussed more below) and that for edge-on discs a combination of rotation and infall is observed kinematically.  

\cite{2012A&A...545A..98C} studied the collapse of magnetized and non-magnetized cores down to the first hydrostatic core. They used  \textsc{radmc-3d} to produce a time series of synthetic SEDs for each core. They found that the SED alone is insufficient to distinguish whether a magnetic field is present. They also found that it is possible to distinguish between a starless core and the first hydrostatic core at mid- to far-infrared wavelengths (specifically at 24 and 70\,$\mu$m) since in the latter there should be a point source emission component at these wavelengths. 

In a follow up paper, \cite{2012A&A...548A..39C} then used these first hydrostatic core models to generate synthetic ALMA observations in the continuum. They determined which bands and configurations would give the greatest insight. Assuming a distance of 150\,pc, they predicted that ALMA would be capable of resolving fragmentation in cores, and furthermore that it should be able to distinguish between magnetised and unmagnetised cores. \cite{2011ApJ...728...78S} also predicted that first hydrostatic cores should be detectable with ALMA in the continuum at distances of 150\,pc. 

\cite{2012ApJ...750...64S} computed synthetic observations of the optically thick CS~$2-1$ and HCN~$1-0$ lines and the optically thin N$_{2}$H$^{+}$~$1-0$ line, based on their dynamical models of a turbulent, gravitationally collapsing cloud. For a gravitationally collapsing spherical core one typically expects to observe a blue asymmetry in an optically thick line. Roughly, this is because with an optically thick line the source of the emission detected by an observer is the gas closest to the observer at a given velocity. For a collapsing core, at a velocity $+v$ the observer therefore sees gas in the upper (lower density) layers of the core moving away from them, and at velocity $-v$ sees matter deeper inside (higher density) the core moving towards them. The blue-shifted emission is hence expected to be brighter. However, \cite{2012ApJ...750...64S} demonstrated that for a realistic geometry, where the core is embedded in a filament, this expected signature may well not be observed from the majority of viewing angles in optically thick lines, owing to the asymmetric distribution of gas around the core. Furthermore, cores in filaments that are actively accreting may even exhibit a red asymmetry. They therefore warned against relying on the idealised blue asymmetric signature when identifying infalling gas observationally. \cite{2014MNRAS.444..874C} built on this work to show that higher J transitions of HCN and HCO$^{+}$ do yield the blue asymmetric signature of collapsing cores much more reliably.

\cite{2014ApJ...783...60M} post-processed the models of \cite{2013ApJ...770...49O} to compare synthetic single dish (sub)mm images (as observed by e.g.\ the James Clerk Maxwell Telescope, JCMT) with those from ALMA. They concluded that the higher angular resolution available to ALMA is essential for detecting previously unseen small-scale structure that is lost to flux averaging when observed by single dish instruments.  

3D magnetohydrodynamic models of cores collapsing to form discs by \cite{2013ApJ...774...82L} show that depending on the initial magnetic field orientation a rotationally supported disc or pseudo-disc with outflowing outer layers can result. Using these calculations as a basis, \cite{2015A&A...577A..22H} computed 2D semi-analytic models from which they produced synthetic continuum images and also moment maps, pv diagrams and line profiles for a range of CO isotopologues (they also computed synthetic observables from the 3D MHD models). Their objective was to determine observational signatures that would distinguish between a rotationally supported disc and an infalling, rotating core. They found that many observables are similar in the two scenarios, or would not be intuitive indicators of one scenario or the other without supporting simulations. However, there are definite signatures to distinguish between the two. For example, the peak velocity as a function of distance from the source centre is flatter if there is only a pseudodisc than if there is a rotationally supported disc.

\cite{2016A&A...593A...6S} studied the observational properties of pre-stellar cores embedded within filaments using dust continuum radiative transfer. This work had a specific focus on the L1689B system \citep[see also, e.g.][]{2001MNRAS.323.1025J, 2002MNRAS.337L..17R, 2004MNRAS.352.1365R}. They found that azimuthal averaging of an axially symmetric core model results in a significant overestimate of the core mass. Furthermore, they found that constant temperature pixel-by-pixel SED modelling produces unreliable column density and temperature estimates.

\cite{2012MNRAS.420L..53O} computed synthetic observations of star forming cores from the \cite{2009ApJ...703..131O} radiation-hydrodynamic calculations of turbulent molecular clouds. They considered both single dish and interferometric (ALMA, CARMA) observations of these models, predicting that starless cores will be featureless to CARMA, except perhaps for an extremely short time ($\sim10$\,kyr). However ALMA was predicted to be capable of detecting substructure at earlier times. \cite{2016ApJ...823..160D} then searched for signs of collapsing starless cores towards Chameleon using ALMA, but of their 56 cores none showed signatures of collapse (statistically at least two were expected and they estimate a 13.5\,per cent chance of zero detections). Although there is a small chance that their sample size is too small, this probably implies that either starless cores are of a different structure to those resulting from the \cite{2009ApJ...703..131O} numerical models (e.g relatively structure-free Bonnor-Ebert spheres), or that star formation in Chameleon is declining and the observed cores are not actually collapsing.

\begin{figure*}
    \centering
    \vspace{-1cm}
    \includegraphics[width=14cm]{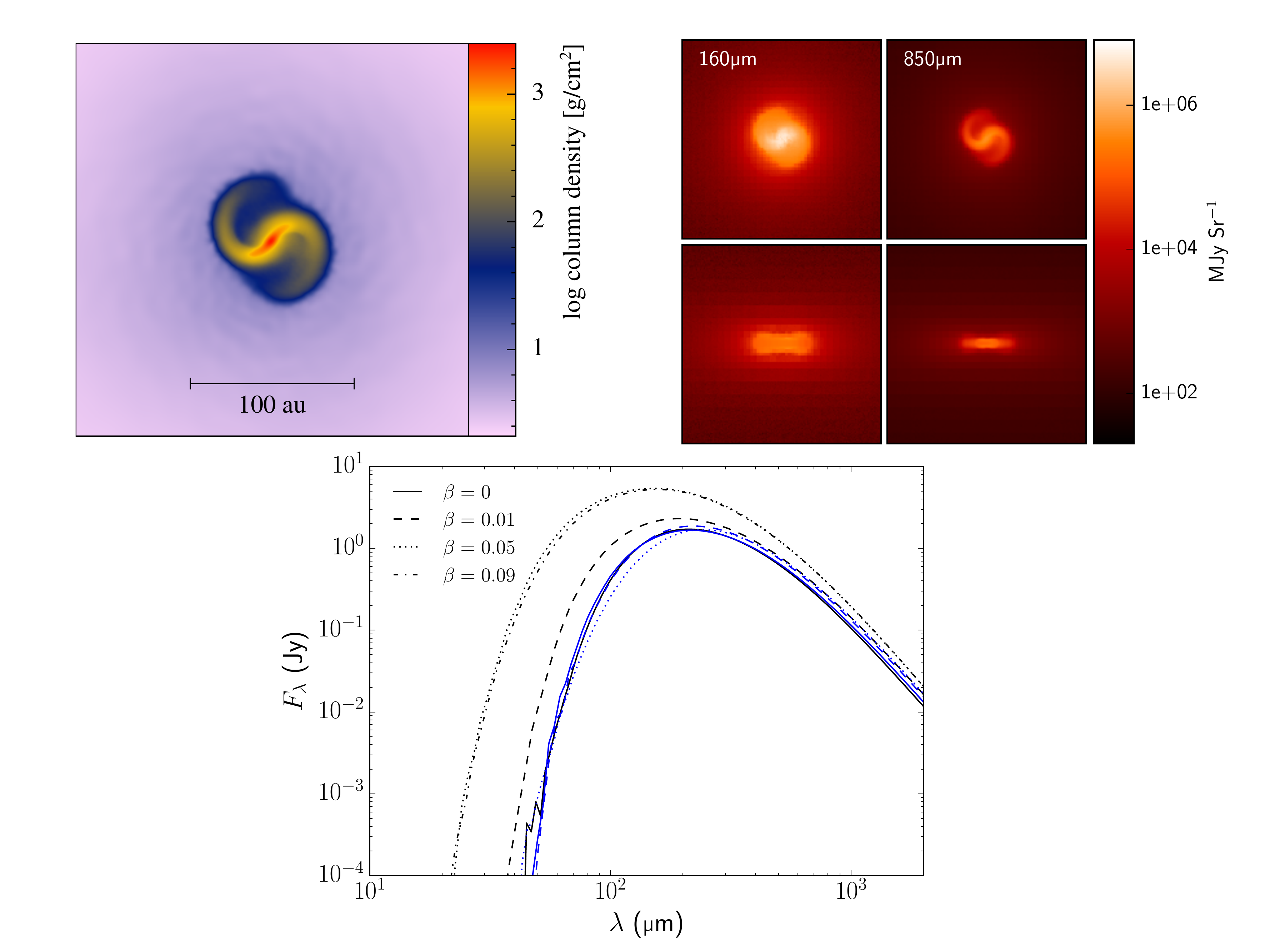}
    \caption{Examples of the work on cores by \cite{2017arXiv171004432Y}. On the top left is a surface density plot of a first hydrostratic core from one of their SPH simulations, which has an initial ratio of rotational to gravitational potential energy $\beta = 0.05$. On the top right are synthetic continuum images of this snapshot at different wavelengths and viewing angles (face on and edge on in the upper and lower panels, respectively). The synthetic SEDs in the bottom panel of this figure illustrate the significant sensitivty of the SED to the core rotation (characterised by $\beta$). {\cite{2017arXiv171004432Y} explored the relative sensitivity of these observables to a range of parameters such as the initial mass, rotation, magnetic field and interstellar radiation field. They found that although there are degeneracies, some real systems are only well described by a small subset of models. } }
    \label{fig:AYoungCores}
\end{figure*}

\cite{2017arXiv171004432Y} post-processed 3D SPH radiation-magnetohydrodynamical calculations of first cores. What separates this from the many prior studies of the SED is that a large parameter space was considered, varying mass, radius, magnetic field, temperature, rotation and the strength of the interstellar radiation field. Furthermore, in the post-processing they also consider different grain size distributions, compositions and observer viewing angles. They found that in their models the peak SED flux predominantly originates from regions exterior to the first hydrostatic core. The SED is also sensitive to the parameters mentioned above; for example, the FIR flux drops off more steeply for a higher mass, weaker interstellar radiation field or a more edge-on inclination. Some of their results, including a simulation snapshot, synthetic continuum images and an illustration of the SED sensitivity to the initial ratio of gravitational to potential energy is shown in Figure \ref{fig:AYoungCores}.  Finally, they compared their models with various candidate first cores using 
SED fitting. They found that some real systems are well described by only a small number of models with many similar parameters, allowing them to constrain the nature of the candidates. For example Aqu-MM1 was interpreted to be closer to edge on, and rapidly rotating. They also concluded that Chamaeleon-MMS1 and Per-Bolo 58 are likely more evolved, having undergone stellar core formation. 

 \cite{2017ApJ...834..201L} computed synthetic polarization maps from 3D ideal MHD simulations of collapsing magnetised protostellar cores. In particular they were interested in comparing their results with polarization maps from the CARMA interferometer \citep{2004SPIE.5498...30W, 2015JAI.....450005H}. They found that the polarization fraction is higher for magnetic fields that are aligned with the plane of the sky, making this a likely explanation for why some cores are observed to have higher polarization fractions. Furthermore, their most strongly magnetised model cores had polarization fractions in excess of those observed, allowing them to constrain the mass-to-magnetic-flux ratio expected for such cores. Also, the outflow angle from their cores varies strongly with time and, for weak fields, is uncorrelated with the magnetic field. For stronger magnetic fields, the outflow and field structure are more aligned. 

There have also been a number of studies that have produced synthetic observations of so-called ``hot cores''. These are short lived ($<0.5$\,Myr), dense ($>10^7$\,cm$^{-3}$), warm ($\sim100$\,K) condensations that might be the precursors of more massive YSOs. Once a protostar forms at the centre of a collapsing core, it heats the surrounding gas and dust. The higher dust temperatures lead to sublimation of the ice mantles surrounding the dust grains, leading to the ejection of a significant abundance of complex molecules back into the gas phase, including species such as NH$_3$, H$_2$O and CH$_3$CN. Since the chemical makeup of the hot core is a function of the dust composition and the nature of the collapse (e.g.\ the time varying temperature and the speed of the collapse), probing the chemical composition of hot cores can give an insight into their history. 

\cite{2007PASJ...59..589O} computed semi-analytic models of cores with an accretion shock bounding the first hydrostatic core, which itself is embedded within an envelope.  Interior to the shocked layer is a post-shock relaxation region, which undergoes strong chemical evolution and has a corresponding plethora of radiative processes and hence dominates much of the emission. However, much of the radiation emitted there is reprocessed and reddened (from $\sim10\mu$m to $\sim100\,\mu$m) in the envelope. Nevertheless H$_2$O emission was still expected to be detectable. 

\cite{2014MNRAS.443.3157C} studied complex organic molecules towards six star-forming cores. They used \textsc{ucl-chem} to compute the chemical evolution of hot cores over time, and to see if ratios of the observed species abundances can help to constrain the masses and ages of the observed cores. From these models, they could conclude that the observed cores were qualitatively ``old'' (>20\,kyr). However, since the formation and destruction pathways of complex organic molecules in astrophysical environments are not well understood from a laboratory perspective, quantitative evaluation was deemed impossible. Instead they included a qualitative comparison with the models, and focused more on the identification of lines from complex organic molecules. 

Finally, \cite{2015A&A...575A..68C} systematically varied the properties of hot cores in 3D chemical and radiative transfer models that combined \textsc{radmc-3d} for the radiation transport and dust temperature calculation and \textsc{saptarsy} \citep{1995ApJ...441..222B, stephan:tel-01514925} for the chemical modelling. They assumed that the gas and dust were thermally coupled, so that the temperature in the \textsc{radmc-3d} dust radiative equilibrium calculation is imposed upon the chemical evolution, an assumption that is probably justified given the high densities in hot cores. Again, in these hot core models complex species are liberated from grains due to high temperatures and shocks and then in the post-shock region the gas cools, reacts and is re-adsorbed onto grain surfaces, changing the chemistry over time. \cite{2015A&A...575A..68C} studied the variation of observables from complex organic molecules and produced a reference framework from which an observed core might be linked into an evolutionary sequence.

\subsubsection{Cloud-cloud collisions and colliding flows}
\label{sec:ccc}

Collisions between clouds are a candidate mechanism for generating the abnormally high densities over a widespread volume expected to be required for the formation of massive star clusters or super star clusters. Without such conditions, feedback may shut off star formation once a handful of massive stars have formed. Dynamical models of cloud-cloud collisions \citep[e.g.][]{1992PASJ...44..203H, 2013ApJ...774L..31I, 2013MNRAS.431..710M, 2014ApJ...792...63T, 2015MNRAS.453.2471B, 2017ApJ...835..137W, 2015MNRAS.453.2471B, 2017MNRAS.465.3483B, 2017arXiv171002285S} support this argument theoretically, but identifying such a process is observationally difficult. Clouds themselves are geometrically complex structures and collisions take place over many millions of years, meaning that the information available to an observer at a given moment in time is very limited. Synthetic observations are therefore required to predict observational signatures of cloud collisions, and also to interpret real systems.

\begin{figure}
    \centering
    \includegraphics[width=7cm]{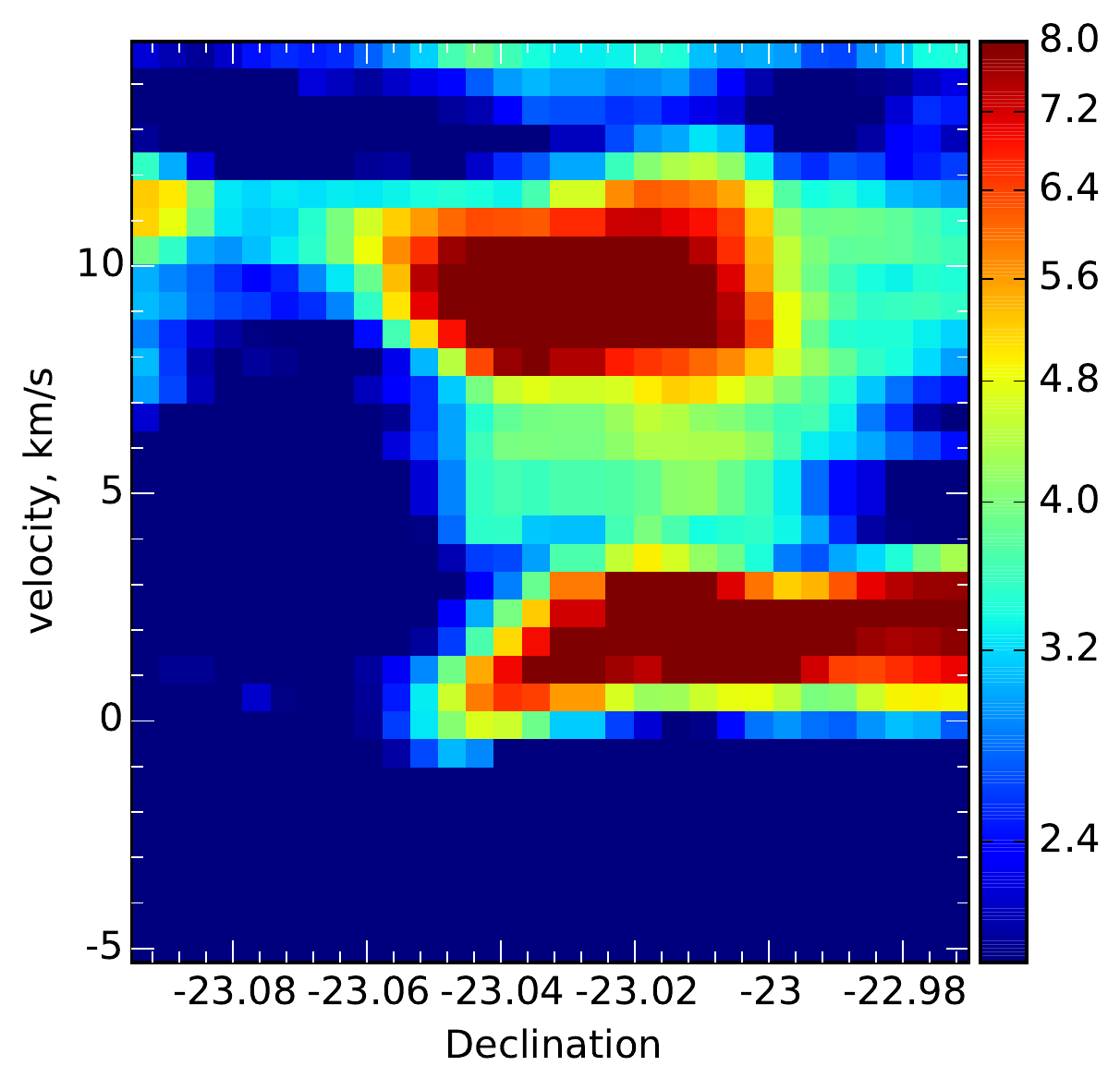}
    \includegraphics[width=7cm]{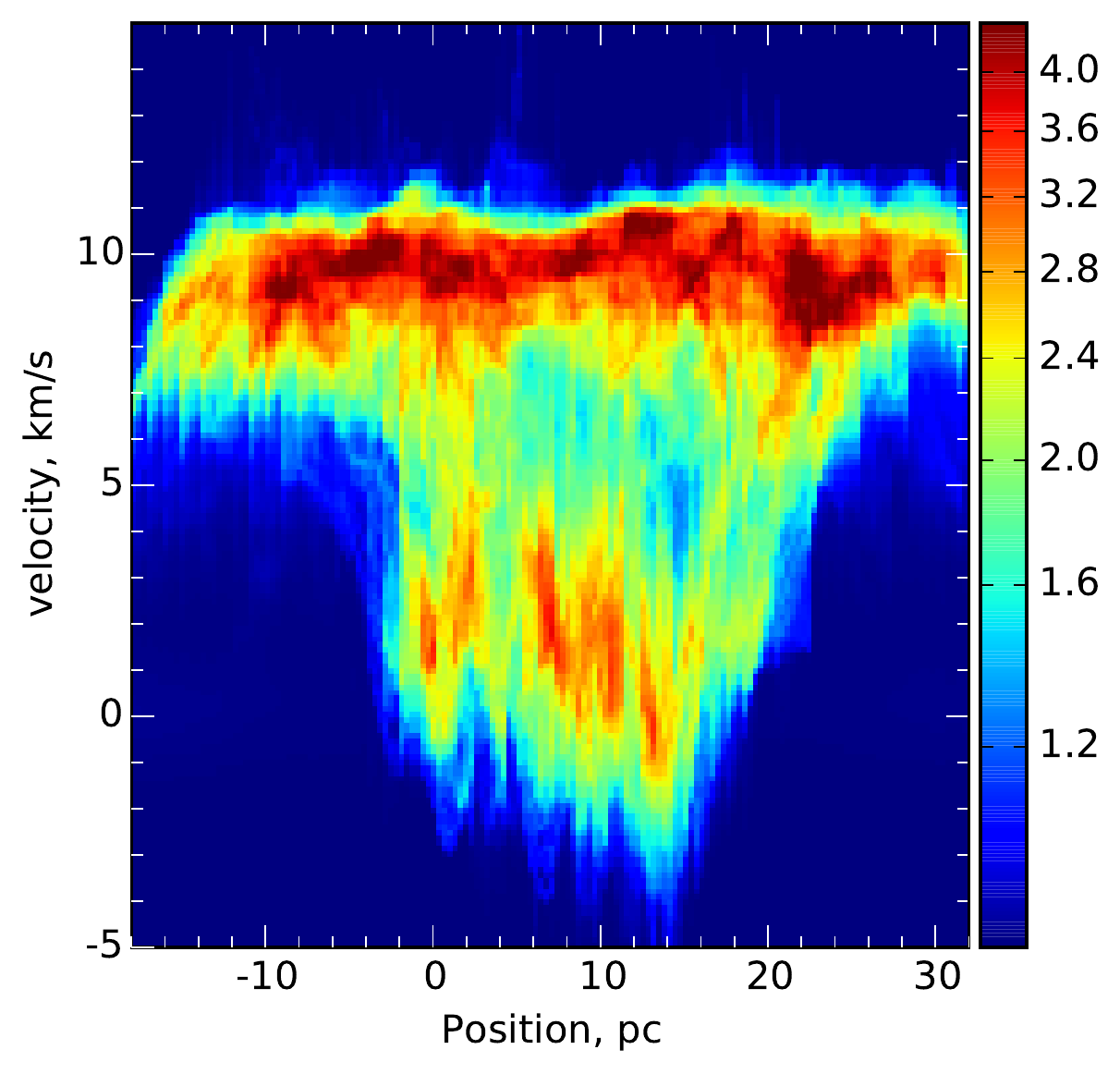}
    \includegraphics[width=7cm]{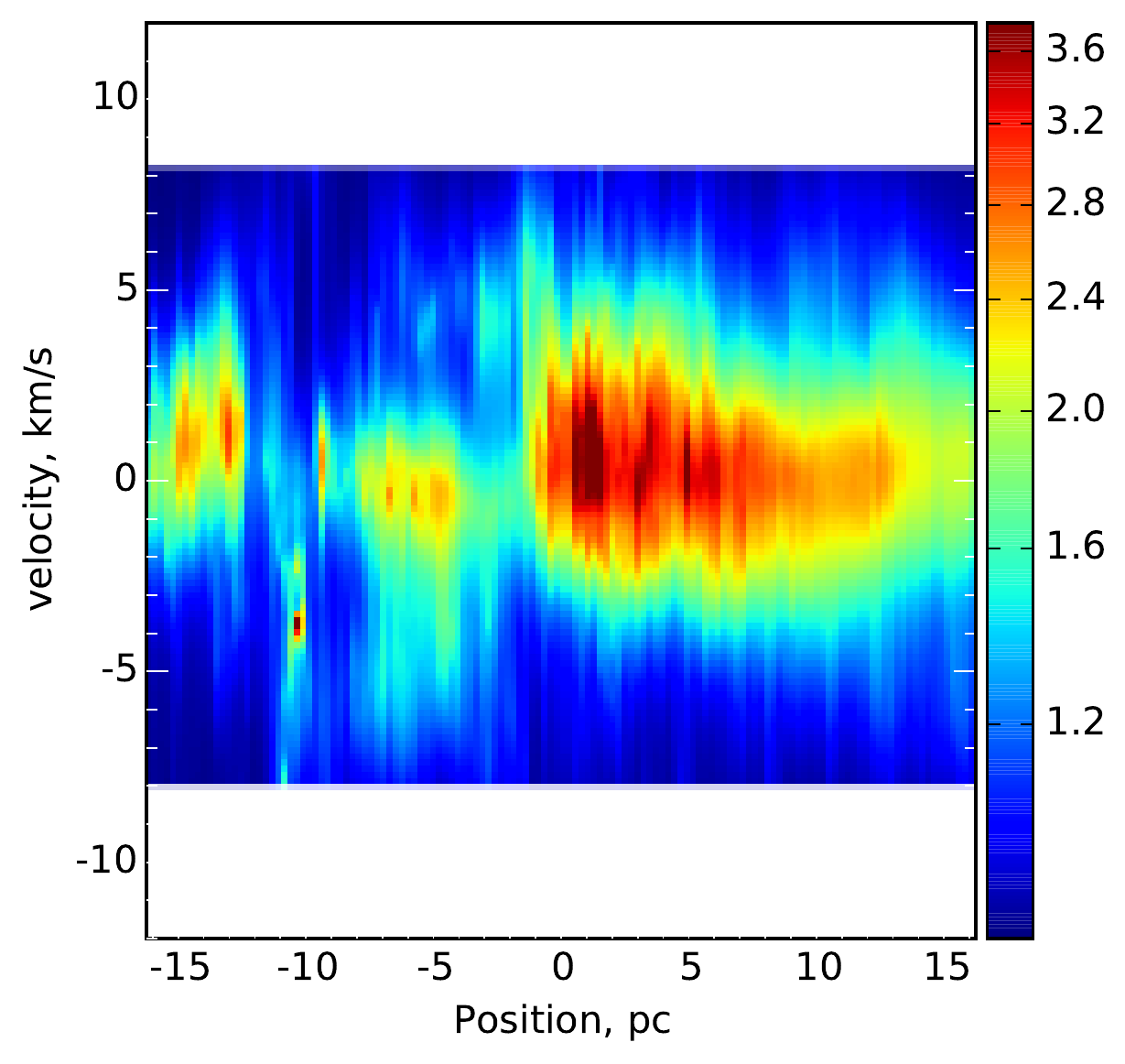}
    \caption{The upper panel is a real CO position-velocity diagram towards M20. The middle is a synthetic P-V diagram from a cloud-cloud collision model and the lower panel from a simulation of an expanding H\,\textsc{ii} region. These Figures are adapted from \cite{2015MNRAS.450...10H}. }
    \label{cccfig}
\end{figure}

\cite{2011A&A...528A..50D} interpreted the star formation history of the Serpens main cluster in terms of colliding clouds. They carried out SPH simulations of colliding cylinders and compared the resulting position-velocity diagrams with observations. For the purposes of this comparison, they assumed that the observed C$^{18}$O emission was an unbiased tracer of the velocity structure of Serpens and that the dust emission was an unbiased tracer of its column density structure. This assumption allowed them to compare the velocities and column densities inferred from the observations directly with the values from the simulation without the need to produce synthetic emission maps, although they did smooth the simulation data spatially and in velocity space to account for the finite resolution of the observations. They concluded that the morphological, kinematic and star formation structure of the region could be explained in terms of one of their collision models. 

\cite{2015MNRAS.450...10H} and \citet{2015MNRAS.454.1634H} produced synthetic CO emission maps based on cloud-cloud collision models from \cite{2014ApJ...792...63T} and \cite{2017arXiv171002285S}. They compared these with similar maps based on models of star formation and radiative feedback in a single cloud carried out by \cite{2012MNRAS.424..377D} and \cite{2013MNRAS.430..234D}. They found that the position-velocity diagrams 
yield a ``broad bridge feature'' (at least over some fraction of viewing angles) which arises in the collision models, but not for an isolated cloud with feedback. This is defined as separate intensity peaks along the line of sight that are connected by intermediate intensity emission at intermediate velocities. A comparison of a real broad bridge with synthetic CO maps of an expanding H\,\textsc{ii} region model and a cloud collision model are shown in Figure \ref{cccfig} \citep[adapted from][]{2015MNRAS.450...10H}.  Additionally, the lifetimes and viewing angle sensitivity of these broad bridge features was studied. Although observing such a signature does not guarantee a collision, it does strengthen the argument for one in the presence of other indicators. Furthermore, in the work to date this signature was determined for head-on collisions only.

\begin{figure}
    \includegraphics[width=0.95\linewidth]{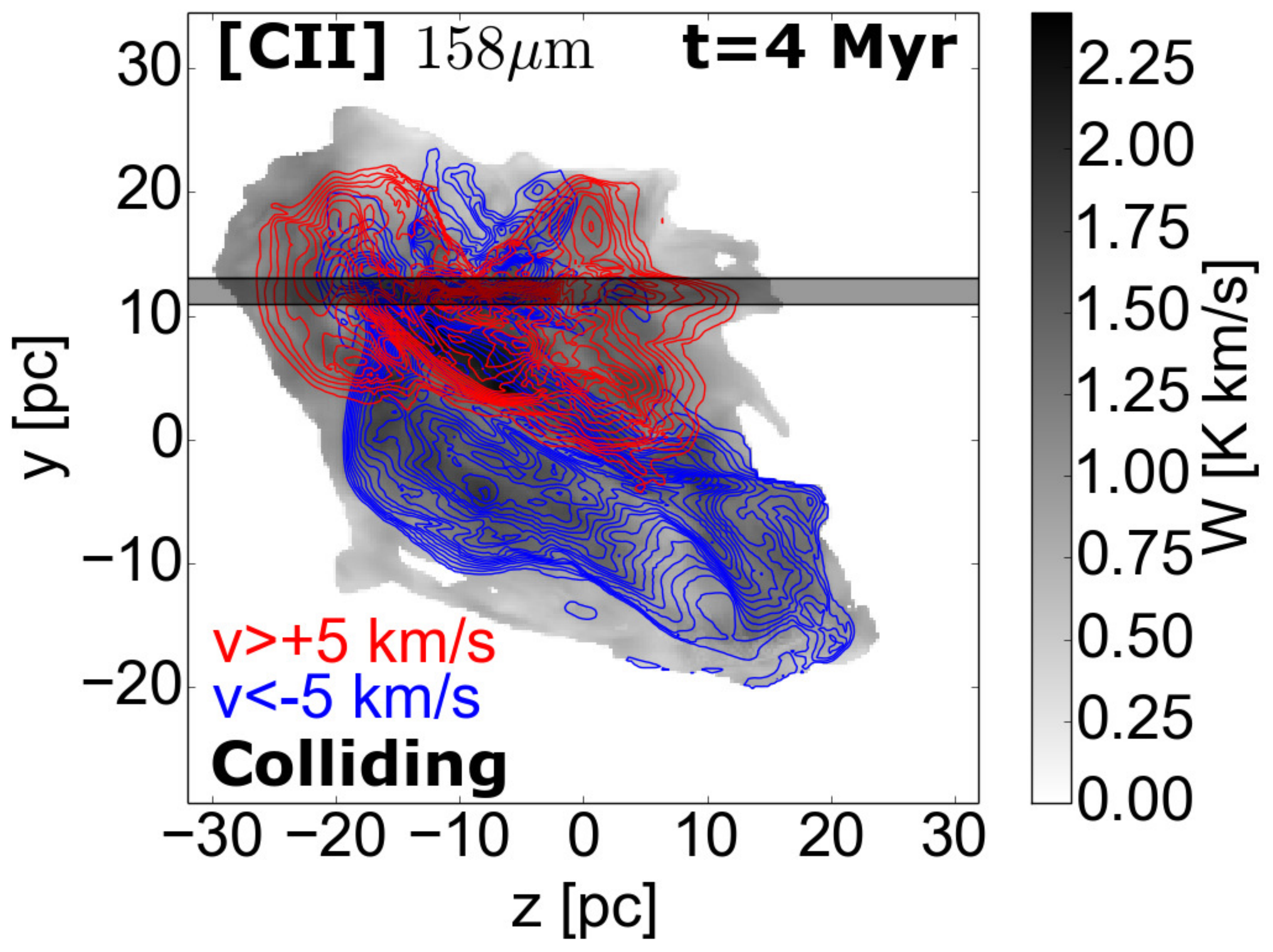}
    \includegraphics[width=0.97\linewidth]{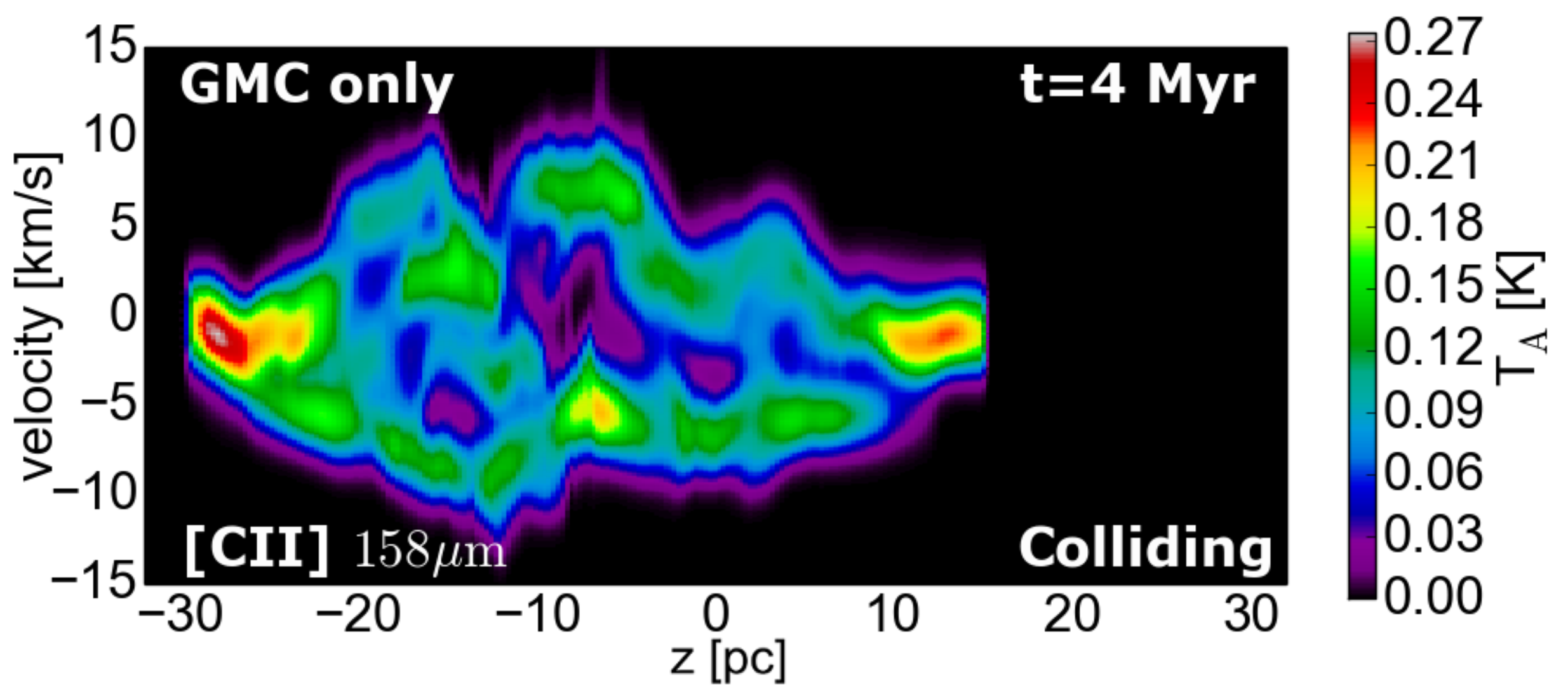}
    \caption{{\it Top panel:} High velocity gas in the [C\textsc{ii}] $158\,\mu$m line of an evolved ($t=4\,{\rm Myr}$) cloud-cloud collision \citep{2017ApJ...841...88W} occurring along the line-of-sight of the reader. The red and blue shifted components overlap, indicating occurence of a cloud collision. The shadowed strip of 2 pc width shows the part of the GMCs, whose p-v diagram is shown in the bottom panel. {\it Bottom panel:} p-v diagram of the 2-pc-wide strip through the dense overlap region. The bridge-effect can be seen in at position $z\sim-8\,{\rm pc}$. Figure taken from \citet{2017arXiv170607006B}}
 \label{fig:ccc_tgb}
\end{figure}

In a series of papers, \cite{2015ApJ...811...56W}, \citet{2017ApJ...841...88W}, \cite{2017ApJ...835..137W} have performed new simulations of cloud collisions, considering a range of different impact parameters and including the effect of magnetic fields. They also post-processed these simulations to study synthetic molecular line images (specifically CO and its isotoplogues) as well as position-velocity diagrams and surface density PDFs. In addition to kinematic signatures like the broad bridges, they discussed the potential for using mid to high-J CO lines as diagnostics of collision. Specifically, the ratio of mid-high J CO lines to lower CO lines is elevated by a cloud-cloud collision. 
By post-processing different snapshots of the \citet{2017ApJ...841...88W} simulations, \cite{2017arXiv170607006B} recently expanded the study of broad bridge features to other lines. They found that the [CII] 158$\mu{m}$ line can trace a collision for a longer period of time than the CO observations examined in prior work because it traces more extended (and hence later colliding) components of the cloud (see Figure~\ref{fig:ccc_tgb}). However, since the brightness of the [CII] emission produced in this scenario will be highly sensitive to the strength of the interstellar radiation field, it is not clear whether one would expect to see this signature for all collisions or merely those occurring in regions with an elevated radiation field strength.

Overall, synthetic observations have helped support the identification (and improve the confidence of identification) of a number of candidate collision sites in recent years \citep[e.g.][]{2015ApJ...806....7T, 2016ApJ...820...26F, 2017ApJ...835..142T, 2017PASJ...69L...5F, 2017arXiv170607164T, 2017arXiv170607964K, 2017arXiv170606956N, 2017arXiv170605871H}. With further signposts, the identification of collisions will become more robust and the overall contribution of collisions to massive star formation will become clearer.  

Colliding flows are similar to cloud-cloud collisions, perhaps distinguished by the latter being the interaction of discrete objects versus a more continuous interacting flow. The expected result is still the same in that the built up layer rapidly undergoes  gravitational collapse \citep[e.g. for dynamical models of this process see][]{2008A&A...486L..43H, 2009MNRAS.398.1082B,  2011MNRAS.414.2511V}.   

\cite{2007ApJ...657..849Y} computed non-LTE synthetic C\,\textsc{ii}, C\,\textsc{i}, and CO observations from calculations of clumps condensing out of a two-phase turbulent medium. They found that line ratios are a signature of the two-phase ISM, for example the CO J=4-3/J=1-0 ratio is greater than unity in the two phase scenario, but less than 0.1 in models of the single phase ISM. 

\cite{2015MNRAS.452.1353H} computed models of molecular cloud formation out of a thermally bi-stable turbulent medium. The resulting clouds were sometimes associated with large-scale infalling flows which synthetic observations predicted would manifest themselves as double peaked H\,\textsc{i} line profiles. 

\cite{2015ApJ...801...77M} compared models of isolated turbulent media with those in which there is a colliding flow between two turbulent media. Although they did not compute synthetic osbervations, they analysed the column density PDF, pv-diagrams and channel maps. They found that a colliding flow changes the PDF from a log-normal distribution to that of a power law. Given that both log-normal and power law distributions are observed, the distinction between two such regions could be due to environmental (i.e.\ in this case collisional) factors.

\subsubsection{Measuring the molecular gas mass}
\label{ssec:traceh2}
One of the simplest properties of a molecular cloud that we might hope to learn from observations is the mass of molecular gas that it contains. However, this turns out to be surprisingly difficult to measure. H$_{2}$ is a very light molecule and moreover has no dipole moment, meaning that its available rotational transitions are quadrupole transitions for which $\Delta J = \pm 2$. Consequently, even the lowest rotational transition of H$_{2}$ has a relative large energy, $\Delta E \sim 0.05$~eV, corresponding to a gas temperature of around 500~K. Therefore, although H$_{2}$ rotational line emission can be used as a tracer of warm molecular gas \citep[see e.g.][]{2016ApJ...830...18T}, it is insensitive to cold molecular gas. Since most of the H$_{2}$ in the ISM of a galaxy like the Milky Way is too cold to excite H$_{2}$ emission \citep{2016MNRAS.462.3011G}, we are forced to rely on other observational tracers of molecular gas.

The most widely-used tracer of H$_{2}$ in molecular clouds is CO line emission. Since efficient formation of CO in the gas-phase requires the presence of H$_{2}$, large concentrations of CO are found only in gas which is H$_{2}$-rich. Moreover, fairly similar amounts of shielding are required to prevent both molecules from being photodissociated by the ISRF, although the fact that H$_{2}$ requires less shielding than CO means that a significant fraction of the H$_2$ in a molecular cloud might be located in a CO-dark molecular phase \citep{2010ApJ...716.1191W}. It is therefore not unreasonable to think that by observing the CO, we might learn something about the H$_{2}$.

There are three main methods for converting from CO emission to H$_{2}$ mass. First, if the cloud is in virial equilibrium, we can use the measured CO linewidth and the area of the region traced by CO to estimate the total mass of the cloud \citep{1988ApJ...333..821M,2006PASP..118..590R}. If we then correct for the presence of helium (simple, since we generally know the He:H ratio fairly accurately) and atomic hydrogen (simple if we have a 21~cm emission map, provided that this emission is optically thin), then we can convert this total mass into an estimate of the H$_{2}$ mass. This method is often used to estimate the H$_{2}$ masses of clouds in resolved extragalactic studies  \citep[see e.g.][]{2010MNRAS.406.2065H} but has the obvious drawback that we are forced to assume virial equilibrium. However, recent work by \citet{2016MNRAS.460...82S} suggests that this method can in some cases yield surprisingly accurate H$_{2}$ masses for clouds that are not in virial equilibrium. \citet{2016MNRAS.460...82S} modelled the chemical and thermal evolution of a variety of turbulent clouds, and then generated synthetic $^{12}$CO and $^{13}$CO emission maps based on these simulations. They then used these maps to validate various different methods for determining the H$_{2}$ mass from the CO emission. They found that even in clouds that were initially not in virial equilibrium, the dense region where most of the CO forms tends to be close to equilibrium, allowing the virial mass estimate to perform better than one might naively expect. One minor concern, however, is that all of the clouds simulated by \citet{2016MNRAS.460...82S} had broadly similar initial conditions: an isolated, spherical overdensity, with an imposed turbulent velocity field. It is clearly important to determine whether the same result holds true for clouds formed from a wider variety of different initial conditions.

The second common method for measuring the H$_{2}$ mass of a molecular cloud involves first determining the $^{13}$CO column density. This can be inferred from the integrated intensity of $^{13}$CO if the excitation temperature of the CO molecules is known. The latter can easily be measured for $^{12}$CO, provided that the emission is optically thick, and hence if we assume that $T_{\rm ex, 12CO} = T_{\rm ex, 13CO} = T_{\rm ex}$, we can calculate the $^{13}$CO column density using the equation \citep{2009tra..book.....W}
\begin{equation}
N(^{13}{\rm CO}) = 3.0 \times 10^{14} \frac{T_{\rm ex} \int \tau_{13}(v) \, {\rm d}v}{1 - \exp(-5.3/T_{\rm ex})}  \: {\rm cm^{-2}},
\end{equation}
where $\tau_{13}(v)$ is the optical depth of the $^{13}$CO $J=1-0$ transition at velocity $v$, which is given by
\begin{equation}
\tau_{13}(v) = -\ln \left[1 - \frac{T_{\rm b}^{13}(v)}{5.3} \left \{
\exp \left(\frac{5.3}{T_{\rm ex}} - 1\right)^{-1} - 0.16 \right \}^{-1}
\right]
\end{equation}
where $T_{\rm b}^{13}(v)$ is the observed brightness temperature of the $^{13}$CO 1-0 line at a velocity $v$. Given $N(^{13}{\rm CO})$, the H$_{2}$ column density then follows if we assume some simple relationship between the $^{13}$CO and H$_{2}$ abundances. Typically, one assumes that $n_{\rm 12CO} / n_{\rm H_{2}} = 8 \times 10^{-5}$, following \citet{1987ApJ...315..621B}, and then adopts a value for the $^{12}$CO/$^{13}$CO ratio that is appropriate for the Galactic environment of the cloud, e.g.\ a value of around 60 for the Solar neighbourhood. 

\citet{2016MNRAS.460...82S} also examined the performance of this H$_2$ mass estimate in their numerical study of turbulent clouds, and showed that it tends to underestimate the true H$_{2}$ mass, by anything from a factor of two or so in Milky Way-like conditions to a factor of ten or more in low metallicity gas, even once one accounts for the lower C and O abundances. They show that the fundamental problem is the assumed relationship between the CO and H$_{2}$ abundances. The value usually adopted assumes that essentially all of the available gas-phase carbon is incorporated into CO. This is not a terrible assumption for GMCs in the Milky Way, although it doesn't account for the CO-dark H$_{2}$ discussed by \citet{2010ApJ...716.1191W}. However, it becomes a much worse assumption in low metallicity conditions, as more of the carbon in those clouds is found in the form of C$^{+}$ or C, and less as CO.

Finally, the other major method used to convert between CO emission and H$_{2}$ mass -- indeed, by far the most common technique -- is the use of an empirical relationship known as the ``X-factor''. Observations have established that in the Milky Way disk, there is a linear relationship between the mean $^{12}$CO $J=1-0$ integrated intensity of a cloud, $W_{\rm CO}$, and its mean H$_{2}$ column density $N_{\rm H_{2}}$:
\begin{equation}
N_{\rm H_{2}} = X_{\rm CO} W_{\rm CO},
\end{equation}
with
\begin{equation}
X_{\rm CO} \simeq 2 \times 10^{20} \: {\rm cm^{-2}} ({\rm K \, km \, s^{-1}})^{-1},
\end{equation} 
with a scatter of around $\pm 30\%$ \citep{2013ARA&A..51..207B}. $X_{\rm CO}$ can be constrained to this high degree of precision in the Milky Way because we have several independent ways to measure cloud column densities (dust extinction, dust emission, gamma-ray emission), which can be combined with information on the H$\,${\sc i} column density from 21~cm surveys \citep[e.g.][]{2006AJ....132.1158S,2016A&A...595A..32B} to yield the H$_{2}$ column density. In other galaxies, however, many of these methods are unavailable, making it much harder to constrain $X_{\rm CO}$ observationally. Therefore, study of how $X_{\rm CO}$ behaves as a function of metallicity or environment is highly reliant on numerical simulations and synthetic observations.

Early attempts to explain the apparent constancy of $X_{\rm CO}$ in nearby molecular clouds involved simple analytical models. A particularly influential example is the ``mist'' or ``droplet'' model of \citet{1986ApJ...309..326D}. This model assumes that the observed CO luminosity of a given patch of gas (e.g.\ an individual molecular cloud in our own Galaxy or a somewhat larger portion of the ISM in a more distant galaxy) is contributed by a large number of non-overlapping clumps. In this case, if the individual clumps are virialized, one can show that the result CO to H$_{2}$ conversion factor scales as
\begin{equation}
X_{\rm CO} \propto \frac{\bar{\rho}}{T_{\rm B}},
\end{equation}
where $\bar{\rho}$ is the mean cloud density and $T_{\rm B}$ is the mean CO brightness temperature of the ensemble of clumps. Since $\bar{\rho}$ and $T_{\rm B}$ both vary only weakly from cloud to cloud within the local ISM 
\citep{2010ApJ...723..492R}, the result is that the values of $X_{\rm CO}$ yielded by this model also vary only weakly. Models of this type have also been discussed by a number of other authors \citep[see e.g.][]{1987ASSL..134...21S,1987ApJ...319..730S,1989ApJ...338..178E}
and they have been used to make predictions for the behaviour of $X_{\rm CO}$ in other galaxies \citep[e.g.][]{1988ApJ...325..389M,1996ApJ...462..215S}. Ultimately, however, the validity of this kind of model depends on the accuracy of the assumptions underpinning it, and these can be controversial -- for instance, there is a long-running argument in the literature over whether or not most molecular clouds are in virial equilibrium \citep[see e.g.][]{1990ApJ...348L...9M,1999ASIC..540...29M,2006MNRAS.372..443B,2011ApJ...738..101G,2013ApJ...779..185K}. There is thus ongoing interest in the use of more sophisticated models to predict the behaviour of $X_{\rm CO}$ as a function of environment.

{ Several different approaches involving synthetic observations have been used to better understand the behaviour of $X_{\rm CO}$. One of the simplest is the use of a PDR code to predict the CO brightness and H$_{2}$ content of clouds as a function of their size and environment. Examples of this kind of model can be found in \citet{1993MNRAS.264..929T}, \citet{1993ApJ...402..195W},  \citet{2006MNRAS.371.1865B}, \citet{2007MNRAS.378..983B} and \citet{2010ApJ...716.1191W}. They represent a significant improvement compared to the simple analytical models discussed above, as they remove the need to make assumptions regarding the molecular gas filling factor or the CO brightness temperature. Nevertheless, one weakness of this approach is that most PDR models of this type approximate molecular clouds as 1D objects (either slabs or spheres), whereas real clouds are observed to be clumpy and turbulent, with geometrically complex structures. This weakness has begun to be addressed in the past few years as advances in computer power have made it possible to model clumpy PDRs in 3D
(see e.g.\ \citealt{2015ApJ...803...37B,2017ApJ...839...90B}; an example of the results of this approach is shown in Figure~\ref{fig:conv}).


Another weakness of the static PDR approach is that these models tell us little about the time evolution of the cloud and hence do not allow us to study the evolution of $X_{\rm CO}$ as a function of time. To address this, one needs to make use of simulations that couple ISM chemistry with three-dimensional hydrodynamics, as already discussed in Section~\ref{sec:composition} above. An increasing number of such models have become available in recent years, differing primarily in the scales on which they model $X_{\rm CO}$ (whole disks or individual clouds), and the accuracy of the methods used to model the chemical evolution of the gas and to produce synthetic maps of CO line emission. 

Models of $X_{\rm CO}$ on the scale of individual clouds that couple hydrodynamics and chemistry have been used to explore the role that dust shielding and turbulence play in setting the value of $X_{\rm CO}$ (\citealt{2011MNRAS.412..337G}, \citealt{2011MNRAS.412.1686S}~2011a,b), and have also been used to quantify the impact of changes in metallicity \citep{2012MNRAS.426..377G}, local star formation rate \citep{2015MNRAS.452.2057C}, and to look at the time evolution of $X_{\rm CO}$ within gravitationally collapsing clouds 
\citep{2016MNRAS.456.3596G,2017arXiv170406487S}. One of the most important results from these models is the finding that at low metallicity, the total CO emission from a given GMC, and hence its value of $X_{\rm CO}$, will be dominated by the contribution coming from a few bright clumps with small angular sizes. Recent ALMA observations of CO emission in low metallicity dwarf galaxies find results that agree well with this prediction \citep{2015Natur.525..218R}.

\begin{figure}
    \centering
    \includegraphics[width=0.98\linewidth]{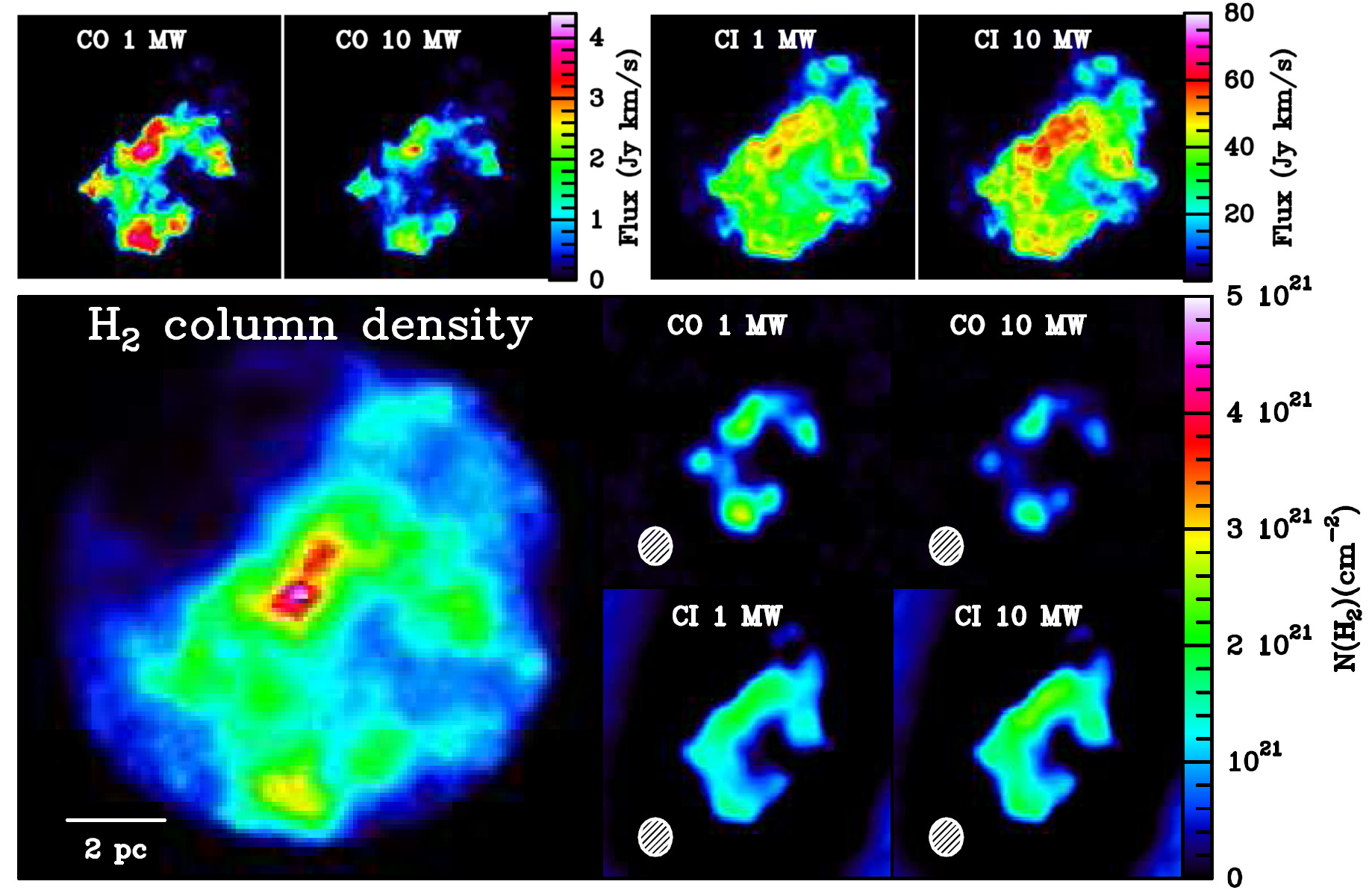}
    \caption{Example showing synthetic ALMA observations of the CO(1-0) and [C\textsc{i}](1-0) lines emitted from a typical Milky Way (MW) GMC. The cloud is irradiated by an MW-average isotropic FUV radiation field and cosmic rays that produce a cosmic ray ionization rate $\zeta_{\rm CR}$ equal to 1 and 10 times the MW average. The large panel on bottom left shows the true H$_2$ column density. The top row shows CO(1-0) (left-pair) and [C\textsc{i}](1-0) (right-pair) in the two different $\zeta_{\rm CR}$ cases. These are the true fluxes in units of (Jy km/s). The bottom four panels show the obtained N(H$_2$) by using the standard CO-to-H$_2$ conversion factor (upper pair) and the C\textsc{i}-to-H$_2$ conversion factor (lower pair). The hatched circle represents the beam size of the telescope. Under MW conditions, both conversion factors return similar estimate for the H$_{2}$ column density. For higher $\zeta_{\rm CR}$, however, the C\textsc{i}-to-H$_2$ method appears to provide a better estimate of N$_{\rm H_{2}}$. In the above example, the \textsc{3d-pdr} code, the \textsc{radmc-3d} code and the \textsc{casa} ALMA simulator have all been used. }
 \label{fig:conv}
\end{figure}

These small-scale studies have been complemented by efforts to model $X_{\rm CO}$ within large ensembles of clouds on the scale of individual galaxies or large portions of galaxies. Examples of this approach include \citet{2011MNRAS.418..664N,2012MNRAS.421.3127N}, \citet{2012ApJ...747..124F}, \citet{2015A&A...575A..56B}, \citet{2016MNRAS.460.2297R} and \citet{2018ApJ...858...16G}. These large-scale models typically do not have sufficient resolution to directly predict the CO content of the gas and hence often rely on sub-grid models calibrated by smaller-scale studies \citep[see e.g.][]{2012ApJ...747..124F}, or simply assume that all of the gas above some density threshold is fully molecular \citep{2015A&A...575A..56B}. The sensitivity of the  results of this kind of model to these simplifications remains to be properly quantified. Large-scale models of this type allow one to explore the influence of the dynamical state of the galaxy -- i.e.\ whether it is evolving in isolation or undergoing a merger -- on its average value of $X_{\rm CO}$, and have been used to explain why $X_{\rm CO}$ appears to be systematically smaller in rapidly star-forming galaxies than in the local ISM \citep[see e.g.][]{2012MNRAS.421.3127N,2015A&A...575A..56B}.

The simulations used to study $X_{\rm CO}$ typically find that it is highly sensitive to metallicity, a result which is also strongly supported by observations \citep[see e.g.][and references therein]{2013ARA&A..51..207B}. There is hence great interest in finding other viable observational tracers of H$_{2}$.} Within our own galaxy, species such as CH or HF have been found to be very effective tracers of H$_{2}$ \citep[see e.g.][and references therein]{2016ARA&A..54..181G}. However, they are generally observed only in absorption, and hence cannot be used to map the distribution of H$_{2}$ over large areas and are of little use for extragalactic observations. Instead, the most promising alternative tracer of H$_{2}$ in extragalactic systems appears to be atomic carbon (C{\sc i}). The $^3P_1\rightarrow\,^3P_0$ transition of C{\sc i} at $492.16\,{\rm GHz}$ was first proposed as a molecular gas tracer by \citet{2000ApJ...537..644G} and \citet{2001A&A...371..433I}, on observational grounds, and by \citet{2004MNRAS.351..147P} on theoretical grounds. More recently, several different studies have used synthetic observations to examine the strengths and weaknesses of C{\sc i} as a tracer of H$_{2}$. 

\citet{2014MNRAS.440L..81O} performed numerical simulations in which they post-processed snapshots from turbulent hydrodynamical runs with {\sc 3d-pdr} to estimate the gas temperature and chemical composition of the gas. They then further post-processed the results with {\sc radmc-3d} to estimate the emission in the C{\sc i} (1-0) and CO $J=1-0$ lines. They found that C{\sc i} was able to trace the low density H$_2$ gas much better than CO and derived a C{\sc i}-to-H$_{2}$ factor for their simulated clouds given by $X_{\rm CI} \sim1.1\times10^{21}\,{\rm cm}^{-2}\,{\rm K}^{-1}\,{\rm km}^{-1}\,{\rm s}$.
Independently, \citet{2015MNRAS.448.1607G} also modelled the distribution of atomic carbon in a turbulent, gravitationally collapsing cloud and produced synthetic C{\sc i} (1-0) and (2-1) emission maps. In agreement with \citet{2014MNRAS.440L..81O}, they also found that C{\sc i} is a better tracer of low density H$_{2}$ than CO. However, they showed that neither tracer works particularly well for lines of sight with $A_{\rm V} < 1$ in a solar metallicity cloud. They derived a value of $X_{\rm CI} \sim1.0\times10^{21}\,{\rm cm}^{-2}\,{\rm K}^{-1}\,{\rm km}^{-1}\,{\rm s}$, within 10\% of the \citet{2014MNRAS.440L..81O} result. Interestingly, a similar value was derived observationally by \citet{2014ApJ...797L..17L} for Vela Molecular Ridge cloud C. Finally, \citet{2016MNRAS.456.3596G} studied the behaviour of $X_{\rm CI}$ in clouds with a wide range of metallicities and UV radiation fields. They found that the $X_{\rm CI}$ scales inversely with metallicity as $X_{\rm CI} \propto {\rm Z^{-1}}$, and also showed that $X_{\rm CI}$ displays far less time variation that $X_{\rm CO}$, particularly in low metallicity systems.

\subsection{The larger-scale ISM}
In this review, we have largely highlighted examples of the use of synthetic observations that involve modelling objects on the scale of molecular clouds or smaller. However, there have also been various efforts to produce synthetic observations of the ISM on much larger scales, ranging from $\sim$kpc scale sub-volumes to models of the entire Galactic disk. In this section, we briefly review some of these efforts. 

We begin by noting that although there have been various efforts to produce synthetic maps of X-ray emission from the hot ISM \citep[see e.g.][]{2012ApJ...756L...3D,2015ApJ...800..102H,2015ApJ...813L..27P}, these are largely outside of the scope of our review, as are related efforts to model X-ray emission from the hot halo of our galaxy. Similarly, we will not discuss attempts to make large-scale synthetic images of ionized gas tracers such as H$\alpha$ \citep[e.g.][]{2017MNRAS.466.3293P}. Instead, our focus here is on efforts to model atomic and molecular emission from the colder and denser phases of the ISM.

Given the central role that molecular clouds play in the formation of stars, CO is clearly one of the key observational tracers that we are interested in modelling on galactic scales. We have already discussed a number of efforts to produce large-scale synthetic CO emission maps, in connection with efforts to understand the connection between $X_{\rm CO}$ and the galactic environment (see Section~\ref{ssec:traceh2}), and we will not repeat that discussion here. Instead, we will discuss a few additional studies that have made use of synthetic CO emission maps to address scientific questions not directly related to $X_{\rm CO}$.

\citet{2014MNRAS.441.1628S} simulated a large patch of a spiral galaxy at high resolution using the {\sc arepo} moving mesh code. They followed the chemical evolution of the gas using a highly simplified chemical network \citep{1997ApJ...482..796N,2007ApJ...659.1317G}. They then post-processed their results to produce synthetic CO emission maps using a curve-of-growth approximation similar to that in \citet{2011MNRAS.412..337G}. They show that molecular hydrogen is found in both the spiral arms and also in extended interarm filaments, which may be related to the ``spurs'' observed in real spiral galaxies \citep[see e.g.][]{2013ApJ...779...42S}. CO emission traces H$_{2}$ very well in the arms, but less well in the filaments, and in some cases structures that are coherent when looked at in H$_{2}$ column density break up into multiple clouds when looked at in CO emission. \citet{2014MNRAS.441.1628S} quantify the amount of CO-dark H$_{2}$ gas formed in their simulation, finding a fraction of around 40\% for typical Milky Way-type conditions. They also explore how this number varies as they change the gas surface density and the strength of the interstellar radiation field. Decreasing the surface density or increasing the radiation field strength both suppress CO formation and dramatically increase the CO-dark H$_{2}$ fraction. However, the CO distribution is far more sensitive to the surface density than to the radiation field strength: a factor of 2.5 change in the former has a large effect than a factor of 10 change in the latter.

\citet{2014MNRAS.444..919P} carried out SPH simulations of a model of the Milky Way galaxy using the {\sc phantom} SPH code \citep{2017arXiv170203930P}. The simulations were performed using a fixed stellar potential, and the chemical evolution of the gas was followed using a similar approach to that in \citet{2014MNRAS.441.1628S}. \citet{2014MNRAS.444..919P} considered a range of different values for many of the important parameters controlling the stellar potential (e.g.\ the bar pattern speed, the orientation of the bar with respect to the observer, the number and pitch angle of the spiral arms). The post-processed each of their simulations using {\sc torus} and produced synthetic $l$-$v$ diagrams that they compared with the observed CO $l$-$v$ diagram \citep{2001ApJ...547..792D}. Their goal was to constrain the details of the stellar potential by determining which set of parameters produces the best fit to the observations. However, they found that none of the combinations that they examined could simultaneously reproduce all of the features observed in the real $l$-$v$ diagram. \citet{2015MNRAS.449.3911P} followed up on this study by using a live stellar distribution in place of a fixed potential, and showed that in this case, a much better fit with the observations could be obtained.

\cite{2015MNRAS.447.2144D} used {\sc torus} to post-process several different SPH simulations of a Milky Way-like galaxy to produce synthetic CO and H$\,${\sc i} emission maps from the point of view of an observer situated at the same position as the Sun. In their simulations, they performed runs with and without stellar feedback and also varied the treatment of radiation shielding and the assumed surface density of the gas. They compared their synthetic maps with CO and H$\,${\sc i} surveys of the Galactic plane (CO: 
\citealt{1998ApJS..115..241H,2010ASPC..438...98M}; H$\,${\sc i}: \citealt{2003AJ....125.3145T}). They show that the no feedback run produces a very poor match to real observations: the scale height of the CO is far too small and the emission close to the midplane is far too bright (see also \citealt{2010MNRAS.407..405D}, who found similar results for H$\,${\sc i} in an earlier study using the same simulation). However, even the more realistic models do a fairly poor job of reproducing the real observations. They significantly under-produce H$_{2}$, and hence produce too much H$\,${\sc i} emission. They also produce CO distributions that are more compact and sparse than that observed in the real Milky Way.

\citet{2016MNRAS.458.3667D} also used {\sc torus} to post-process SPH simulations of a similar model galaxy, but in this case used the high resolution simulations of \citet{2015MNRAS.447.3390D} and explored the relationship between the clouds they identified in their synthetic CO emission map using the {\sc SCIMES} algorithm \citep{2015MNRAS.454.2067C} and the actual molecular structures present in their simulation. They found that CO emission was a good tracer of the densest molecular clouds, particularly if observed with high sensitivity and angular resolution, but tended to miss low density molecular structures. Consequently, large coherent structures in H$_{2}$ could often be broken up into collections of smaller clouds when observed in CO.

Finally, \citet{2017arXiv170703650S} have recently carried out simulations of the Milky Way's bar and Central Molecular Zone (CMZ) using {\sc arepo}, with the same simple chemical and thermal model as in \citet{2014MNRAS.441.1628S}. They show that the combined action of thermal instability and the ``wiggle'' instability \citep{2004MNRAS.349..270W,2014ApJ...789...68K,2017MNRAS.471.2932S}leads to the infall of gas onto the CMZ becoming asymmetric despite the symmetric potential\footnote{To be more precise: the infall rate averaged over tens of Myr remains symmetric, but the amount of infall occurring at any given instant becomes asymmetric.}, thereby providing a natural explanation for the observed asymmetry of dense gas in the CMZ \citep{1988ApJ...324..223B}. \citet{2017arXiv170703650S} also produce approximate CO $l$-$v$ diagrams using an optically thin approximation and show that these closely resemble real CO observations of the CMZ region.

\begin{figure}
    \centering
    \includegraphics[width=0.98\linewidth]{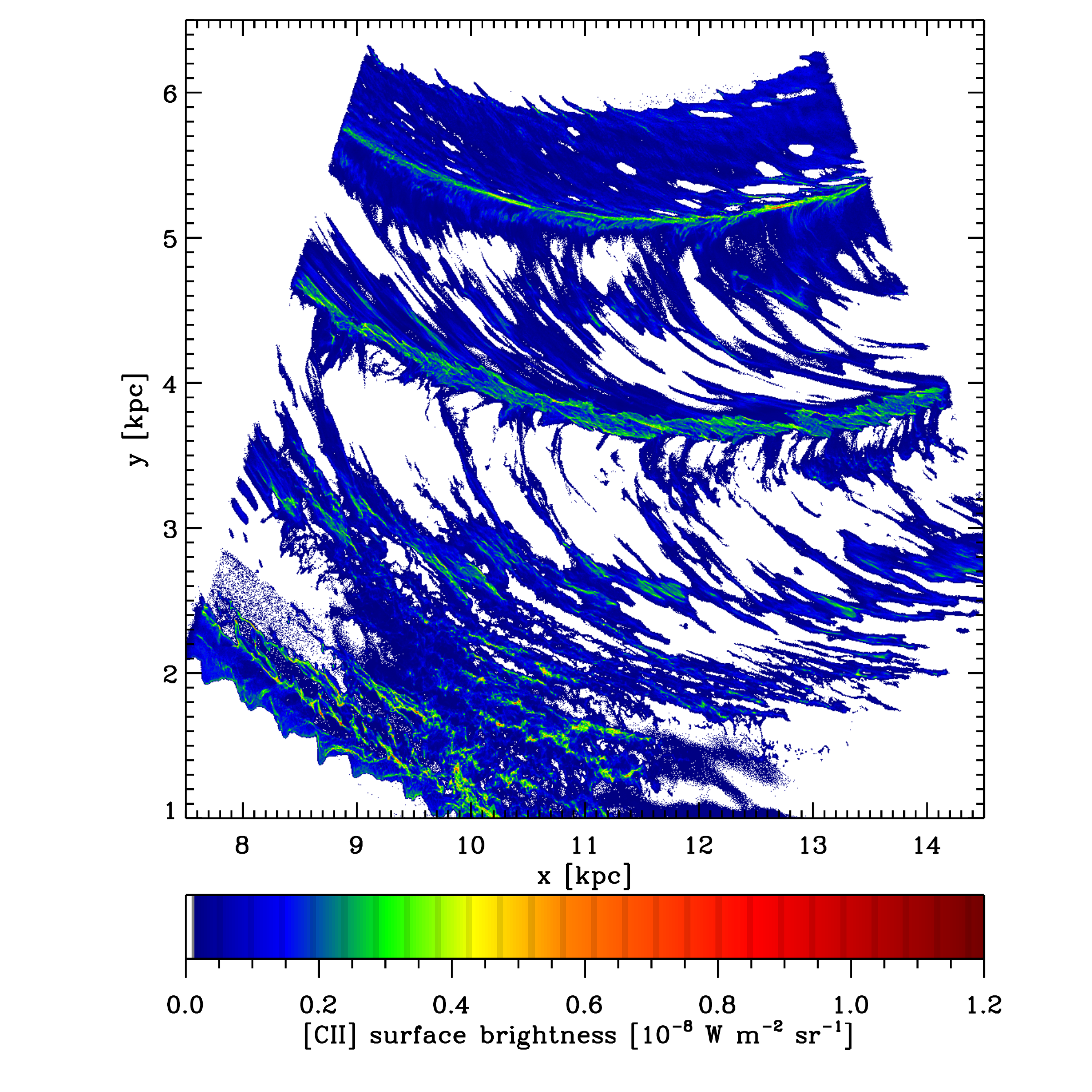}
    \caption{Synthetic [C\,{\sc ii}] surface brightness map based on the ``Milky Way'' simulation of \citet{2014MNRAS.441.1628S}, computed using the optically thin approximation described in \citet{2016MNRAS.462.3011G}. Regions shown in white have surface brightnesses $I_{\rm CII} \leq 10^{-10} \, {\rm W \, m^{-2} \, sr^{-1}}$. (This figure is Figure~10 of \citealt{2016MNRAS.462.3011G} and is used with permission of the authors).} 
 \label{fig:CII}
\end{figure}

Besides CO, another highly interesting tracer of the ISM on galactic scales is fine structure emission from ionized carbon. The [C\,{\sc ii}] 158~$\mu$m line is the brightest emission line observed in local star-forming galaxies \citep{2017ApJ...845...96C}, and has attracted considerable interest as a tracer of the star formation rate and a potential tracer of CO-dark molecular gas. A considerable amount of the observed [C{\sc ii}] emission originates from CO-dark molecular gas \citep[e.g.][]{2014A&A...561A.122L,2016A&A...593A..42T,2017ApJ...839..107P}. Recent simulations by \citet{2017MNRAS.464.3315A} using multi-phase 3D radiative transfer and Bayesian statistics, showed that $75\pm5\%$ of the C{\sc ii} emission in Galactic conditions arises from molecular regions. Two recent attempts have been made to produce synthetic [C\,{\sc ii}] emission maps based on large-scale simulations of the ISM. \citet{2015EAS....75..385F} post-processed simulations from the SILCC project \citep{2015MNRAS.454..238W}, using {\sc radmc-3d} to produce synthetic [C\,{\sc ii}] datacubes. They used these to study the source of the [C\,{\sc ii}] emission, finding that much of it was produced in molecular gas at the boundaries of dense molecular clouds. However, a follow-up analysis using much higher resolution simulations {found instead that most of the [C\,{\sc ii}] emission comes from regions dominated by atomic hydrogen \citep{Franeck2018}. This study also showed that very high resolution ($\Delta x \leq 0.1$~pc) is necessary to produce numerically converged maps of [C\,{\sc ii}] emission.}

\citet{2016MNRAS.462.3011G} post-processed the {\sc arepo} simulations of \citet{2014MNRAS.441.1628S} to produce synthetic [C\,{\sc ii}] and [O\,{\sc i}] emission maps using an optically thin approximation (see e.g.\ Figure~\ref{fig:CII}). As these simulations include no star formation, the resulting emission maps are lower limits on the emission that one would actually expect to measure, representing only the contribution from neutral gas far from regions of active star formation. \citet{2016MNRAS.462.3011G} investigated whether the [C\,{\sc ii}] and [O\,{\sc i}] lines were a good tracer of the H$_{2}$ column density, finding only a weak correlation between the [C\,{\sc ii}] integrated intensity and $N_{\rm H_{2}}$, and almost no correlation between the [O\,{\sc i}] integrated intensity and $N_{\rm H_{2}}$. They also examined whether these lines would be detectable in nearby galaxies using the upGREAT and FIFI-LS instruments on SOFIA, and concluded that at least some of the emission should be detectable given a reasonable time on source. 
  
In addition to these studies, there have also been a number of efforts to produce synthetic [C\,{\sc ii}] emission maps for high redshift galaxies 
\citep[e.g.][]{2015ApJ...813...36V,2017ApJ...846..105O}, but these are outside of the scope of this review and so we will not discuss them here.
  
Finally, there have been several efforts to produce synthetic H\,{\sc i} maps on $\sim$~kpc scales in simulations of entire galaxies or large portions of galaxies. An early effort in this direction was make by \citet{2000ApJ...540..797W}, who post-processed their 2D simulations of an LMC-type model galaxy to produce synthetic H\,{\sc i} images. They showed that large kpc-scale holes in the distribution of H$\,${\sc i} emission, similar to those observed in the LMC, were difficult to produce with supernova feedback but were plausibly a result of the non-linear evolution of the gas in the stellar potential. 

More recently, \citet{2010MNRAS.406.1460A,2012MNRAS.422..241A}, \citet{2010MNRAS.407..405D} and \citet{2015MNRAS.447.2144D} have all used the {\sc torus} radiative transfer code to produce H\,{\sc i} emission maps based on SPH simulations of disk galaxies carried out by Dobbs and collaborators. 
\citet{2010MNRAS.406.1460A} examine the morphology of the H\,{\sc i} emission as seen by an observer outside the galaxy and show that it agrees fairly well in a qualitative sense with observations of the grand design spiral M81. The other three studies examine the H\,{\sc i} distribution from the perspective of an observer within the disk. As already noted, \citet{2010MNRAS.407..405D} find that in the absence of stellar feedback, the scale height of the H\,{\sc i} is much too small, while \citet{2012MNRAS.422..241A} show that the inclusion of feedback produces a scale height in much better agreement with observations. Finally, the results of \citet{2015MNRAS.447.2144D} have already been discussed above, in connection with CO.

Synthetic H\,{\sc i} emission maps have also been produced by \citet{2014ApJ...786...64K}, based on the stratified box simulations of the ISM carried out by \citet{2013ApJ...776....1K}. These simulations model a $512 \times 512$~pc square patch of the ISM and account for the effects of galactic rotation, star formation, supernova feedback in the form of momentum input, and radiative heating and cooling. Unlike most of the SPH simulations discussed above, these stratified box simulations do not track the chemical evolution of the gas; instead, they assume when computing their H\,{\sc i} maps that all of the cold gas is in the form of atomic hydrogen. \citet{2014ApJ...786...64K} examine the distribution of spin temperatures and H\,{\sc i} opacities in their model in some detail and show that although H\,{\sc i} is optically thin along the majority of their lines-of-sight, there is a subset of lines-of-sight with H\,{\sc i} optical depths $\tau_{\rm HI} > 1$, with a few reaching values as high as $\tau_{\rm HI} \sim 10$. However, the absence of chemistry in their models means that they are likely over-estimating the H\,{\sc i} opacity along lines-of-sight that pass through dense gas clouds, as we would expect that in reality a significant fraction of this dense gas would actually be H$_{2}$. 

\subsection{PDRs}
PDR physics has already featured throughout this review. For example since the bulk of the ISM is in the PDR phase it is important for gas mass estimates (see section \ref{ssec:traceh2}). In this section we therefore focus our attention on some illustrative results that are applied to famous PDRs in Orion: the Orion Bar, Orion Veil and the Horsehead.

The Orion Bar and Veil are both located in close proximity to the Trapezium cluster. The Veil refers to the PDR in the foreground of the Trapezium stars, while the Bar is offset along the line of sight and is at least partially viewed more edge-on (discussed below). The Horsehead is a famous pillar in another part of the Orion molecular cloud complex, sculpted by the nearby $\sigma$ Ori stellar group. These systems are all interesting because they are close enough for as to study both the physical structure (density, temperature, magnetic field) and stratified chemical structure simultaneously.

There are many chemical objectives in the study of PDRs. For example, new detections of species and understanding whether the composition is dominated by pure gas-phase chemistry, or liberation of species from grains by processes such as thermal evaporation, photo-desorption or sputtering in shocks. Furthermore, PAHs are difficult to detect in high extinction zones, so gas-phase chemistry might yield an insight into the pre-destruction PAH properties of the medium. The proximity of the Orion PDRs to cold, dense gas elsewhere in the Orion cloud complex allows the two regimes to be easily compared. Moreover, as we will discuss further below, some PDR studies are also geared towards understanding the physical structure and evolution of PDRs, rather than their chemistry.

The Orion Bar PDR is one of the most studied individual astrophysical clouds to date and is almost certainly the most studied PDR. It is viewed approximately edge on and is relatively close at $\sim410$\,pc, making it the ``prototypical'' PDR. The list of papers computing theoretical emission measures with a focus on the gas properties is huge, including
    \cite{1985ApJ...291..722T}, \cite{1989ApJ...344..791B}, \cite{1995A&A...303..541J},
    \cite{2000ApJ...540..886Y}, \cite{2000A&A...364..301W}, \cite{2009ApJ...693..285P},
    \cite{2010A&A...518L.116H}, \cite{2011MNRAS.416.1546A}, \cite{2011A&A...530L..16G,2016Natur.537..207G,2017A&A...601L...9G},
    \cite{2012ApJ...752...26M}, \cite{2013A&A...550A..96N,2015A&A...578A.124N,2017A&A...599A..22N},
     \cite{2013A&A...560A..95V}, and
    \cite{2015A&A...575A..82C}.
    All of these studies are 1D and typically aim to reproduce the emission features of some subset of the large variety of atomic and molecular species detected towards the Orion Bar. Regretfully, we cannot discuss all of these studies here without making our extremely long review even longer, and so here we just focus on a few examples. 

    \cite{1985ApJ...291..747T,1985ApJ...291..722T} applied their influential PDR model to the PDR behind the Trapezium cluster (which includes the Orion Bar). They used observed and modelled C, O and CO lines to constrain the density and temperature in the PDR, assuming an incident UV field of $10^5$\,G$_0$ emitted by the Trapezium stars from a distance 0.1-0.2\,pc. The best agreement with observed emission in these lines implied that the PDR is fairly uniform and dense ($10^{5}$\,cm$^{-3}$) material at around 500\,K. 

    \cite{1995A&A...303..541J} observed 18 different molecules toward the Orion Bar, including CO, HCO$^+$, HCN, and SiO, which they interpreted using 1D and 2D numerical models. From their 1D calculations, they concluded that in order to match the column density and temperature structure, the zone of peak emission has to be more edge-on along the line of sight, with the other components being more face on. They verified this with 2D Monte Carlo radiative transfer calculations using the approach discussed in \citet{1994ASPC...58..421S,1996A&A...307..271S}.     
    
    \cite{2009ApJ...693..285P} use \textsc{cloudy} to compute the hydrostatic structure of the Orion Bar, accounting for threading by a magnetic field. They calculated the conditions across the entire H$^{+}$--H--H$_{2}$ interface. 
    The magnetic field can influence the PDR by changing the flux of cosmic rays (either decreasing it via exclusion, or channeling them along the field lines to increase the flux) and can also modify the hydrostatic structure if there is a significant component of the field oriented parallel to the bar, as this provides magnetic pressure support against compression. \cite{2009ApJ...693..285P} showed that magnetic pressure plays an important role in the structure of the H\,\textsc{i} region in the Bar. They arrived at their best fit models by comparing emission lines and line ratios from their models with those observed. 
    
    The potential importance of magnetic fields in PDRs was also highlighted observationally by \cite{2004ApJ...609..247A} and \cite{2016ApJ...825....2T}, who studied the Orion Veil. The Veil has 2 components, distinguished by their line-of-sight velocity in H\,\textsc{i}. \cite{2004ApJ...609..247A,2006ApJ...644..344A} used \textsc{cloudy} models to determine the density and temperature of the two components. They also included magnetic fields in the hydrostatic structure calculation, finding that energetically speaking, component A is dominated by the magnetic field. \cite{2016ApJ...819..136A} recently showed that H$_2$ emission from the Orion Veil is almost exclusively from component B, which interestingly is the closer of the two components to the Trapezium stars (and hence is more strongly irradiated). The reason for this is that component B is denser, meaning that the inner layers are well shielded. 

    To our knowledge, there has so far only been one attempt to model the Orion Bar in 3D, which was made recently by \cite{2017A&A...598A...2A}. Their approximate 3D calculation is built up from a series of 1D spherical models, with each sphere being isotropically irradiated by a local FUV flux. Each spherical clump can also have its own size and mass. They use this series of 1D models to build up a 3D picture of the PDR structure and produce integrated emission and synthetic spectral line profiles. They found that they required the dense clumps to be removed near the PDR surface (i.e.\ the medium must be smoother there) in order to fit the observed line-integrated intensities and spatial distribution of lines from CO, HCO$^+$ and [C\textsc{ii}].

    As mentioned above, a key question in the study of PDRs surrounds the importance of grain destruction in determining their chemical structure (through e.g.\ photo-desorption). Several studies of the Orion Bar and the Horsehead have attempted to address this question. For example, \cite{2009A&A...508..737P} used PDR models to argue that observed deuterium fractionation ratios in the Orion Bar, and in particular the DCN/HCN ratio, are consistent with gas-phase chemical calculations, and do \textit{not} require the liberation of species from grain surfaces (as is the case in hot cores). They use this to argue that sublimation is a secondary effect, at least in the parts of the system where DCN is abundant.  
    
\cite{2009A&A...494..977G} present new HCO and H$^{13}$CO$^{+}$ 
observations towards the Horsehead, which is comprised of a PDR and denser core. Interestingly they find high HCO abundances in the PDR, but not in the denser core. They suggested that this could either be because HCO is liberated in photo-desorption events, or if chemical pathways utilising carbon radicals such as CH$_2$ to produce HCO are important. \cite{2011A&A...534A..49G} observed H$_2$CO towards the Horsehead. They found that the observed H$_2$CO abundance in the core can be explained by their chemical models, but that in the PDR, the abundance is too high for their models to reproduce. They therefore concluded that photo-desorption is important in the PDR for setting the observed H$_2$CO abundance. \cite{2013A&A...557A.101G} also find that CH$_3$CN abundances in the PDR are too high to be explained by gas-phase chemical models, suggesting ice-mantle destruction is the explanation.

Overall then, there is significant evidence for chemistry in PDRs that is influenced by the partial destruction of grains. The dust/PAH properties of PDRs are therefore also of considerable interest, where observable. \cite{2012A&A...541A..19A} used the \textsc{DustEm} code \citep{2011A&A...525A.103C} to compute synthetic observations to compare with \textit{Herschel} and \textit{Spitzer}/IRAC continuum observations of the Orion Bar. They find that they require the PAH abundance to be lower than the canonical ISM value in order to agree with the observed 3.6\,$\mu$m flux, suggesting photon-induced destruction and/or coagulation as processes via which this can occur. This is a result that is nicely consistent with the findings of the gas-phase studies discussed earlier.

{To date there has only been limited study of PDRs in low metallicity regimes, which will of course be important for understanding PDRs over cosmic time and in different galactic environments. Chemical pathways, for example the importance of grain surface reactions, are expected to vary with metallicity. Furthermore, in a cosmological sense at earlier times when metallicity is lower, the star formation rate and hence cosmic ray ionisation rate is also higher.  \cite{2015MNRAS.450.4424B} probed the chemical evolution of single zone models, both with and without UV irradiation, finding that large OH abundances persist down to very low metallicities. They proposed that the cold dense ISM in the era of Population II star formation may have been atomic rather than molecular \citep[see also][]{2012MNRAS.426..377G,2012ApJ...759....9K}. Other examples of PDR studies in the low metallcity regime include \cite{1999ApJ...513..275B} and \cite{2012A&A...544A..84P}}. 

\section{\Large Part 4: Future developments}
\label{sec:future}

Having reviewed methods and applications of synthetic observations, we now turn our attention towards the future.

\subsection{Some astrophysical problems for synthetic observations to address}

We start our consideration of future developments by considering some examples of problems that synthetic observations will be key for addressing in the near future. 

\begin{enumerate}

    \item \textbf{Using position-position-velocity (PPV) data to interpret the 3D structure of clouds.} Tools such as \textsc{scouse} \citep{2016MNRAS.457.2675H} are allowing us to gain new, more intuitive insights into PPV maps. However, when we view a 3D distribution of gas on position-position-velocity axes, our brains naturally try and interpret the data as a spatial distribution. It is therefore
    perhaps unsurprising that in recent years there has been speculation on the link between PPV data and the 3D spatial (PPP) distribution of gas. Can PPV data give us an insight into the 3D spatial structure of gas? A good way to test this correlation is to apply algorithms for estimating 3D spatial structure to synthetic observations of systems where the 3D structure is actually known. \\
    
    \item \textbf{Properly interpreting line ratio measurements from unresolved systems} As we have already discussed in Section~\ref{sec:linerat}, one difficulty we face when using emission line ratios as diagnostics of the physical conditions in an observed gas distribution is that if the smallest scale we can resolve is larger than the characteristic scale on which the emission properties change, then we may end up
    averaging over different regions with very different line ratios. If so, then the line ratios that we measure may not correspond to those actually found in any part of the gas distribution, making any inferences we draw based on them highly suspect. Synthetic observations can help us to understand when this is likely to be a problem, and which tracers are most robust against this kind of confusion. \\

    \item{\textbf{Understanding the degeneracy between \\ non-equilibrium chemistry and initial conditions}. There are now a number of theoretical papers finding that the chemical composition of the ISM is out of equilibrium, and that equilibrium is not necessarily reached on a dynamically relevant timescale (see e.g.\ \citealt{2017arXiv170406487S} for a recent example). If this is the case, then the chemical composition of a cloud at the time it is observed is a function of its initial chemical composition, which is generally not known. This is potentially a serious problem if we want to use chemical-dynamical models to interpret observations. However, the impact of departures from chemical equilibrium on {\em observational} tracers of molecular clouds remains to be fully explored. In particular identifying the relative sensitivity of emission from different tracers to departures from equilibrium will be key. This point will require a combination of diverse observations, chemical-dynamical models and synthetic observations, making this one of the more challenging problems presented here.} \\ 
    
    \item \textbf{Measuring ionising photon escape fractions.} The radio continuum flux from H\,\textsc{ii} regions can be used to diagnose the exciting ionising flux. This diagnostic is often used to make an estimate of the properties of the stars responsible for exciting an H\,\textsc{ii} region. However, some of these photons -- perhaps a large fraction --  escape into the wider ISM, making such an estimate a lower limit. If one independently knew the spectral classification of the stars responsible for exciting an H\,\textsc{ii} region, this could be compared with the radio continuum flux to estimate the escape fraction, but this is a highly difficult and uncertain measurement. 
    
    \cite{2010ApJ...709..791B} concluded that H\,\textsc{ii} regions are more cylinder like than bubble like, since little/no molecular gas is detected in the fore/back ground of the bubble. They proposed that the size of bubbles are roughly the size of the thickness of the filamentary cloud in which they are embedded, since once the H\,\textsc{ii} region breaks out into the wider ISM, the expansion stalls. 
    
    Given the above two points, one would expect a relationship between the radio flux discrepancy and the fraction of the sky that the H\,\textsc{ii} region sees covered by molecular gas (which is a function of the H\,\textsc{ii} region/cloud size). Exploring this observationally will be very challenging. For example \citep{2016A&A...592A..47N} demonstrated that the uncertainties in spectral types dominates this kind of analysis at this stage. Synthetic observations may help us to deduce a method by which we can more robustly infer (or probe statistical correlations related to) the escape fraction of ionising photons.  \\
        
    \item{\textbf{Finding the best tracers of total gas mass in different environments}
    A key problem in modern astrophysics is calculating total gas masses from a limited number of tracers, such as e.g.\ CO or continuum emission. The emission observed from the tracer has to be translated into a total mass via an uncertain, and potentially variable conversion factor. As discussed in Section~\ref{ssec:traceh2}, there is a growing body of work searching for optimal tracers and trying to understand the limitations of routinely employed conversion factors (e.g.\ when is [CI] preferable to CO?). Synthetic observations, in addition to chemical evolutionary models, will be essential for determining  optimal tracers and their sensitivity to different environments.\\}
    
    \item \textbf{Understanding geometric and microphysical degeneracies} The modelling and interpretation of real 3D systems is typically carried out in terms of 1D models and column density estimates. Comparison with multidimensional models will be important for understanding and minimizing the uncertainties in these models and measurements. This is particularly important for systems such as PDRs where self-shielding effects are important and the chemical evolution is not purely determined by local conditions. {However, even with a 3D model it is still a comparison with a 2D projection, so degenerate distributions along the line of sight could match the observations well. Furthermore, there are degeneracies in the microphysics of models. For example SED fitting can be degeneratively sensitive to the grain size distribution, abundance and composition.  A thorough understanding of degeneracies in the modelling process, both from geometrical and microphysical (e.g.\ grain properties) effects will be important. }\\
    
    \item{\textbf{Robustly detecting fragmenting discs about massive YSOs}. As discussed in this review, massive YSOs are expected to continue accreting through a disc, which itself is likely to be massive enough to be gravitationally unstable. However, the discs themselves are proving to be difficult to detect, let alone any sub-structure within them. Synthetic observations will be required to robustly interpret the next generation of massive YSO observations, to possibly identify fragmentation and hence the production of further lower mass stars. \\}
    
    \item{\textbf{Testing more observational diagnostics} As we have discussed throughout this review, a number of observational diagnostics used to infer the conditions in (or classify) astrophysical systems have now been tested using synthetic observations. Doing so is key to getting a better handle on uncertainties. Moving forward it will be important to continue testing and improving those observational diagnostics that have not yet been assessed, as well as future observational techniques. \\}
    
    \item{\textbf{Exploring the possibility of using synthetic data to train machine learning tools.} Machine learning tools generally require training sets, based on which they go on to automatically perform some future operation or classification. Where real data is in limited supply, synthetic training sets could potentially fill the gap. Recently, \cite{2017arXiv171103480X} computed synthetic near infrared (Spitzer) continuum images based on MHD models of wind-blown bubbles and combined them with real data to provide a training set for the machine learning algorithm \textsc{brut}. They concluded that the additional synthetic training data significantly improved the automated detection of bubbles in the ``citizen science'' Milky Way Project\footnote{\url{https://www.zooniverse.org/projects/povich/milky-way-project}} dataset. This is potentially a powerful use for synthetic observations in the near future.\\}
    
    \item{\textbf{Improving measurements of the star formation rate}
    A number of different observational diagnostics exist of the star formation rate on galactic scales, ranging from measurements of the UV continuum emission from young stars to measurements of the total infrared flux \citep[see e.g.][for a recent review]{2013seg..book..419C}. However, many of these diagnostics are intended for use only on large scales, and break down when applied to small spatial scales, or when the star formation rate is highly variable in time. Synthetic observations will be invaluable for quantifying when and where different diagnostics become unreliable, and also for suggesting new possibilities that are more robust on small scales \citep[e.g.][]{2017ApJ...849....2K}.} \\
    
    \item {Understanding the influence of{ stellar feedback on the ISM, molecular cloud formation and evolution and star formation} 
    Stellar feedback has a significant effect on the surrounding ISM, blowing bubbles, dispersing gas and potentially triggering new waves of star formation. However regions where feedback operates are geometrically complex, viewed from a single perspective and feedback itself operates over of order the lifetime of a massive star. Furthermore when identifying triggered star formation, the ages of stars are inferred from their discs, which themselves are observationally complex. Synthetic observations will hence be key to continue to improve our understanding of the action of feedback on star formation and the ISM.  
    \\}
\end{enumerate}

\subsection{Upcoming technical developments}
\label{sec:technicaldevs}
Several areas in our technical implementation and processing of synthetic observations are in need of improvement, including the following:

\begin{enumerate}
    \item \textbf{Improved microphysical-dynamical coupling.} 
    There is a growing realisation that the emission properties of certain systems are strongly coupled to the dynamics, in a way that post-processing does not necessarily capture. For example, if the chemical evolution of a system is out of equilibrium, then post-processing of a static snapshot is inappropriate since the chemical composition (and hence emission) is sensitive to the history of the system.
    Furthermore, the chemistry itself may also be sensitive to physical processes such as radiation transport and magnetic fields. Understanding and accounting for this sensitivity where necessary will be an important continuing development in the future. \\
    
    \item \textbf{Reproducibility of complex calculations}
    
    The complexity of modern microphysics, radiative transfer and dynamics codes makes calculations difficult to reproduce. For example in the \cite{2007A&A...467..187R} PDR code comparison project a lot of effort was required just to get participants running the same calculations \citep[similar difficulties arose in the][radiative transfer benchmarking tests]{2002A&A...395..373V, Pinte:2009}. There is therefore the concern that published results, for example from complex chemical networks, might be difficult to reproduce. Making codes publicly available goes some way towards alleviating this issue and guarantees (with version control and proper referencing to the version used in a given paper) that a calculation can always be re-run. However this doesn't make a calculation easy to reproduce with a different code. Listing every detail of the scheme in every paper is inefficient, but standardising networks for certain problems, or hosting configurations online may help to alleviate the issue. \\
    
    \item \textbf{Better understanding of reduced networks.} 
   
    Reduced chemical networks are a powerful and widely used tool for including complex microphysics in dynamical applications. However,
    any reduced network has a limited range of validity, and there is a danger that when these networks are used by researchers without an astrochemical background, they may be used for applications outside of the range of conditions in which they are valid. It is important in the future to better quantify the range of environments in which particular reduced networks are valid, and to better communicate this information to
    non-astrochemists. In addition, the field would also benefit from exploration of the use of automated methods for deriving reduced networks ``on-the-fly'' during simulations (e.g.\ the Computational Singular Perturbation method of \citealt{CSP94}). \\

    \item{\textbf{Determination of the best analysis methods for different astrophysical scenarios and observing regimes}}
    
    In Section~\ref{sec:bestmethod}, we discussed the fact that many techniques exist for analysing (synthetic) observations, but that the suitability of these techniques for different applications and observing regimes is currently uncertain. We have a well-defined set of techniques and a large number of astrophysical applications and observing regimes. There is therefore an excellent opportunity here for groups to ``crank the handle'' and run large numbers of calculations in the spirit of \citet{2014ApJ...783...93Y}, \citet{2016ApJ...833..233B} and \citet{2017MNRAS.471.1506K} to assist in the endeavour of determining the applicability of different techniques and sensitivity to different parameters. \\ 
    
    \item \textbf{Better understanding of resolution requirements}

    Simulations that have enough resolution to represent the gas dynamics in a converged fashion nevertheless may have insufficient resolution to properly capture the microphysics important for determining the line or continuum emission, or may have the resolution in the wrong place (e.g.\ a model of an expanding H$\,${\sc ii} region may have good resolution in the dynamically important shock front at the edge of the ionized region, but poor resolution in the H-H$_2$ transition zone in the surrounding PDR). The resolution required to make robust synthetic observations of particular tracers is often unclear, and although there have been a few recent efforts to quantify it in particular cases \citep[e.g.][]{2017arXiv170406487S,2017ApJS..233....1K}, much more work remains to be done on this. \\
    
    \item \textbf{Incorporation of sub-grid scale effects}
    
    Sometimes, the resolution required to accurately model the full range of scales important for determining the emission from a particular tracer is so large that there is no reasonable prospect of achieving it in a 3D calculation in the forseeable future. In this case, the use of a sub-grid scale model to represent emission coming from unresolved scales is essential. A few efforts along these lines have already been made \citep[e.g.][]{2012ApJ...747..124F}, but considerable work remains to be done to develop and validate suitable sub-grid models. \\
    
    \item \textbf{Dialogue with astrochemists}
        
    As astrochemistry in dynamical applications becomes more commonplace, astrophysicists cannot rely purely on standard reduced networks.
    Ultimately, communication with astrochemists and proliferation of astrochemical knowledge throughout the community will also be important.
    We discuss this more in Section~\ref{sec:communication} below. \\
        
    \item \textbf{Future instrumentation}
    
    Tools like \textsc{casa} that incorporate interferometric effects in our synthetic observations permit us to model observations with telescopes such as ALMA. It is vital to ensure that similar bespoke modelling capabilities are available for upcoming facilities such as the Square Kilometer Array (SKA), the James Webb Space Telescope, and beyond.  \\
    
    \item \textbf{Understanding uncertainties and the sensitivity to initial conditions}
    
    {
    The chemical networks and radiative transfer required for both dynamical models and synthetic observations have intrinsic uncertainties that are generally not acknowledged (or are themselves unknown). For example, the uncertainty inherent to microphysical data (such as rate coefficients) and due to the sensitivty of the reaction network should be quantified \citep[see e.g.][for efforts in this regard]{2006A&A...451..551W, 2010A&A...517A..21W, 2017MolAs...6...22W}. Furthermore, astrochemistry is inherently non-linear, so the results can be very sensitive to the input parameters such as abundances \citep[see e.g.][]{2001A&A...370.1044S, 2013ChRv..113.8710A}. This also needs to be understood and quantified, or at least more generally acknowledged. }
    
\end{enumerate}

\subsection{Communication}
\label{sec:communication}

The field of synthetic observations is a highly interdisciplinary one, which can involve modelling gas dynamics, chemistry and thermal physics, radiation transport, the impact of the telescope and instrument on the observed radiation, plus whatever further analysis an observer would carry out on the data. Communication between experts in these different fields, and the avoidance of field-specific jargon, is therefore essential. In the remainder of this section, we highlight some 
positive existing examples and make some suggestions as to how the 
field of synthetic observations could benefit from active, constructive and well-documented communication. It is important to stress that our goal here is not to be prescriptive, but rather to start a discussion in the field over how we can improve the way in which we communicate with each other.

\subsubsection{Communication platforms}
\label{sec:communication_platform}

    Many fields of research could benefit from a portal where methods (observational and theoretical) can be informally presented and discussed, as well as their potential limitations and biases. {In addition, such a resource would be useful for distributing tools and results, discussing ideas and organising meetings. At present, some resources available are:}

\begin{itemize}
    \item [a] \textbf{Synthetic Observation Community Facebook group}\footnote{\url{https://www.facebook.com/groups/1787903124817908}}\\
This group offers a place for live discussion, advertisement of papers, requests for advice etc.\ One disadvantage is that joining it requires one to have a Facebook account, which some astronomers may prefer to avoid.
    \item [b] \textbf{LinkedIn group}\footnote{\url{https://www.linkedin.com/groups/8558988}}\\
    Similar to the above, but using the LinkedIn platform.
    \item [c] \textbf{Synthetic observations mailing list}\\
A mailing list for the discussion of topics relevant to the production and use of synthetic observations. To sign up, email \url{koepferl@usm.lmu.de} with the subject "synthetic observation community"
    \item [d] \textbf{Star formation newsletter}\\
Although this newsletter (mentioned already above) focuses on the topic of star formation, it does provide a location to promote or discuss synthetic observation papers that are relevant to this topic.
    \item [e] \textbf{Astrochemical newsletter}\\
    This newsletter\footnote{\url{http://acn.obs.u-bordeaux1.fr}}, edited by Pierre Gratier, Marcelino Agundez, Mathieu Bertin, Edith Fayolle and Valentine Wakelam, focuses on astrochemistry, but again provides a location to discuss synthetic observation papers closely related to this topic.
\end{itemize}

\subsubsection{Conferences and workshops}
\label{sec:communication_workshop}
To the best of our knowledge, there have been no large conferences dedicated to the topic of synthetic observations of star formation or the ISM. However, the field is starting to attract some attention in the form of splinter sessions at large general astronomy meetings such as the European Week of Astronomy and Space Science (EWASS) -- see e.g.\ the S3 session at the 2017 EWASS meeting\footnote{\url{http://eas.unige.ch/EWASS2017/session.jsp?id=S3}}. Meetings of this kind are important for communicating the power of synthetic observations to larger groups who may be unfamiliar with the methods/capabilities. They are also important for communicating key results, such as limitations of observational techniques that have been discovered using synthetic observations. 

Workshops are gatherings of smaller groups (e.g.\ fewer than 30) that often have a lot of time for open discussion or active work rather than being purely dedicated to talks (so-called ``unconference" sessions). The enhanced contact time between individuals and more engaging format make them an ideal means of driving forward real progress in new directions, particularly in an interdisciplinary sense. For example, workshops between astrochemists and those wanting to include chemistry in dynamical codes, or between real observers and those wanting to simulate observations from a given instrument, would be excellent examples of future small-scale meetings. {Useful venues for workshops include the Lorentz Centre\footnote{\url{http://www.lorentzcenter.nl/}} (Leiden, NL) and the RAS Burlington House\footnote{\url{https://www.ras.org.uk/about-the-ras/room-hire}} (London, UK), the latter of which is free for sessions organised by RAS fellows. }

As we have already mentioned, \textit{communication between astrophysicists and astrochemists} is hugely important for the field of synthetic observations. The chemical composition of the ISM plays a crucial role in determining many of its observable properties (e.g.\ the strength of the various molecular rotational and vibrational lines), and in many regimes is set by highly complicated, potentially vast, chemical networks. The importance of understanding how to best model this chemical complexity, and the extent to which we can trust the results, cannot be overstated. An excellent recent meeting in this regard was the intermediate-sized ``Current and future perspectives of chemical modelling in astrophysics'' meeting\footnote{\url{http://www.hs.uni-hamburg.de/astromodel2017/}} which brought together laboratory chemists, astrochemists and somewhat astrochemically green astrophysicists with an interest in applying chemical models to their simulations. This engagement drove an improved understanding of the different groups' interests, needs and the resources available (including their reliability). This valuable meeting is expected to continue every 2 years, with the next one likely to take place in Heidelberg in 2019. 

\textit{Communication amongst astrophysicists}, especially between theorists and observers, is obviously also vital. The 2016 workshop\footnote{\url{https://lorentzcenter.nl/lc/web/2016/802/info.php3?wsid=802&venue=Snellius}} at the Lorentz Center called ``Comparing Apples with Apples: Concordance Between Simulations and Observations of Star Formation" focused exactly on this topic and brought together experts in all of the different sub-fields (hydrodynamical modelling, chemistry, radiative transfer, observations, etc.) touch on by the field of synthetic observations. In that meeting, the different limitations of simulations, observations and synthetic observations were discussed together with the pitfalls which can arise when coming from different backgrounds. 
Ideally, meetings of this type will also be repeated on a 2 to 3 year cycle.

\subsubsection{Transparency and reproducibility}
\label{sec:communication_papers}
The peer review system remains the industry standard for quality control in astrophysics. However, as our capabilities become more and more sophisticated, the interdisciplinary nature of the field increases and it becomes more likely that subtle but essential aspects of a paper may be neglected. For example, code comparison projects have illustrated that even when multiple groups think they are running the same model, often they are not \citep{2007A&A...467..187R, 2015MNRAS.453.1324B}. This is because there are often a large number of assumed, internal, code parameters that are not specified in a model's description. Moving forward it will become more and more essential that the language and style of astrophysical publications adapt to facilitate transparency and reproducibility, and/or that supplementary information (such as the exact code version used and initial conditions for a model) are freely accessible. 

Below we list what we consider to be a set of best practices for accurately communicating the methods used when constructing synthetic observations. As with the rest of this section, we do not consider this to be the last word on the topic, but instead hope to provoke a wider discussion in the synthetic observation community. \\

\textbf{Best practices for synthetic observation papers}\\

When we write our papers, clarity in the following is essential:
\begin{itemize}
    \item The set of parameters used
    \item Full descriptions of units and how things are classified (e.g.\ the Habing field $G_0$, defined here in section \ref{sec:linerat}, is often unintuitive to those unfamiliar with it)
    \item Transparent assumptions (e.g.\ in a slab-symmetric PDR code, is the illumination isotropic or does it come from only one direction? This is not always clearly stated, but leads to clear differences in behaviour; see e.g.\ \citealt{2007A&A...467..187R})
    \item The methods used. In particular, producing synthetic observations that are bespoke to some instrument can be impossible if the reduction pipeline is unclear
    \item The precise version of any software used -- this is particularly important if the program is one that is being actively developed
    \item The statistical tools used
    \item The caveats, both in terms of the technical aspects of the model itself (e.g.\ limitations due to finite resolution or missing physics)
    and more broadly (i.e.\ astrophysically). 

\end{itemize}

Furthermore, in view of the interdisciplinary nature of the field of synthetic observations, we encourage authors to be more didactic than they would be if only addressing specialists in their particular part of the field.

\subsubsection{Making synthetic observations available}
\label{sec:communication_dataset}

Following some proprietory period almost all modern observations are stored in publicly accessible science archives. There is no reason that this could not also be the case for synthetic observations so that they can be downloaded and used by others. In Section \ref{sec:apps} we have already given some examples of the huge pool of existing synthetic observational data that would have lasting use for the community when interpreting future observations. Of course using archival synthetic data will likely not be as good a match to real data as would be possible with a bespoke model, but it makes such a comparison much more accessible. 

Unfortunately, making synthetic observations available in a useful fashion is not as simple as merely putting it on a webpage. One important issue that needs to be considered is how to best communicate the methods used to produce the data and any applicable caveats. That is, the associated metadata is going to be very important. 
For example, metadata is required to help interested parties to find the right data. As this review has hopefully made clear, synthetic observations of star formation and the ISM have been carried out on a very wide range of different scales, including many different types of physics, making it difficult to conceive of any single, simple organisational scheme. This is in contrast to the highly successful databases of simulation results made available by the teams responsible for large cosmological simulations such as Illustris \citep{2015MNRAS.447.2753T, 2015A&C....13...12N} or EAGLE \citep{2016A&C....15...72M}, as in this case much of the data can be organised in the form of a catalogue of dark matter haloes that lists their properties plus those of any associated gas, stars, etc. 
A further problem is that even if we had such a simple organisational scheme, much of the current work on synthetic observations is being carried out by relatively small groups, who do not necessarily have the resources to develop and maintain large simulation databases. Finally, in some cases there is also the issue of data volume. A large 3D simulation of the ISM that includes a magnetic field, radiation and non-equilibrium chemistry can easily produce hundreds of gigabytes or even multiple terabytes of data. Many research groups do not have the local computational infrastructure necessary for making this quantity of simulation data available over the Internet. 

Efforts are currently underway to tackle many of these issues, particularly within the framework of the International Virtual Observatory Alliance (IVOA); see e.g.\ the detailed list of resources they make available online at: \url{http://wiki.ivoa.net/twiki/bin/view/IVOA}. Here, we highlight just a few examples of useful resources:

\begin{itemize}
    \item \textbf{Interstellar Medium and Jets services} \\
    The Interstellar Medium and Jets services\footnote{\url{https://ism.obspm.fr}} are a set of online resources for modelling and interpreting observations of atomic and molecular line emission from the ISM and from protostellar jets and outflows. It is supported by the VO-Theory group at the Paris Astronomical Data Centre, and currently consists of three main resources: an online version of the Meudon PDR code\footnote{\url{https://ism.obspm.fr/?page_id=33}} \citep{2006ApJS..164..506L} plus a database of pre-computed models; an online version of the Paris-Durham shock code\footnote{\url{https://ism.obspm.fr/?page_id=151}}; and the StarFormat database\footnote{\url{http://starformat.obspm.fr/starformat/projects}}, which contains the results of detailed numerical simulations of molecular cloud formation and star formation. Data from these services is made available in a VO-compliant fashion, and indeed has helped to validate the VO-Theory standards developed by the IVOA.

    \item \textbf{CAMELOT} \\
    CAMELOT\footnote{\url{https://camelot-project.herokuapp.com}} -- the Cloud Archive for MEtadata, Library \& Online Toolkit -- is an attempt to provide a platform to simplify the comparison of simulations and observations of molecular clouds and star-forming regions by making it easier to find the appropriate dataset to compare to \citep{2016ascl.soft05006G}. The central idea is to bridge the divide between inherently 2D observational data and 3D simulations by cataloguing both types of data in terms of three key physical parameters than can be measured in both simulations and observations -- the surface density ($\Sigma$), the velocity dispersion ($\sigma$), and the radius ($R$). Data on these parameters for simulated or observed regions can be uploaded, queried or visualised, allowing simulations and observations located in a similar part of parameter space to be identified. It is important to note that CAMELOT itself does not host any of the raw data from simulations or observations, and hence it is essentially a specialized search engine rather than a fully-fledged data repository. 
 
    \item \textbf{Open-source data storage}\\
    As previously mentioned, one of the problems involved with making synthetic observation data generally accessible is the large size of some of the datasets involved. This can make it impractical to host the data locally, motivating the use of some kind of external repository. Notable examples in this context are Zenodo\footnote{\url{https://zenodo.org}}, dataverse\footnote{\url{https://dataverse.org/}} and yt-hub\footnote{\url{https://girder.hub.yt}}, which allow the upload of large astrophysical datasets and associated metadata. Zenodo is particularly interesting in that it generates a digital object identifier (DOI) for each upload, making it possible to cite the data directly in scientific publications. Zenodo also offers different levels of data access, ranging from fully public to more conservative configurations such as embargoed, restricted and closed access. This enables uses such as making the raw data available only to the referee while the paper describing it is undergoing review and then opening the access once the paper is accepted. {In principle Zenodo could conveniently host raw models right through to synthetic observations.}
\end{itemize}

{It is worth noting that using common file formats wherever possible (e.g. FITS or HDF5) makes for easier exchange and analysis of data, and hence postprocessing by some secondary tool. This will be important for any science archive.}


\subsubsection{Acknowledging critical supporting research }
\label{sec:communication_scientificResearch}

Throughout this review we have demonstrated the power of synthetic observations, but also the breadth of knowledge that is required to properly generate, manipulate and interpret them. The scientific paper that results in an improved understanding of some astrophysical system is standing on the shoulders of astrochemists (who produced the microphysical data) and software developers (who produced the tools used to visualise, process and produce the data, as well as databases in which to store it), amongst others. A problem in the current academic paradigm is the limited formal acknowledgement of these vital contributors to the process. For example, career progression depends on large numbers of papers and citations, which are not necessarily compatible with years spent developing a numerical tool. We do not have any easy solutions to this problem, but feel that it is important to stress that it is a problem, and moreover one that will only grow as the resources that we depend on in the field of synthetic observations become more complex and require more focused development.

\section{\Large Summary and conclusions}
\label{sec:summary}
We have attempted here to tackle the mammoth task of reviewing the production and use of synthetic observations in astrophysics. Given the enormous potential extent of such a review, we have limited our attention to problems regarding star formation and the ISM.

In this review we have defined synthetic observations as a prediction of the manner in which an astrophysical source will appear to an observer. We have discussed the many reasons for generating them, which include testing observational diagnostics, making more observationally meaningful predictions from numerical models, and making more robust interpretation of observations. 

We have reviewed the extensive variety of different methods used to generate synthetic observations in different regimes, including photoionized gas, PDRs and XDRs, cold molecular clouds, and astrophysical shocks.  We also included a discussion of instrumentational effects, for example capturing the reduced sensitivity to large scale structure inherent to high resolution (extended configuration) interferometry. Furthermore we discussed a large number of methods and tools for comparing synthetic and real observations, including their limitations (primarily coming from the lack of testing in their application and hence uncertainty in their suitability for various problems). 

The applications of synthetic observations to problems in star formation and the ISM are extremely broad. 
We have reviewed a number of such applications and hope to have illustrated the power and growing importance and popularity of synthetic observations in this field of research.

Finally, we have discussed some future developments that we expect to see in the field. These include a series of questions that synthetic observations are expected to be important for answering. We also discussed technical developments that will be important for driving the field forwards, including items such as improving the coupling between gas dynamics and chemical composition, creating and understanding robust reduced reaction networks, understanding the sensitivity of different statistical diagnostics  in different regimes, and ensuring reproducibility as calculations become more complex. We also see increased access to synthetic observations, for example in science archives analogous to real observational ones, as being a significant future step. As our calculations become more sophisticated, it will become harder (and hence more important) to ensure that they are transparent and reproducible. Since synthetic observations span such a range of expertise (from observational astronomy, through numerics to laboratory chemistry) communication, collaboration and proper acknowledgement of supporting research (e.g.\ software development and astrochemical research) will also be crucial.

Overall, this review aims to provide a useful summary of the field for both those who are new to the field of synthetic observations and looking to gain a foundation, as well as to those who are already applying them and looking to drive our capabilities and understanding forward. We also aim to promote interaction between distinct but complementary disciplines such as astrochemistry and astrophysics.  

\section*{Acknowledgements}
{The authors extend thanks to the two anonymous referees for taking the time and effort to give a thorough and valuable assessment of this substantial review. We also thank members of the community who contacted us with comments. }

This work greatly benefitted from the Lorentz Center meeting ``Comparing Apples with Apples: Concordance Between Simulations and Observations of Star Formation'', which took place in August 2016. In particular, many of our recommendations in Section~\ref{sec:communication} stem from discussions that took place during this meeting. 

TJH is funded by an Imperial College Junior Research Fellowship. SCOG acknowledges support from the Deutsche Forschungsgemeinschaft in the Collaborative Research Centre (SFB 881) ``The Milky Way System'' (subprojects B1, B2, and B8) and in the Priority Program SPP 1573 `\
`Physics of the Interstellar Medium'' (grant number GL 668/2-1). SCOG also acknowledges support from the European Research Council via the ERC Advanced Grant STARLIGHT (project number 339177). CMK acknowledges support from the Bayrischen Gleichstellungsf\"orderung (BGF).
 
The authors thank Jonathan Mackey, Stella Offner, Thomas Peters, John Ilee, Catherine Walsh, Alexander Tielens and Valentine Wakelam for useful discussions and comments on the manuscript. The authors also thank Barbara Ercolano, Jonathan Mackey, Katherine Johnston, Dominique Meyer, Rowan Smith, Hannah Dalgleish, Zhi-Yu Zhang, L\'aszl\'o Sz\"ucs, Will Henney, Jane Arthur and Alison Young for sharing figures with us.

\bibliographystyle{mnras}\biboptions{authoryear}

\bibliography{refs}

\label{lastpage}
\end{document}